\def\wt{\widetilde}
\newcommand{\Rho}{{\mbox{\sf P}}}
\newcommand{\sRho}{\scalebox{.7}{{\mbox{\sf P}}}}

\def\ts{\textstyle}
\def\w{{\hspace{.1mm}{\rm w}\hspace{.1mm}}}
\def\cal{\mathcal}

\def\be{\begin{equation}}
\def\ee{\end{equation}}
\def\beq{\begin{equation}}
\def\eeq{\end{equation}}
\def\bea{\begin{eqnarray}}
\def\eea{\end{eqnarray}} 

\def\eqn#1{(\ref{#1})}

\def\nn{\nonumber}

\documentclass[letterpaper, 12pt, oneside]{book}

    \usepackage[left=1.5in, right=1in, top=1in, bottom=1in, dvips, pdftex]{geometry}
    
    \usepackage{fancyhdr} 
    \usepackage{amsmath,amsthm, amsfonts,amssymb} 
    \usepackage{braket}
    \usepackage{graphics}
    \usepackage{color}
    \usepackage{psfrag} 
    \usepackage{epsfig} 
    \usepackage{setspace} 
 \usepackage{lscape}
\usepackage{multirow}
\usepackage{amssymb,amsmath}
\usepackage{amstext}
\usepackage{amscd}
\usepackage{slashed}

    \usepackage[linktocpage,bookmarksnumbered,
                colorlinks=true,urlcolor=blue,      
               citecolor=blue,linkcolor=blue,      
                pdftitle={--INSERT TITLE HERE--}, 
                pdfauthor={ABRAR SHAUKAT}, 
                pdfsubject={UC Davis Ph.D. Doctoral Thesis},%
                pdfkeywords={UC Davis Ph.D. Doctoral Thesis}]{hyperref}
    \usepackage{breakurl}

    \theoremstyle{definition}

    \theoremstyle{remark}

    \numberwithin{equation}{chapter}
    \numberwithin{figure}{chapter}

    \newcommand{\n}{\vspace{12pt}} 
      
    \newcommand{\newchapter}[3] 
    {                           
      \chapter[#2]{#3}
      \chaptermark{#1}
      \thispagestyle{myheadings}
    }

\begin{document}
    
    
     
 
    \pagenumbering{roman}
    \pagestyle{plain}

    %
    %

\singlespacing

~\vspace{-0.75in} 
\begin{center}

  \begin{large}
    {\bf Unit Invariance as a Unifying Principle of Physics}
  \end{large}\\\n
  By\\\n
  {\sc ABRAR SHAUKAT}\\
  B.S. (UNIVERSITY OF CALIFORNIA, DAVIS) [2004]\\[4mm]
  DISSERTATION\\\n
  Submitted in partial satisfaction of the requirements for the degree of\\\n
  DOCTOR OF PHILOSOPHY\\\n
  in\\\n
  PHYSICS\\\n
  in the\\\n
  OFFICE OF GRADUATE STUDIES\\\n
  of the\\\n
  UNIVERSITY OF CALIFORNIA\\\n
  DAVIS\\\n\n
  
  Approved:\\\n\n
  
  \rule{4in}{1pt}\\
  ~ANDREW WALDRON (Chair)\\\n\n
  
  \rule{4in}{1pt}\\
  ~STEVEN CARLIP \\\n\n
  
  \rule{4in}{1pt}\\
  ~NEMANJA KALOPER\\
  
  \vfill
  
  Committee in Charge\\
  ~2010
  
\end{center}
   \begin{quote}
 { \it If you would be a real seeker after truth, it is necessary that at least once in your life you doubt, as far as possible, all things. } 
\begin{center} Rene Descartes \end{center}
\end{quote}

\vspace{1in}

{\singlespacing
\begin{center}
{\bf Miracle} \\[4mm]
Air, wind, water, the sun \\
all miracle. \\[6mm]

The song of Red-Winged Blackbird \\
miracle. \\[6mm]

Flower of Blue Columbine\\
miracle. \\[6mm]

Come from nowhere\\
Going nowhere\\
You smile\\
miracle.\\[6mm]
{\it --Nanao Sakaki--}
\end{center}
}
\vspace{1in}


   ~\\[1in]
\centerline{To a plethora of phenomenon that awaits\\ explanation--not necessarily through science }
       
    %
    %
        

    ~\\[7.75in] 
\centerline{
  \copyright\ ABRAR SHAUKAT,
  2010. All rights reserved.
}


    %
    %
    

    \doublespacing

    %
    %
    
    \addcontentsline{toc}{chapter}{Table of Contents}
\begin{singlespacing}
  \tableofcontents
\end{singlespacing}

    %
    %
    
\phantomsection
\addcontentsline{toc}{chapter}{List of Figures}
\begin{singlespacing}
  \listoffigures
\end{singlespacing}

    %
    %
    
\phantomsection
\addcontentsline{toc}{chapter}{List of Tables}
\begin{singlespacing}
  \listoftables
\end{singlespacing}

    %
    %
    
    %
     \section*{Abstract} 
\addcontentsline{toc}{chapter}{Abstract}

A basic principle of physics is the freedom to locally choose any unit system when describing physical quantities.  Its implementation amounts to treating Weyl invariance as a fundamental symmetry of all physical theories.  In this thesis, we study the consequences of this ``unit invariance" principle and find that it is a unifying one.  Unit invariance is achieved by introducing a gauge field called the scale, designed to measure how unit systems vary from point to point.  In fact, by a uniform and simple Weyl invariant coupling of scale and matter fields, we unify massless, massive, and partially massless excitations.  
As a consequence, masses now dictate the response of physical quantities to changes of scale.  This response is calibrated by certain ``tractor Weyl weights".  Reality of these weights yield Breitenlohner-Freedman stability bounds in anti de Sitter spaces.  Another valuable outcome of our approach is a general mechanism for constructing conformally invariant theories.  In particular, we provide direct derivations of the novel Weyl invariant Deser--Nepomechie vector and spin three-half theories as well as new higher spin generalizations thereof.  To construct these theories, a ``tractor calculus" coming from conformal geometry is employed, which keeps manifest Weyl invariance at all stages.   In fact, our approach replaces the usual Riemannian geometry description of physics with a conformal geometry one.
Within the same framework, we also give a description of fermionic and interacting supersymmetric theories which again unifies massless and massive excitations.  


    \newpage
    
    %
    %


    \section*{Acknowledgments} 
\addcontentsline{toc}{chapter}{Acknowledgements}
First and foremost, I would like to thank my adviser, Andrew Waldron, for his constant guidance and support.  In an age where a healthy student to adviser interaction is perishing, he diligently and delightfully worked with me and kept me motivated throughout the last three years. Without his constant nudging, I wouldn't have been able to finish the work presented.  I would like to thank him for patiently teaching me the price of moving $\mu$ past $\nu$ and $\psi$ past $\chi$. His incisive and critical comments were essential to my scientific growth and significantly influenced my understanding of both math and physics--especially the Yang-Mills theories.  

Many thanks to Rod Gover for his collaborations with me, for explaining the subtleties in tractor calculus,  and for always lending a helping hand.  I am grateful to Professor Dmitry Fuchs for generously helping me understand complicated topics in differential geometry.  I would also like to thank both David and Karl for engaging me in active discussions that lead me to understand the mathematical intricacies hidden behind physics equations.  I am indebted to Derek Wise for explaining Klein geometry, homogeneous spaces, and Palitini's formalism. 

I would like to take this opportunity to express my gratitude to Steve Carlip for polishing my presentation skills and for claryfying many topics of quantum gravity in the GR meetings.  Through his critique of students' presentations including my own, I learned the art of scientific presentations: spend more time motivating the topic than discussing equations!!  I am also grateful to Nemanja Kaloper for providing me an opportunity to give a talk in the Joint Theory Seminar and for agreeing to be on the thesis committee on short notice. 

I would like to thank my mother for her infinite loving support of my passions; my deepest gratitude to her for instilling the seeds of new curiosities in me and then watering, nurturing, and caressing them.  Many thanks to my father for spending valuable time with me and patiently answering my questions--especially those silly questions that mark the onset of childhood.  I would also like to thank my wife for providing me uninterrupted stretches of time to work on my thesis. 

I am indebted to my friends who not only provided me the emotional support but also the much needed distractions beneficial for the pursuit of science.  Their unabated support throughout these years have been invaluable.

    \newpage
       
    %
    %

    \pagestyle{fancy}
    \pagenumbering{arabic}

    %
    %
    
    \chapter{Introduction}
\begin{quote}
{\it It is increasingly clear that the symmetry group of nature is the deepest thing that we understand about nature today.}
\flushright{Steven Weinberg}
\end{quote}

Symmetry has played a vital role in deepening our understanding of physics.  Continuous symmetries explain the emergence of conserved quantities
and conservation laws.  The prediction of the existence of new particles and even antiparticles were a result of the far reaching consequences of symmetries.  Symmetry arguments determine the allowed possibilities for particle decays. In fact, symmetry has been and can be used as a guiding principle to develop physical models.  Symmetry, therefore, is indispensable to physics and dictates that nature at its fundamental level is simple and beautiful.

Symmetry can be of two types: global or local.  Global symmetries have constant parameters, while the parameters of local symmetries depend on space-time points.  Local symmetries are called gauge invariances and generically imply constraints.  Hence, a theory with manifest gauge invariance contains redundant degrees of freedom.  For instance, consider electromagnetism in four dimensions where a four component  photon field~$A_{\mu}$ reduces to two physical degrees of freedom upon fixing completely a $U(1)$ gauge.  Although this two component photon field suffices to explain most of the experimental data, its Lorentz invariance is not manifest.  Manifest Lorentz invariance is achieved by four component photon field.  Hence, a four component photon field captures the symmetry and locality principles of physics, even though a two component photon field already describes the physically observed degrees of freedom.


Another example of a theory with manifest local symmetry is the theory of gravity with general coordinate invariance. Although gravity can be written in terms of a trace-free, transverse, and spatially symmetric tensor with two independent components $g_{ij}{}^{TT}$, gravity, in this formalism, is not manifestly diffeomorphism invariant.  Eight additional redundant degrees of freedom are required to write it in a manifestly diffeomorphism invariant manner, which amounts to using a ten component metric tensor $g_{\mu\nu}$. Mathematically, manifest diffeomorphism invariance is achieved by using tensor calculus on Riemannian manifolds.  Upon coordinate transformations, tensors transform in a way encapsulating manifest diffeomorphism invariance. 

Similarly, theories can be constructed to make other local symmetries manifest.  In this thesis, our primary focus is a local scale symmetry reflecting the freedom to locally choose a unit system.
Instead of writing physics in a preferred unit system, we can write it in a manifestly locally ``unit invariant" way.  In particular, we survey all locally unit invariant theories and study their physical consequences.  This allows us to construct massless, massive, and even partially massless theories using a single geometrical framework.  It is accomplished by using mathematical machinery called tractor calculus, which was developed by conformal geometers to study conformal geometry ~\cite{ Eastwood:1987ki, Bailey:1990qn, MR1463509, Eastwood:1997}.  Tractors are to unit invariance what tensors are to diffeomorphism invariance.  In $d$-dimensions, tractors require $d+2$ components to keep manifest unit invariance.  The idea of tractor calculus is to arrange fields as $SO(d,2)$ multiplets transforming under certain conformal group gauge transformations  in contrast to arranging them as $SO(d-1,1)$ multiplets with (local) Lorentz group in tensor calculus. 

Conformal geometry is intimately related to what is often called, in physics, Weyl invariance.  A theory is Weyl invariant if the action does not change when the underlying metric transforms as
\be
g_{\mu\nu} \mapsto \Omega^2 g_{\mu\nu} \, .
\ee
Here one may in principle also allow the fields to transform as any function of $\Omega$, $g$, and $\Phi$:
\be
\Phi \mapsto f(\Omega, g_{\mu\nu}, \Phi_i)  \, .
\ee
This Weyl invariance is the harbinger of the local unit invariance which we discuss in the next section. 


\section{Rigid and Local Unit Invariance}

Classically, all physical theories respect a rigid scale invariance, the fundamental symmetry in physics that prevents us from adding two physical quantities with different dimensions. For example, we can not add lengths to areas because they transform differently under this symmetry. The scaling properties of physical quantities are encoded in their weights (which label representations of $SO(1,1)$--the non-compact dilation group.)  Under this rigid scale or rigid unit invariance symmetry, the fields and the dimensionful couplings of the theory transform as
\be
\Phi_i \mapsto \Omega^{w_i} \Phi_i\, ,  \qquad  \qquad \lambda_{\alpha} \mapsto \Omega^{w_{\alpha}}\lambda_{\alpha} \, ,
\ee
where $\Omega$ is the rigid parameter, and $w$ is the weight of the object.  Then, in an action principle, rigid unit invariance  requires
\be
 S[\Phi_i;\lambda_\alpha]=S[\Omega^{w_i}\Phi_i;\Omega^{w_\alpha}\lambda_\alpha]\, .\label{units}
\ee
Since the couplings of the theory can be rendered dimensionless upon multiplying by an appropriate power of~$\kappa (\sqrt{8 \pi G}c^{-2})$, only this single dimensionful coupling is needed.

The equation \eqn{units} simply says that physics is independent of the global choice of unit system.  It amounts to the statement that all physical observables are rescaled by the same amount throughout space-time. But what happens if the physical quantities are rescaled by different amounts depending on their space-time location? In other words, should  physics  be independent of the  local choice of unit system as well? The answer is clearly in the affirmative--at least classically. However,  physics does not seem to allow for this freedom.  Mathematically, Weyl invariance seems not to be a manifest symmetry of nature.   But neither are $U(1)$ or general coordinate invariance  or other symmetries, if one picks a particular gauge.   Perhaps physical theories have been written in a special Weyl gauge thus breaking the Weyl symmetry! In fact we will show that theories with dimensionful couplings are the gauge fixed versions of certain Weyl invariant ones. 
 
\begin{figure}[h]
\centering
\includegraphics[width=140mm, height=70mm]{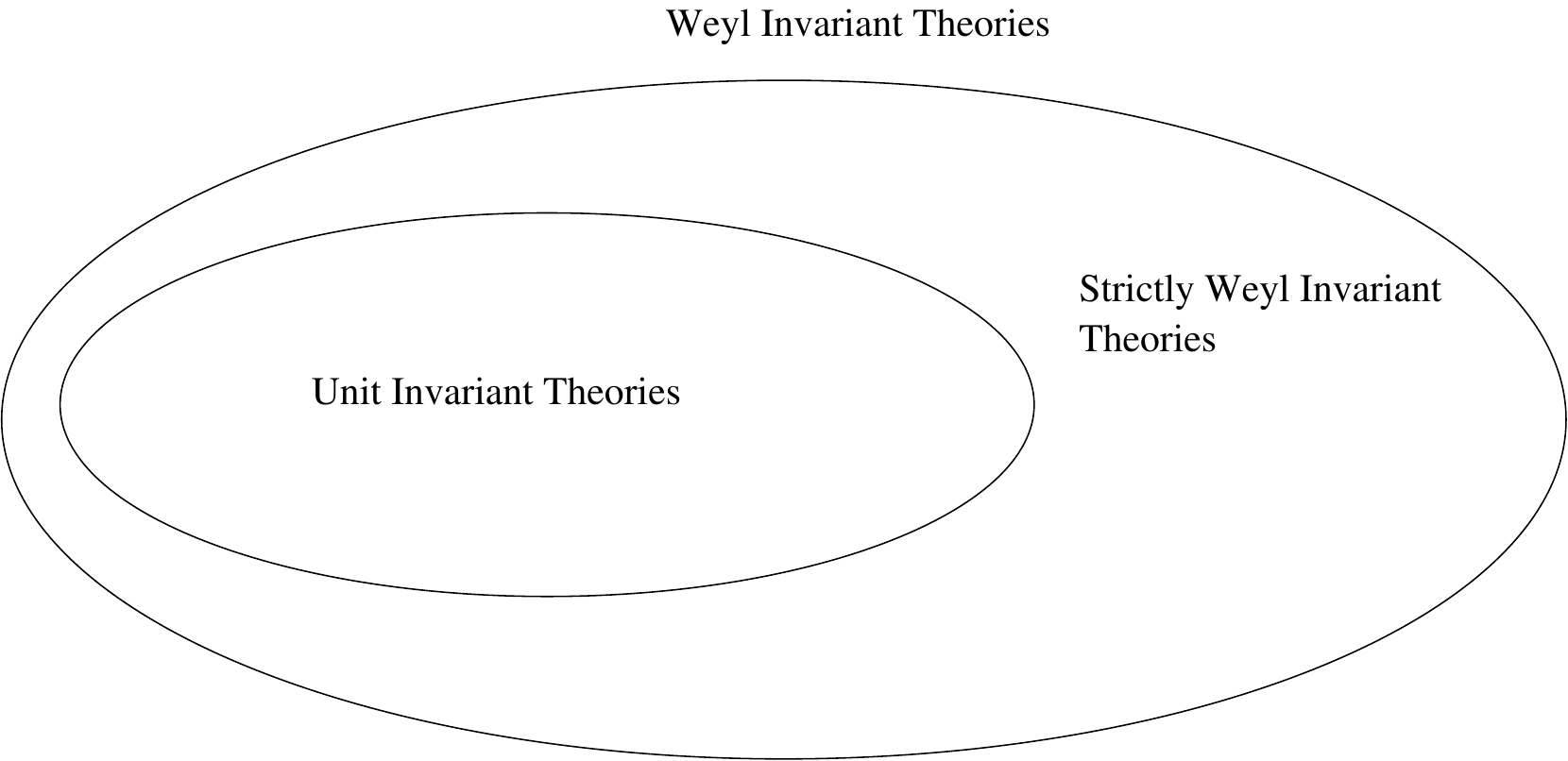}
\caption{Two types of Weyl invariant theories: unit invariant and strictly Weyl invariant theories.}
\label{strict_Weyl}
\end{figure}

The main idea of the thesis is to restore Weyl symmetry  by introducing a non-dynamical gauge field, $\sigma$.  
 We call $\sigma$ the scale, in concordance with the mathematics literature and also 
because of its geometric interpretation, which is simply as a local Newton's ``constant'' encoding how the choice of unit system varies over space and time.  
A theory is (locally)unit invariant if the action does not change when the underlying metric, the scale, and the fields all transform as
\bea
g_{\mu\nu} &\mapsto& \Omega^2 g_{\mu\nu} \, \\[2mm]
\sigma \mapsto\,  \Omega \, \sigma\,, &\qquad& \Phi \mapsto f(\Omega, g_{\mu\nu}, \Phi_i)  \, .
\eea
Here, one has to allow the fields to transform as any function of $\Omega$, $g$, and $\Phi$ such that the action is invariant.  If no such function exists, then the 
theory is not unit invariant.  Notice that unit invariance is a special case of Weyl invariance.  There is yet another kind of Weyl invariance which  we call ``strict Weyl invariance."
The theories that do not depend on the scale $\sigma$ for their Weyl invariance are called strictly Weyl invariant theories, which is graphically depicted in Figure~\ref{strict_Weyl}.
 In the physics literature, the gauge field $\sigma$ is also called a Weyl compensator or dilaton and was first employed 
by Weyl himself and later by Deser and Zumino in a manner slightly different than Weyl's but close to ours.~\cite{Zumino:1970tu, Deser:1970hs}.

\section{Coupling to Scale}
\label{Compensator}
Typically, the theories of physics do {\it not} respect Weyl invariance.  Nevertheless, there are some Yang-Mills theories that are Weyl invariant, notably the conformal gravity and  supergravity theories of~\cite{Kaku:1977pa} which are based on Yang--Mills theories of the conformal, or superconformal groups.  In Chapter~\ref{TractorConnection}, we show that there is an intimate relation between the ``gauging of space-time algebras'' methodology of~\cite{Kaku:1977pa,Kaku:1977rk,Townsend:1979ki} and the tractor calculus techniques that are presented in the thesis.  

Other non-Weyl invariant theories can be turned into  Weyl invariant ones by using Weyl compensators~\cite{Zumino:1970tu,Deser:1970hs}, because the scale, $\sigma$, transforms in a way to cancel the terms left over from the metric rescaling.  Choosing a particular $\sigma$ then amounts to picking a gauge in which the equations are simpler but not Weyl invariant.  This choice of gauge\footnote{ It is worth nothing that, for Weyl symmetry, we can always gauge away $\sigma$ locally without breaking the locality of the theory--a manuever that is not always possible for Lorentz symmetry.} is similar to the one made in electromagnetism when a Coulomb gauge is picked to simplify the equations at the cost of a broken $U(1)$ invariance. 

Restoring symmetry is more subtle than breaking it by a gauge choice! As discussed earlier, it entails an introduction of new redundant degrees of freedom.  In this section, we discuss how Weyl compensating method naturally couples an extra scalar field $\sigma$ to restore Weyl invariance.  To be more precise, it is the unit invariance that $\sigma$ restores--the special case of Weyl invariance.  As a warm up, we will recast gravity in a Weyl compensated way followed by a more involved example of a Weyl compensated scalar field theory.

\subsection{Weyl Compensated Gravity} 
\label{gravity}
Consider  the Einstein-Hilbert action  
\be
S_{\rm EH}(g_{\mu\nu})=-\frac1{2\kappa^2}\int \! d^dx\,\sqrt{-g} \, R \, \label{EH} ,
\ee
with coupling $\kappa$.  While it is clearly diffeomorphism invariant,  it is not  Weyl invariant.
But it can be turned into a Weyl invariant one via coupling to scale.  Instead of working with a particular metric and coupling $(g, \kappa)$, we work with a double equivalence class of metrics and couplings
\be
[ g_{\mu\nu}, \sigma] = [\Omega^2 g_{\mu\nu}, \Omega \sigma] \, .
\ee  In the original action, we replace 
$g_{\mu\nu} \mapsto \sigma^{-2} g_{\mu\nu}$ and remove the coupling $\kappa$ producing a Weyl compensated action
\be
S(g_{\mu\nu},\sigma) = \kappa^2 S_{\rm EH}(\sigma^{-2}g_{\mu\nu})= -\frac{1}{2}\int d^dx\, \sqrt{-g}\, \sigma^{-d}\Big(\,(d-2)(d-1)[\nabla_\mu\sigma]^2+R \, \sigma^2\Big)\, . \label{conf_inv_act}
\ee
This in fact, up to a simple field redefinition, is the standard action for a conformally improved scalar field. Notice, it is Weyl invariant only when both the metric $g_{\mu\nu}$ {\it and}  the scale  $\sigma$ transform 
\be
\label{Wtrans}
g_{\mu\nu} \mapsto \Omega^2 g_{\mu\nu}\, ,\qquad 
\sigma \mapsto \Omega \sigma  \,   .
\ee
Hence, the {\it r\^ole} of the scale or ``Weyl compensator/dilaton" $\sigma$ is to guarantee Weyl or unit invariance.  It can be checked explicitly that we recover~\eqn{EH} upon identifying the dimension one field $\sigma = \kappa^{\frac{2}{d-2}}$ with a constant coupling. This can always be achieved for it amounts to picking a gauge for the transformation~\eqn{Wtrans}.

There is, however, an alternative trick to represent~\eqn{conf_inv_act} in a manifestly Weyl invariant way.  From $\sigma$ and the metric, we first build a new $(d+2)$-dimensional vector~$I^M$
\be
I^M = \ \begin{pmatrix}
\sigma \\[1mm] \nabla^m \sigma \\[1mm] -\frac 1 d [\Delta +\frac{R}{2(d-1)}]\sigma\, \end{pmatrix} \, ,
\ee
which, we call the scale tractor.  Under Weyl transformations~\eqn{Wtrans}, the scale tractor transforms as 
\be
I^M \mapsto   U^M{}_N I^N \, , \label{tractor}
\ee 
where the $SO(d,2)$ matrix $U$ is given by 
\be U=
\begin{pmatrix}
\Omega&0&\;0\;\\[2mm]
\Upsilon^m&\delta^m_n&\;0\;\\[2mm]
-\frac12\Omega^{-1}\,\Upsilon_r\Upsilon^r&-\Omega^{-1}\Upsilon_n&\Omega^{-1}
\end{pmatrix}\, , \qquad \Upsilon_\mu = \Omega^{-1}\, \partial_\mu \Omega\, .
\label{Um}
\ee
Parabolic $SO(d,2)$ transformations of this special form will be called ``tractor gauge transformations.'' 
Objects that transform like~\eqn{tractor} are called weight zero tractors and the scale tractor $I^M$ is a special case of this. 
Miraculously, the action~\eqn{EH} can be recasted in terms of the scale tractor, which takes the manifestly unit invariant form:
\be
S(g_{\mu\nu},\sigma)=\frac{d(d-1)}2\int \frac{\sqrt{-g}}{\sigma^d}\,  I^M \eta_{MN} I^N \label{Isq} ,
\ee
where $\eta_{MN}$ is the block off-diagonal, $SO(d,2)$-invariant, tractor metric 
\be
\eta_{MN}=\begin{pmatrix}0&0&1\\0&\eta_{mn}&0\\1&0&0\end{pmatrix}\,
.\label{eta} \ee 
The scale tractor is doubly special:  It introduces the Weyl compensator or a ``scale'' $\sigma$ in the theory and in some sense controls the breaking of unit invariance. Holding the scale constant both picks a 
metric (see Section~\ref{Choice_Scale}) and yields the dimensionful coupling~$\kappa$ which allows dimensionful masses to be calibrated to dimensionless Weyl weights.

\subsection{Unit Invariant Scalar Field}
\label{Compensated_Scalar}
Having presented the manifestly Weyl invariant, tractor, formulation of gravity, we now add matter fields and first focus on a
single scalar field~$\varphi$. The standard ``massless'' scalar field action 
\be
S=\--\frac12 \int \sqrt{-g}\,\nabla_\mu \varphi  \, g^{\mu\nu} \,  \nabla_\nu \varphi 
\ee
is not Weyl invariant but can easily be reformulated Weyl invariantly using the scale $\sigma$. From $\sigma$ we build the one-form
\be\nn
b=\sigma^{-1}d\sigma\, ,
\ee
which under Weyl transformations changes as
\be\nn
b\mapsto b+\Upsilon\,,
\ee
where $\Upsilon = \Omega^{-1} d\Omega$.
Assigning the weight $w$ to the scalar~$\varphi$ 
\be\nn
\varphi\mapsto \Omega^w \varphi\, ,
\ee
then the  combination $\wt \nabla_\mu = \nabla_\mu - w b_\mu$ acting on $\varphi$ transforms covariantly
\be\nn
\wt \nabla_\mu \varphi\mapsto \Omega^w \wt \nabla_\mu \varphi\, .
\ee 
Hence we find an equivalent, but manifestly Weyl invariant, action principle
\be\label{trivial}
S=-\frac12\int  \frac{\sqrt{-g}}{\sigma^{d+2w-2}} \ \wt \nabla_\mu\varphi \, g^{\mu\nu}\,  \wt\nabla_\nu \varphi\, .
\ee
Of course, this is just the result of the ``compensating mechanism'', whereby, for any action involving a set of  
fields $\{\Phi_\alpha\}$ and their derivatives, replacing $\Phi_\alpha\mapsto \Phi_\alpha/\sigma^{w_\alpha}$
and $g_{\mu\nu}\mapsto \sigma^{-2}g_{\mu\nu}$ yields an equivalent Weyl (unit) invariant action.  Again, there is a way to rewrite the action
in a manifestly unit invariant way by introducing a new tractor operator called the Thomas-$D$ operator.  We will provide that reformulation in Section \ref{Scalar}.

The set of Weyl compensated flat theories that are unit invariant by virtue of the above Weyl compensator trick, does {\it not} map out the entire space of possible scalar
theories, even at the level of those quadratic in derivatives and fields.  In this thesis, we will use tractor calculus to find unit invariant theories that 
are not captured by the naive Weyl compensation mechanism in addition to the ones that are.

\section{General Organization} 
The next chapter starts by providing physical and mathematical motivation for studying conformal manifolds.  The rest of the chapter is devoted to mathematical preliminaries providing a generic method to describe a conformal manifold with a concrete example of a conformal sphere. We establish a notion of conformal flatness, review the conformal group, and discuss its relation to the Lorentz group.  Lastly, we develop a calculus on conformal manifolds called tractor calculus.

In Chapter~\ref{TractorConnection} we treat both hyperbolic and conformal spaces as homogeneous spaces and demand that the Yang-Mills
transformations coincide with the geometrical ones.  This leads to the derivation of the Levi-Civita connection and the
tractor connection which provides a notion of parallelism for tractors.  The reader can safely skip this chapter if not interested 
in the detailed derivation of the tractor connection and tractor gauge transformations. 

In Chapter~\ref{Bosonic}, we utilize the tractor calculus developed in the previous chapter to construct bosonic theories.  The theories include scalar, vector, and spin-two theories with masses
related to their Weyl weights.   At special weights, these theories reduce to the conformally improved scalar, the conformally invariant spin-one theory of Deser-Nepomechie, partially massless theories, and a new conformally invariant spin-two theory. We recover the Breitenlohner-Freedman bounds in AdS.  General higher
spin theories are also considered and described in terms of tractors.  

In Chapter~\ref{Fermionic}, we review the theory of tractor spinors in order to describe fermions and supersymmetry.  A tractor spinor, which is of fundamental importance for the construction
of Fermi theories is defined and its tractor transformations given.  Using the tractor spinor setup, we show the Weyl invariance of the massless Dirac equation in any dimensions.
Moreover, we construct spin one-half Dirac equation as well as the spin three-half Rarita Schwinger equation. 

Chapter~\ref{SUSY} contains all elements of the previous chapters as we write a supersymmetric theory involving both bosons and fermions.
Our starting point is the massive tractor supersymmetric theory in AdS.  Next, we add interactions and write an interacting Wess-Zumino model 
with arbitrary interaction terms. 

The last chapter is devoted to conclusion with future outlook.   We give reasons that Weyl anomalies can be resolved or at least better understood using our tractor techniques and conformal geometry.    We further speculate that physics with two times can be neatly recasted in terms of tractors with a deeper insight into the role and meaning of second time dimension. 


We end the thesis with an extensive set of appendices that include a compendium of Weyl transformations, tractor component expressions, and tractor identities.  Doubled reduction techniques useful for producing a $d$-dimensional massless theory from a $d+1$-dimensional massive one is also included as an appendix.  These appendices should serve as a toolbox for a reader interested in performing calculations pertaining to local scale transformations.



    \chapter{Conformal Geometry and Tractor Calculus}
\label{Conformal}

\begin{figure}[h]
\centering
\includegraphics[width=120mm, height=120mm]{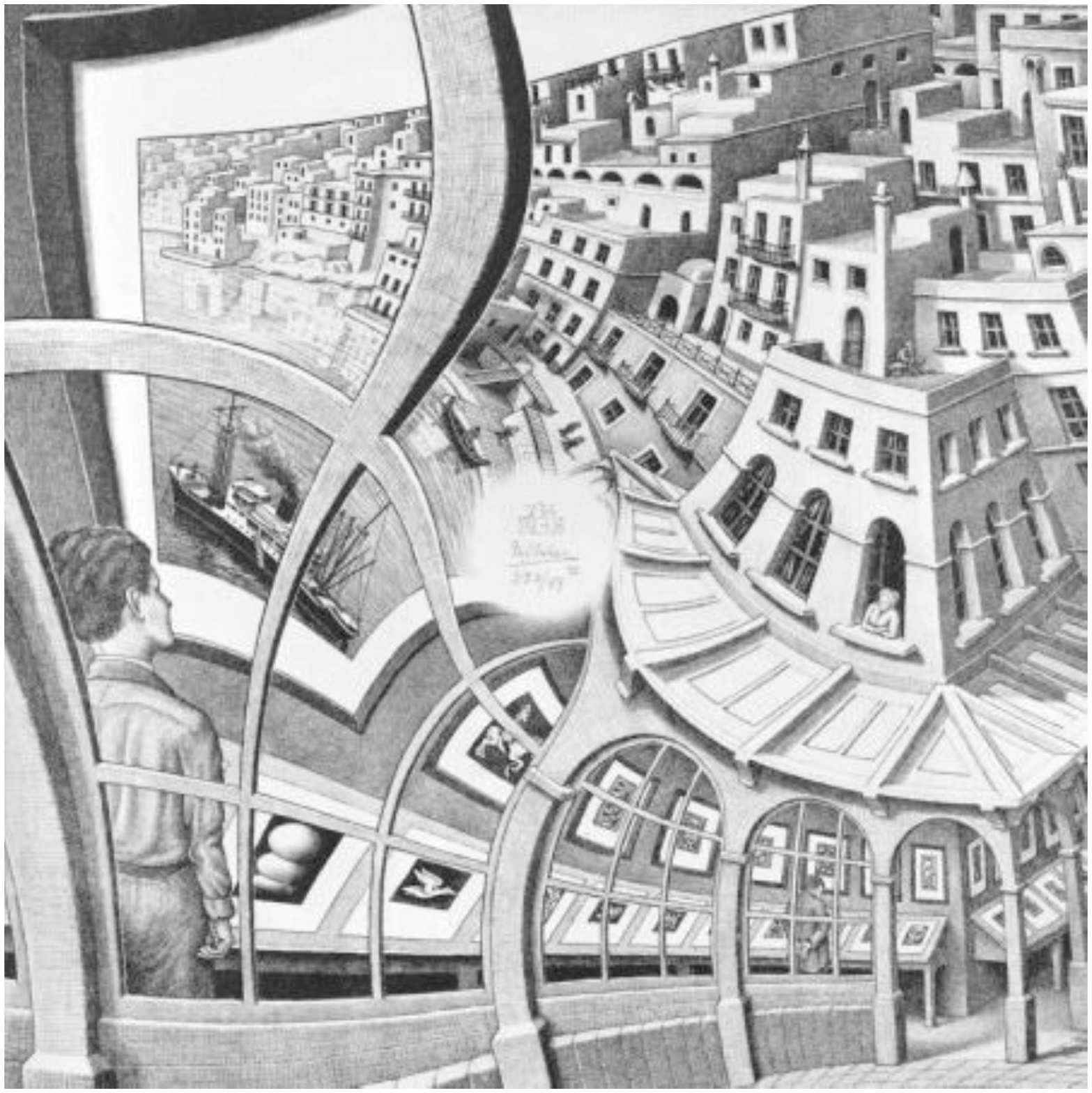}
\vspace{-1.5cm}
\begin{center}
Escher's conformal mapings
\end{center}
\end{figure}

After Einstein introduced a connection to describe the effects of a gravitational field, Weyl wondered if a similar connection can be used to describe the effects of other forces of nature such as electromagnetism. Weyl's real goal, however, was even more grandiose: he wanted to describe both general relativity(GR) and electromagnetism(EM) in a single geometrical framework.  It was shown, long before, that Maxwell's equations were invariant under a certain change of electromagnetic potential.  Weyl believed that EM respected local scale symmetry, which is true in four dimensions. Unlike diffeomorphism invariance, local scale invariance does not preserve the length of vectors as they move in space-time. Marrying EM with gravity then, for Weyl, amounted to finding the geometry that naturally harbored scale invariance in addition to diffeomorphism.  

Weyl's starting point was  mathematical in nature, and entirely motivated by the uneasiness he had towards Riemannian geometry.  In Riemannian geometry, the non-vanishing curvature implies that the direction of a vector on parallel transport around loops changes compared to the original vector,  while its norm remains constant. Weyl contended that the norm of a vector itself should change around loops and not be independent of its space-time location.  Mathematically, the condition of parallel transport implies a condition of integrability
\be
\nabla_{\mu} \xi^{\rho} = 0 \,   \Rightarrow R_{\mu\nu}{}^{\rho}{}_{\sigma} \xi ^{\sigma} = 0 \,
\ee
for the direction of a vector while no such condition exists for its norm.  Weyl wanted a similar integrability condition on the norm as well.
Weyl's critique to Riemannian geometry can be summed up by the following lines taken from his paper~\cite{weyl:1918}
\begin{quote}
{\it The metric allows the two 
magnitudes of two vectors to be compared, not only at the same 
point, but at any arbitrarily separated points. A true infinitesimal 
geometry should, however, recognize only a principle for trans- 
ferring the magnitude of a vector to an infinitesimally close point 
and then, on transfer to an arbitrary distant point, the integrability 
of the magnitude of a vector is no more to be expected that 
the integrability of its direction. }
\end{quote}
For a detailed discussion of Weyl's ideas and their influence on gauge theory, we refer the reader to~\cite{Straumann:2005hj}.

\begin{figure}[h]
\centering
\includegraphics[width=100mm, height=100mm]{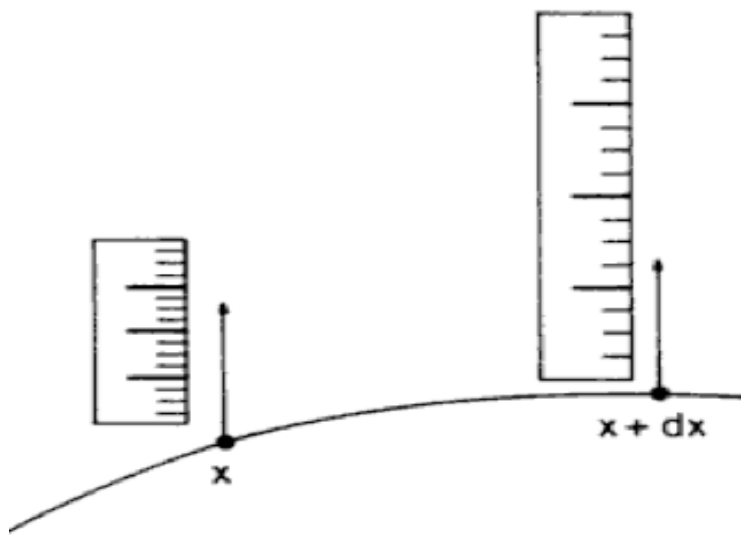}
\caption{The scale factor $\sigma(x)$ in Weyl's gauge theory is shown by the change in length of the meter stick from x to x+ dx.~\cite{M}\label{scalefig}}
\end{figure}

In Riemannian geometry, the norm is given by 
\be
l^2 = g_{\mu\nu} \xi^{\mu}\xi^{\nu} \,,
\ee
where $\xi^{\mu}$ is a parallel vector. The total derivative of this expression gives
\be
2ldl = (\partial_{\rho}\,g_{\mu\nu}dx^{\rho})\,\xi^{\mu}\xi^{\nu} +2 \,g_{\mu\nu}\, d\xi^{\mu} \xi^{\nu} =( \nabla_{\rho} g_{\mu\nu}) \xi^{\mu}\xi^{\nu} dx^{\rho} \,,
\ee
which, in Riemannian space, vanishes due to the metric compatibility, and hence the length of the vector remains constant.   Weyl realized that Riemannian geometry must be modified to allow the possibility of a varying norm.  Instead of a Riemannian metric, he considered an alteration
\be
\hat g_{\mu\nu} = e^{2\epsilon\lambda(x)} g_{\mu\nu} \approx(1+2\epsilon\lambda)g_{\mu\nu} \, .
\ee
Notice that even though $g_{\mu\nu}$ obeys the metric compatibility condition with respect to the old Levi-Civita connection, the new metric $\hat g_{\mu\nu}$ clearly violates it.  Using the new metric, to first order in $\epsilon$
\be
\widehat dl =  \epsilon( \nabla_{\rho} \lambda)\hat l\, dx^{\rho}   \,,
\label{dl}
\ee
which can be conveniently written as
\be
\widehat dl = b_{\mu} dx^{\mu}  = \sigma^{-1} \partial_{\mu} \sigma dx^{\mu}  \,.
\ee
The scale field $\sigma$ is analogous to a measuring stick whose size depends on its location in space-time.  Therefore, the vector norm measured against  it changes as shown in  Figure~\ref{scalefig}.
 
Hence, by rescaling the metric $ g_{\mu\nu} \mapsto \Omega^2(x) g_{\mu\nu}$, Weyl achieved a varying norm.  The geometry generated by the class of metrics $[g]$, where two metric are equivalent if they differ by a scale factor $\Omega^2(x)$, is known as conformal geometry which we discuss next.

\section{Conformal Manifolds}
\label{Cmanifolds}
Riemannian geometry is the study of a Riemannian manifold that consists of a smooth manifold  $M$ equipped with a positive definite metric $g$.  Conformal geometry, is the study of a conformal manifold $M$ equipped with a conformal class of metrics $[g_{\mu\nu}]$, where  $[g_{\mu\nu}]$ is an equivalence class of Riemannian metrics  
differing by local scale transformations
\be
[g_{\mu\nu}]=[\Omega^2 g_{\mu\nu}]\, ,
\label{rescaling}
\ee
for any non-vanishing smooth scalar function $\Omega^2(x)$. It is immediately clear that conformal manifolds are equipped with a well defined notion of angle but not length.  

\begin{figure}[h]
\centering
\includegraphics[width=100mm, height=100mm]{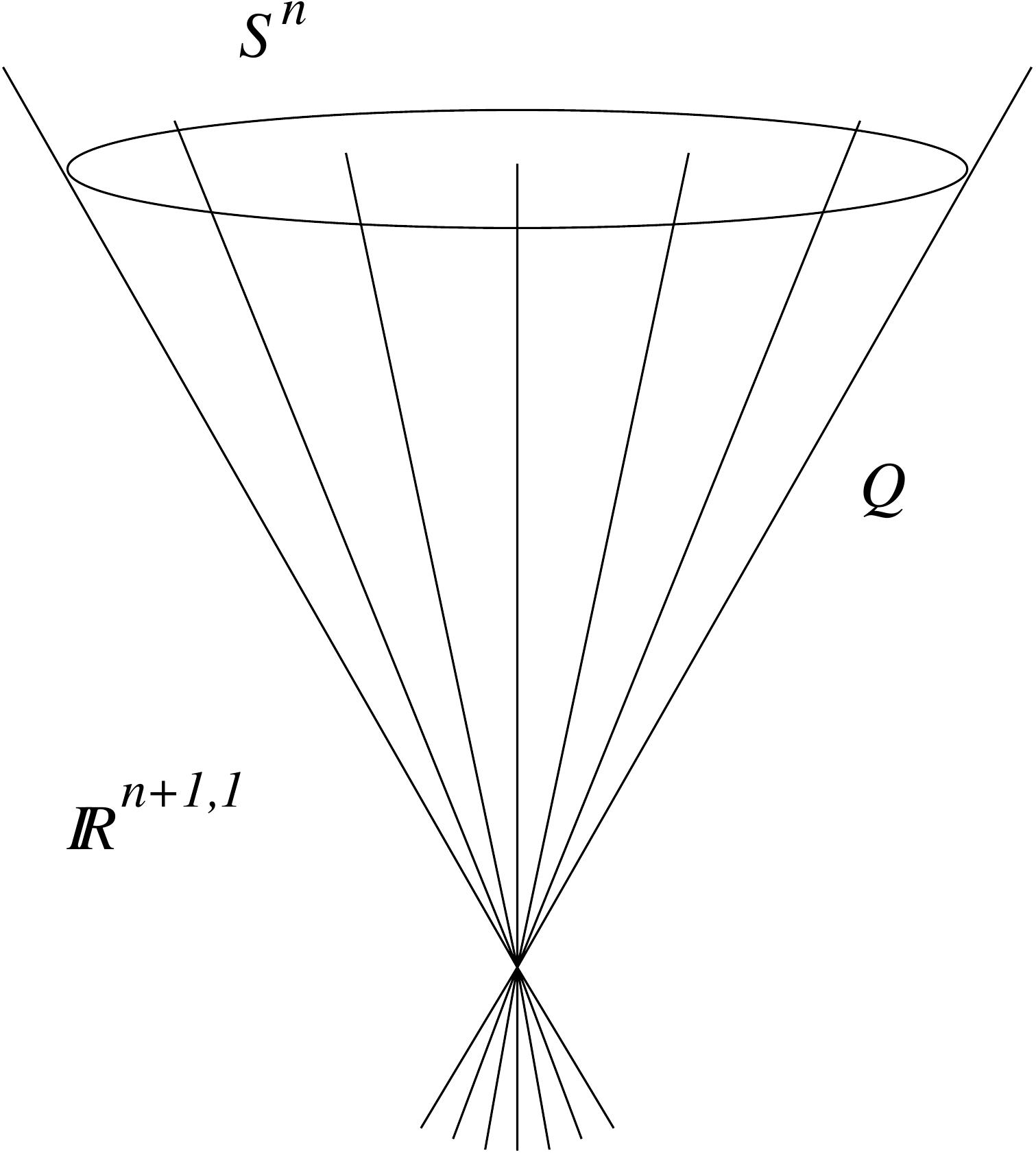}
\caption{The conformal manifold $M$ realized as rays $\xi$ of the cone $Q$
in the ambient space $\wt M$.\label{rays}}
\end{figure}

The most basic example of a conformal manifold is the sphere $M=S^d$ with a conformally flat
class of metrics.  A very useful way to view this manifold is as the set of null rays in a Lorentzian
ambient space of two dimensions higher. Let us explain this point in some detail: first we introduce a $d+2$ dimensional 
ambient space $\wt M={\mathbb R}^{d+1,1}$ with Lorentzian metric
\be
\wt {ds}^2
=-(dX^0)^2+\sum_{m=1}^{n} (dX^m)^2 +(dX^{d+1})^2\equiv dX^M h_{MN} dX^N.
\label{ambient metric}
\ee
This metric enjoys a homothetic Killing vector which equals the Euler operator
\be
X\equiv X^M\frac{\partial}{\partial X^M}\, ,\qquad [X,\wt{ds}^2]=2\wt{ds}^2\, ,
\label{canonical}
\ee
and derives from a homothetic potential
\be
H=\frac 12 X_M X^M\, ,
\qquad
X_M=\frac{\partial H}{\partial X^M}\, .
\ee
The potential $H$ can be used to define a cone $Q$ by the condition
\be
H=\frac 12X^M X_M=0\, .
\ee
The space of light like rays is the sphere $S^d$.  In particular, we identify point $x$ on the sphere $S^d$ with null rays $\xi=[\xi^M(x)]$.
This is depicted in Figure~\ref{rays}.

A ray is described by a null vector $\xi^M(x)$ up to overall rescalings so
\be
[\xi^M]=[\Omega\, \xi^M]\, .
\ee
Given a smooth choice of null vectors representing rays $\xi^M(x)$, 
we can pull the ambient metric~\eqn{ambient metric} back to the sphere
\be
ds^2=d\xi^M d\xi_M=\frac{\partial \xi^M}{\partial x^\mu}
\frac{\partial \xi_M}{\partial x^\nu} dx^\mu dx^\nu\, .
\ee
However, using the null condition $\xi^M\xi_M=0$ we see that a {\it different choice of null vectors} $\Omega\,  \xi^M$ but {\it equivalent rays}, pulls back to the metric $\Omega^2 ds^2$. 
In this way we produce a conformal manifold with conformally equivalent class of metrics $[ds^2]=[\Omega^2 ds^2]$. 

\begin{figure}[h]
\centering
\includegraphics[width=100mm, height=100mm]{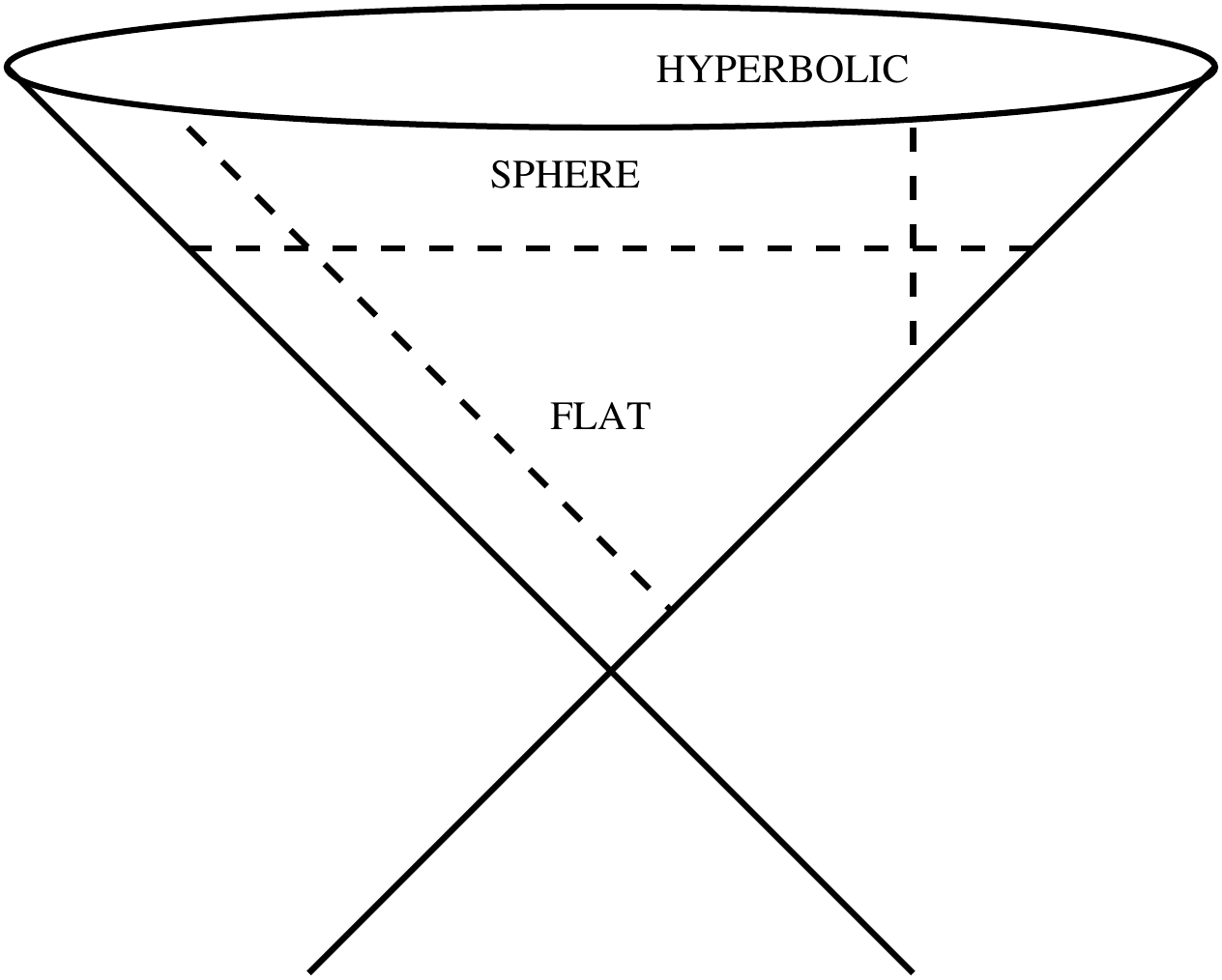}
\caption{Flat, spherical, and hyperbolic conformal manifolds realized as different slices of the cone. \label{slices}}
\end{figure}

To be completely explicit, adopting standard spherical coordinates $x^\mu=(\theta,\varphi^i)$
on $S^d$ the choice of rays
\bea
\xi^0\ &=&1\, ,\nn\\[2mm]
\xi^{d+1}&=&\cos\theta\, ,\nn\\[2mm]
\xi^{1}\ &=&\sin\theta\cos\varphi^1\, ,\nn\\[2mm]
&\vdots&\nn\\
\xi^m\ &=&\sin\theta\sin\varphi^1\ldots\sin\varphi^m\, ,
\eea
yields the canonical, conformally flat metric on $S^d$ Lorentzian
\be
ds^2=d\theta^2 + \sin^2\!\theta\,\sum_{m-1}^d\Big(\prod_{i=1}^{m-1}\sin^2\varphi^i\Big)(d\varphi^m)^2\, .
\ee
This particular choice of rays produces the conformal sphere.  However, there are other choices of rays that generate conformally flat and hyperbolic manifolds.  Choosing a particular set of rays then amounts to slicing the cone in special ways as shown in Figure~\ref{slices}.

This is the ambient approach to conformal manifolds and can be employed to obtain general conformal manifolds in arbitrary dimensions with arbitrary metric signatures.  In particular, if a Lorentzian $d$-dimensional conformal manifold is desired, we take ${\mathbb R}^{d,2}$ as the ambient space.  Generally, the ambient space will possess two extra dimension, one of space and one of time, compared to the conformal(base) manifold it describes.  


Having defined conformal manifolds, we need a notion of conformal flatness or more generally an invariant theory on them.  A manifold $M$ is conformally flat if, among the conformal class of metrics, it contains a flat metric--in the Riemannian sense. In other words, there exists a positive function $\Omega^2(x)$ such that  any metric in the conformal class can be rescaled to the flat one:  $\eta_{\mu\nu} = \Omega^2(x)g_{\mu\nu}$.  There is, however, another equivalent way to detect conformal flatness.

In Riemannian geometry, flatness is determined by vanishing of the Ricci scalar.  Conformal flatness, on the other hand, is measured by vanishing of the Weyl tensor in $d \geq 4$, which is related to the Riemann tensor by
 \be
R_{\mu\nu\rho\sigma}=W_{\mu\nu\rho\sigma}+\Rho_{\mu\rho}g_{\nu\sigma}
-\Rho_{\nu\rho}g_{\mu\sigma} -\Rho_{\mu\sigma}g_{\nu\rho}
+\Rho_{\nu\sigma}g_{\mu\rho}\, , 
\label{Curvature_Identity}
\ee where $W_{\mu\nu\rho\sigma}$ is
the trace-free Weyl tensor and $\Rho_{\mu\nu}$ is the $rho$-tensor.  The Weyl tensor is a conformal invariant and does not depend on the choice of a metric from the conformal class.  In three dimensions, the Weyl tensor vanishes and
instead, the Cotton tensor, which is the curl of the $rho$-tensor, measures the obstruction to conformal flatness.  The $rho$-tensor(sometimes called the Schouten tensor) is the trace
adjusted version of the Ricci tensor:
\be \Rho_{\mu\nu}=\frac1{d-2}\,
\Big(R_{\mu\nu}-\frac12\frac1{d-1}\, g_{\mu\nu}\, R\Big)\, ,\qquad
\Rho\equiv \Rho^\mu{}_\mu = \frac{R}{2(d-1)}\, . 
\ee 
It is worth nothing that in two dimensions, every conformal metric is locally conformally flat because there always exists a smooth non-vanishing function $\Omega^2(x)$ such that  any metric in the conformal class can be rescaled to the flat one.

At this point,  we have described a conformal manifold and defined the notion of conformal flatness on it.  Next, we review the symmetry group of the conformally flat manifold, which facilitates the construction of tractor calculus--the natural calculus on a conformal manifold.  This group is the conformal group, $CO(d)$, which is intimately related to the Lorentz group $SO(d,2)$.  
\subsection{Conformal Group \nopunct} 
The conformal group consists of the Poincar\'{e} group, dilations, and conformal boosts.  Dilations are scaling transformations that send 
\be
x^{\mu} \mapsto \lambda x^{\mu} \, ,
\ee
while conformal boosts transform the coordinate
\be
x^{\mu} \mapsto \frac{x^{\mu} - b^{\mu} x^2}{1-2b.x + b^2x^2}\, .
\ee
This group in $d$-dimensions has 
\be
\frac{(d+2)(d+1)}{2}
\ee
 generators.  In four dimensions, it consists of 15 elements: ten Poincar\'{e}, four conformal boosts, and one dilation. The transformations generated by the conformal group are called 
conformal transformations that are the special coordinate transformations that rescale the metric while preserving the angles as well as the light cone structure.

It is worth nothing that dilations and conformal boosts are responsible for rescaling the metric, while the Poincare transformations leave the metric invariant.  Explicitly, under coordinate transformations, the metric $g_{\mu\nu}$ changes by 
\be
\delta  g_{\mu\nu} = - 2\nabla_{(\mu} \xi _{\nu)}\,,
\label{poincare}
\ee
where $\xi= x^{'} - x$.  The invariance of the metric demands the following {\it Killing equation}~\cite{Blago}:
\be
- 2\nabla_{(\mu} \xi _{\nu)} = 0 \, .
\label{killing}
\ee
But our goal is to find the coordinate transformations that rescales the metric by a conformal factor: 
\be
g^{'}_{\mu\nu}(x^{}) = \Omega^2(x)  g_{\mu\nu} \,, \qquad \Omega(x) = e^{\lambda(x)} \,.
\ee
To first order, the change in the metric is given by
\be
\delta g_{\mu\nu} = 2 \lambda g_{\mu\nu}\,.
\ee
Equating it with~\eqn{poincare} yields the {\it conformal Killing equation}:
\be
 \nabla_{(\mu} \xi _{\nu)} =  - \lambda g_{\mu\nu} \, .
\label{killing1}
\ee
Expanding $\xi^{\mu}$ in a power series in $x$, one immediately finds the solutions to be
\be
\xi^{\mu}(x) = \epsilon^{\mu} + \omega^{\mu}{}_{\nu} x^{\nu} -\lambda x^{\mu} + (b^{\mu} x^2 - 2\,b . x \,x^{\mu}) \, .
\ee
The first two terms,  $\omega^{\mu\nu}$ and $\epsilon^{\mu}$, are easily recognizable as the parameters of the Poincar\'{e} transformations (Lorentz rotations and translations).  They are the solutions of the homogenous equation~\eqn{killing1} where $\lambda =0$. The rest are the parameters of dilation (D) and conformal boost (K) respectively. Together they form the parameters of the conformal group.  

Explicitly, the generators are given by
\bea
i P_m=\partial_m  ,&
iM_{mn}=X_m\partial_n-X_n\partial_m , &
iK_m=2x_m\,x \cdot \partial  - x^2   \partial_m ,
\nn\\[5mm]
&\!\!   iD=x \cdot \partial\, .& 
 \label{transambient operators}
\eea
Notice, we have deliberately put an ``i" in front of the generators to guarantee that they coincide with their quantum mechanical counterparts.  They generate the conformal algebra given below:
\bea
[iD,P_m]\ \ \ &=&\ \  -P_m\, ,\nn\\[2mm]
{}[iP_m,K_n]\ \ &=& \ \ 2\delta_{mn}D- 2M_{mn} \, ,\nn\\[2mm]
{}[iM_{mn},M_{rs}]&=&\delta_{nr}M_{ms}-\delta_{ns}M_{mr}-\delta_{mr}M_{ns}+\delta_{ms}M_{nr}
\, ,\nn\\[2mm]
{}[iD,K_m]\ \ &=&\ \ \ K_m\, .
\eea
To understand the relationship between conformal and Lorentz groups, firstly note that the Lorentz group in ($d+2$)-dimensions consists of  $d+1$ boosts and $\frac{d(d+1)}{2}$ rotations, a total of  $\frac{(d+2)(d+1)}{2}$ elements--a number that exactly matches the number of elements in $CO(d)$. This is not a coincidence; in fact, it  hints at a relationship between the ambient Lorentz group and the conformal group that we are about to explore. 

We return to the ambient space, ${\mathbb R}^{d+1,1}$, on which we allow the Lorentz group $SO(d+1,1)$ to act in the usual way, which
preserves the metric 
\be
\eta_{MN}=\left(\begin{array}{ccc}
-1&0&0\\0&\delta^{mn}&0\\0&0&1\end{array}\right)\, .
\label{eta}
\ee
The $d+2$ dimensional Lorentz algebra $so(d+1,1)$ obeys
\be
[M_{MN},M_{RS}]=\eta_{NR}M_{MS}-\eta_{NS}M_{MR}-\eta_{MR}M_{NS}+\eta_{MS}M_{NR}\, .
\ee
It is now highly advantageous to employ light cone coordinates in which the
indefinite ambient space metric equals
\be
\eta_{MN}=\left(\begin{array}{ccc}
0&0&1\\0&\delta^{mn}&0\\1&0&0\end{array}\right)\, .
\label{eta}
\ee
In this basis the generators $M^M{}_{N}$ of $so(d+1,1)$ decompose as
\be
M^M{}_{N}=\left(\begin{array}{ccc}-D&-\frac1{\sqrt2}P_n&0\\[4mm]- \frac1{\sqrt2}K^m&
M^m{}_{n}&\frac1{\sqrt2}P^m\\[4mm]0&\frac1{\sqrt2}K_n&D \end{array}\right)\, .
\label{matrix of generators}
\ee
The notation chosen here anticipates that the  {\it r\^oles} of the ambient Lorentz generators
from the viewpoint of the underlying $d$-dimensional manifold $M$ are translations~$P_m$, dilations~$D$, rotations~$M_{mn}$, and conformal boosts $K_m$.   Again, there are two extra generators in addition to the generators of the Poincar\'{e} group:   dilations~$D$ and the conformal boosts $K_m$.  The relation between the ambient Lorentz group and the conformal group should be clear by now.  In fact, the Lorentz group $SO(d+1,1)$ {\it is} the conformal group  $CO(d)$ of a $d$-dimensional base manifold. 

\subsection{Conformally Invariant Theories}
Apart from its mathematical significance, the conformal group has many applications in physics. It was first introduced by Bateman~\cite{Bateman} who showed that the Maxwell's equations are invariant under it.  In 2-dimensions, the conformal group is infinite dimensional and the string action on a world sheet is conformally invariant.  The Yang-Mills equations that describe the weak and strong interactions are also invariant under the conformal group.  A few other theories~\cite{Hortacsu:2001bp}, notably Dirac's massless theory and conformally invariant theories of Deser and Nepomechie~\cite{Deser:1983tm, Deser:1983mm} are invariant under conformal transformations as well.


The theories of Deser and Nepomechie were constructed in two steps:  (i) The models were coupled Weyl invariantly  to an arbitrary, conformally flat, metric when both the metric and the physical fields transformed.  (ii)  The metric was held constant while only the physical fields transformed, which turned the original Weyl invariance  into rigid conformal invariance.  This clever maneuver achieved lightlike propagation by
imposing the sufficient (but not necessary) condition of rigid conformal invariance.

In this thesis we extend this method to incorporate both massless and massive theories (lightlike or not) in
curved backgrounds and as a consequence uncover a relationship between mass and Weyl invariance.
This relationship relies on an elegant description of Weyl invariance in terms of mathematical objects called tractors which we present next.

\newpage

\section{Tractor Calculus}
\label{Tractor_Calculus}

\begin{quote}
{\it I admire the elegance of your method of computation; it must be nice to ride through these fields upon the horse of true mathematics while the like of us have to make our way laboriously on foot.}
\flushright{ Einstein, to Levi-Civita on tensor analysis}
\end{quote}
Tractor calculus is the calculus on conformal manifolds just like tensor calculus is the calculus on Riemannian manifolds.  Tractors are to Weyl invariance what tensors are to diffeomorphism invariance. Tensors can be classified by their transformation properties under changes of coordinate systems.  Tractors are classified by their tensor type and ``tractor weights'' under Weyl transformations.  In formal language, just as one talks of sections of tensor bundles, one can define tractors as sections of certain tractor bundles.  Tensors are employed to construct diffeomorphism invariant theories while tractors are used to build Weyl invariant ones.  

Amidst many similarities, there is, however,  a stark difference between tensors and tractors: unlike tensors, tractors use $d+2$ components to describe a theory in d-dimensions. It should not come as a surprise since the conformal group, $CO(d)$ in $d$ dimensions is $SO(d,2)$. In fact, tractors arrange the field content of physical theories as $SO(d,2)$ multiplets transforming under 
\be U=
\begin{pmatrix}
\Omega&0&\;0\;\\[2mm]
\Upsilon^m&\delta^m_n&\;0\;\\[2mm]
-\frac12\Omega^{-1}\,\Upsilon_r\Upsilon^r&-\Omega^{-1}\Upsilon_n&\Omega^{-1}
\end{pmatrix}\, , \qquad \Upsilon_\mu = \Omega^{-1}\, \partial_\mu \Omega\, .
\label{UM}
\ee
which we encountered before in~\eqn{Um}.  Tractors provide us with tools to facilitate calculations while constructing Weyl invariant theories.

In tractor calculus one studies multiplets with gauge
transformations given by the matrix $U(\Omega)$, or, in tighter
language, sections of tractor bundles with parallel transport defined
by the connection ${\mathcal A}_{\mu}$. For example a tractor vector field of
weight\footnote{The term ``conformal weight'', or just
  ``weight'' is often used in mathematics literature where
  Weyl invariance is often called conformal invariance. 
  } 
  $w$ is
a system consisting of functions $T^+$, $T^-$ and a vector field $T^m$ that, under the rescaling of the metric ($g \mapsto \Omega^2g$)  is required to satisfy
\bea T^M\equiv
\begin{pmatrix}
T^+\\T^m\\T^-
\end{pmatrix}&\mapsto&\Omega^w U^M{}_N T^N = \Omega^w 
\begin{pmatrix}
\Omega T^+\\
T^m + \Upsilon^m T^+\\
\Omega^{-1}[T^--\Upsilon_n T^n -\frac12 \Upsilon_n\Upsilon^n T^+]
\end{pmatrix}\, . \nn\\\label{tractorvector}
\eea 
 We often say the tractor components
$T^+$, $T^m$ and $T^-$ are placed in the the top, middle and bottom
slots, respectively. When two tractor quantities are multiplied, their weights are added. 


Next, we present a notion of parallel transport on tractor bundle.  The ``tractor connection" is given by a clever combination of the vielbein, Levi-Civita connection, and the $rho$-tensor:
 \be {\cal D}_\mu= \partial_{\mu} + \mathcal{A}_{\mu}=
\begin{pmatrix} \partial_{\mu} &-e_{\mu n}&0\\[2mm]
\Rho_{\mu}{}^m&\nabla_{\mu}&e_{\mu}{}^m\\[3mm]
0&-\Rho_{\mu n}&\partial_{\mu} 
\end{pmatrix}\, .
\label{Tconnection}
\ee
Under metric rescalings, this connection transforms  as
$ {\cal D}_\mu\mapsto U {\cal D}_{\mu} U^{-1}$
with $U$  given by~\eqn{UM}. It is fundamental to tractor calculus.  Acting on a tractor vector, it yields 
\be
\mathcal{D}_\mu \left(\begin{array}{c}T^+ \\[1mm]T^m\\[1mm] T^-\end{array}\right)=
\left(
\begin{array}{c}
\partial_\mu T^+ - T_\mu\\[1mm]
\nabla_\mu T^m - \Rho_\mu^m T^+ + e_\mu{}^m T^-\\[1mm]
\partial_\mu T^- - \Rho_{\mu\nu} T^\nu
\end{array}
\right)\, ,
\label{Tractor_derivative}
\ee
 where $\nabla_\mu$ is the Levi-Civita connection. This formula then extends to give a covariant
gradient operator on arbitrary tractor fields via the Leibniz rule.

Having defined the tractor connection on a conformal manifold, our next step is to calculate its curvature.  The curvature associated with the tractor connection~\eqn{Tconnection} is given by
\be
{\cal F}_{\mu\nu}= [{\cal D}_{\mu}, {\cal D}_{\nu}]  =  \begin{pmatrix}
0&0&0\\[2mm]
C_{\mu\nu}{}^m&W_{\mu\nu}{}^m{}_n&0\\[2mm]
0&-C_{\mu\nu n}&0
\end{pmatrix}\, .
\label{Tcurvature}
\ee
Notice that the curvature consists of the Cotton tensor, $C_{\mu\nu}{}^m$,  and the Weyl tensor $W_{\mu\nu}{}^m{}_n$--the two tensors that measure conformal flatness.  Hence, it is a central operator for measuring conformal flatness; vanishing tractor curvature implies a conformally flat connection.  Under metric transformations, the curvature transforms in a way similar to tractor connection.  Explicit transformations are provided in Appendix~\ref{Compend}.

Under Weyl transformations, the tractor metric remains invariant:
\be
 U^M{}_R{}U^N{}_S\eta^{RS}=\eta^{MN}  \,.
\ee
This coincide with our intuition from the ambient perspective.  Although the metric on the base manifold changes, it doesn't affect
the ambient metric.  The tractor metric 
\be
\eta_{MN}=\begin{pmatrix}0&0&1\\0&\eta_{mn}&0\\1&0&0\end{pmatrix}\,
,\label{eta} \ee 
is, therefore, a weight zero, symmetric, rank two tractor tensor
that is parallel with respect to ${\cal D}_\mu$.  Using the last fact
it is safe in calculations to use the tractor metric and its inverse
to raise and lower tractor indices $M,N,\ldots$ in the usual fashion,
and this we shall do without further mention. Along similar lines, it
is easy to see that 
\be
X^M\equiv\begin{pmatrix}\ 0\ \\0\\1\end{pmatrix} 
\ee 
is a (Weyl invariant) weight one tractor vector which corresponds to components of the homothetic killing vector~\eqn{canonical}.   It is also called the canonical tractor 
which may be used to ``project out'' the top slot of a tractor vector field.  Moreover, it is a null operator: $X_M X^M = 0$.

Weight $w$ tractors can be mapped covariantly to $w-1$ ones by the Thomas $D$-operator.  It acts on weight $w$ tractors by 
\be
D^M\equiv\begin{pmatrix}(d+2w-2)w\\[2mm](d+2w-2){\cal
  D}^m\\[2mm]-({\cal D}_\nu{\cal D}^\nu + w \Rho)\end{pmatrix}\label{D}
\ee 
and is extremely important since it contains the tractor covariant derivative as well as the tractor Laplacian.  It is important not to confuse the
symbol $D^M$ with the tractor connection ${\cal D}^m$
(contracted with an inverse vielbein) which appears in the middle slot
of the Thomas $D$-operator. Although $D^M$ is {\it not} a covariant
derivative, nevertheless it can often be employed to similar
effect.  Acting on a tractor vector of weight $w$, the Thomas $D$-operator \be
D^MV^N = \begin{pmatrix}(d+2w-2)w V^N \\[2mm](d+2w-2){\cal
  D}^m V^N\\[2mm]-({\cal D}_\nu{\cal D}^\nu + w \Rho)V^N\end{pmatrix} \, ,
\ee
where we evaluate the middle and the bottom slot using~\eqn{Tractor_derivative}.  The result is given
in Appendix~\ref{Tractor_components}.

Let us denote the Laplacian on scalars by $\Delta \equiv {\cal
  D}_\nu {\cal D}^\nu$.  Acting on a weight $w=1-\frac d2$ scalar
$\varphi$, the Thomas $D$-operator takes a simpler form
\be D_M
\varphi= \begin{pmatrix}0\\[1mm]0\\[1mm]-\Big(\Delta - \frac{d-2}2 \,
  \Rho\Big)\ \varphi\end{pmatrix}\, , 
  \label{SD}
\ee 
where the operator in the bottom slot is the conformally invariant
wave operator (or in Riemannian signature, the Yamabe operator).
Hence~$D^M\!\!$~$\varphi$~$=$~$0$ yields the equation of motion of a conformally
improved scalar $\Delta \varphi = \frac{d-2}2\, \Rho \varphi$.  It is
also important to note that $D^M$ is a null operator 
: \be D^M D_M = 0\, .\label{null} \ee
Acting on a weight one scalar field $\sigma$, the $D$-operator  produces  the scale tractor 
\be
I^M = \frac{1}{d}D^M\sigma \, .
\ee
Both $I$ and $D$ are of fundamental importance and will be utilized heavily to construct physical theories. 

Finally, we define a further tractor operator built from the canonical tractor ($X_M$) and the Thomas $D$-operator called the Double $D$-operator
\be\label{XDD}
(d+2w-2)D^{MN}=X^N D^M - X^M D^N\, \,
\ee
which obeys
\be
D^{MN}=
[X^M, D^N] + (d+2w)\eta^{MN}   \, . \label{identity} 
\ee
Explicitly, in components it is given by
\be\label{DD}
D^{MN}=
\begin{pmatrix}
0&0&w\\[1mm]
0&0&{\cal D}^m\\[1mm]
-w&-{\cal D}^n&0
\end{pmatrix}\, .
\ee
It is Leibnizian and involves only a single covariant derivative:
\be
D^{MN} (H_RT^R) = (D^{MN} H_R) T^R  + H_R (D^{MN}T^R)\, .
\ee
This operator will be of central importance especially when constructing fermionic theories

In this thesis, we not only construct tractor equations of motion,
but also action principles.  To that end we need to know how to
integrate the Thomas $D$-operator by parts. Notice that the
$D$-operator does not satisfy a Leibniz rule; this is to be expected
because its bottom slot is a second order differential
operator.  Nonetheless, at commensurate weights, an integration by
parts formula does hold. For example, if $V^M$ is a weight $w$ tractor
vector and $\varphi$ is a weight $1-d-w$ scalar then (see {\it e.g.} \cite{Gover:2001b})
\be \int \sqrt{-g} \, V_M D^M \varphi = \int \sqrt{-g}\,
\varphi D^M V_M\, .\label{parts} 
\ee 
Notice, there is no sign flip in
this formula and that the integral itself is Weyl invariant because
the metric determinant carries Weyl weight~$d$. An analogous formula
holds for tractor tensors. The double-$D$ operator is Leibnizian, and its formula for integration by parts is the standard one.
Moreover, upto surface terms
\be
\int \sqrt{-g} D_{MN} \Xi^{MN} = 0 \, 
\ee
for any $\Xi^{MN}$ of weight zero.

\subsection{Choice of Scale}
\label{Choice_Scale}
The final piece of tractor technology we will need is how to
handle choices of scale.  Based on our physical principle, we take conformal geometry as the starting point.  
On a conformal manifold,  there is no preferred metric. Still, we would like to pick a representative metric because
it allows us to work with Riemannian or Pseudo--Riemannian structures with their corresponding Levi-Civita
connections, in terms of which we can write out tractor formulas.  Picking a representative metric in conformal geometry serves a similar purpose as picking a coordinate system
does in differential geometry. But how does one pick a representative metric?

Adjoining  the conformal class of a weight $w=1$, non-vanishing, scalar $\sigma$ to this
with equivalence 
\be
[g_{\mu\nu},\sigma]=[\Omega^2 g_{\mu\nu},\Omega \sigma]\, ,
\ee
we can use $\sigma$ to uniquely (up to an overall constant factor)
pick a canonical metric $g_{\mu\nu}^0$ from this equivalence class by requiring
the accompanying representative scalar $\sigma^0$ to be constant for that
choice.  For instance, if our conformal class is given by
\be
[- dt^2+dx^2, x] \, ,
\ee
where $g$ is the two-dimensional flat metric and $\sigma = x$, then choosing $\Omega= x^{-1}$  renders $\sigma$ constant 
and turns the old flat metric into an AdS metric given by
\be
ds^2 = \frac{-dt^2+dx^2}{x^2} \,.
\ee
This is the canonical metric determined by this choice of scale.   Generically, the metric corresponding to a constant $\sigma$ will equal
\be
g^0 = \sigma^{-2} g \, ,
\ee
where $g$ is the old metric associated with an arbitrary $\sigma$.

The scale tractor evaluated at $\sigma^0$ plays 
a distinguished {\it r\^ole}. 
 Explicitly, it is given by
 \be
I_0^M=\frac 1d\ D^M \sigma^0 = \begin{pmatrix}
\sigma^0 \\[1mm]0 \\[1mm] -\frac 1 d  \Rho  \sigma^0\, \end{pmatrix} \, .\label{I}
\ee
In what follows, we will drop ``0" and display formulas in this frame where $\sigma$ is constant.
To obtain its correct tractor transformation law, one must
  first transform $\sigma$ as a weight one scalar and subsequently
  evaluate the derivatives in $D^M$.
 The scale tractor can also be used to detect conformally Einstein manifolds.  Notice that
\be
{\cal D}_{\mu} I^M = \sigma \begin{pmatrix}0\\[2mm]\Rho_\mu{}^m-\frac 1 d\, e_\mu{}^m\Rho \\[2mm] -\frac 1 d \,  \partial_\mu \Rho \end{pmatrix}
\ee
so that $I^M$ is parallel if and only if $g_{\mu\nu}$ is an Einstein
metric\footnote{The space of solutions to the requirement of parallel
  $I^M$ can be enhanced to include almost Einstein structures by
  allowing zeroes in $\sigma$.  at conformal
  infinities~\cite{Gover:2004at, Gover:2008al}.}.  
  Hence, at arbitrary scales, when
$g_{\mu\nu}$ is conformally Einstein, the scale tractor is a parallel
tractor.  Picking a constant $\sigma$ then amounts to choosing a preferred Einstein metric from the conformal class of Einstein metrics.
It also follows that $[D^M,I^N]=0$. If in addition to $I^M$
parallel, one has vanishing Weyl tensor, then $g_{\mu\nu}$ is
conformally flat and it follows that Thomas $D$-operators
commute:
\be
[D^M, D^N] = 0 \, .
\label{D_commute}
\ee
Many of the computations in this thesis pertain to arbitrary
conformal classes of metrics, but as the spin of the systems we study
increases, we typically restrict ourselves to conformally Einstein, or
even conformally flat metrics. 

 Recall in Section~\ref{gravity}, we gave an action principle for gravity in terms of the scale tractor, but the equations of motion
were not worked out.  A strong temptation is to pose 
\be
 {\cal D}_{\mu} I^M = 0  \, ,
\ee
as the equations of motion.  But a closer analysis shows that its middle slot is trace free which leaves the constant value of $\Rho$ undetermined.  
So, the correct equations of motion at constant scale are given by
\be
 {\cal D}_{\mu} I^M = 0  \, , \qquad I^2 = 0 \,.
\ee
The second equation fixes $\Rho$ and in the presence of matter, it can be modified rather easily.   In fact setting $I^2 = \lambda$, a constant, amounts to choosing the value of the cosmological constant. 
 
We have now assembled enough tractor technology to construct physical
models. We refer the reader to the
literature~\cite{Gover:2001, Bailey:1994, Gover:2009vc, Gover:2002ay, Cap:2002aj} for a more detailed and motivated account of tractors.  The next chapter provides detailed derivation of the tractor connection in the context of conformal gravity.  The less mathematically inclined reader
can skip the next chapter and go directly to Chapter~\ref{Bosonic} where we start applying tractor calculus to build physical theories.

     \chapter{Cartan Connections}
\label{TractorConnection}
In the previous chapter, the reader probably noticed the frequent appearance of the tractor connection.  Most of the tractor operators,  namely the scale tractor, the Thomas $D$-operator,  and the double $D$-operator are defined in terms of it.  In addition, the tractor connection provides a notion of parallel transport and its curvature establishes a notion of conformal flatness.  The tractor connection, therefore, is the central operator in conformal geometry and tractor calculus.  But what exactly is it? And more importantly, how does it arise in conformal geometry?  This chapter aims at answering these questions.  

There are several approaches to motivate the tractor connection.  We have already motivated it by writing the conformally Einstein condition as a parallel requirement on the scale tractor.  Another way to obtain the tractor connection is from the Cartan Maurer form of $so(d+1,1)$ pulled back to a conformal manifold viewed as a homogenous space of $SO(d+1,1)/P$ where P is a parabolic subgroup of the conformal group.  However, this only works for the conformally flat case.  So instead we view the conformal group $SO(d+1,1)$ as a space-time algebra on a similar footing as the Poincar\'{e} group.    Then, demanding that the $SO(d+1,1)$ Yang-Mills transformations agree with the transformations of conformal gravity, the unwanted independent fields turn into dependent ones.  The remaining fields then describe conformal geometry and the Yang-Mills connection becomes the tractor connection.  This is a general program known as the gauging of space-time algebras~\cite{Kaku:1977pa,Kaku:1977rk,Townsend:1979ki}.  The details will become clear shortly.  Mathematically, this scheme has its germ in homogenous spaces (see Appendix~\ref{Homogeneous}) and Cartan's~\cite{MR0046972} theory of moving frames.  Hence, before considering th tractor connection, let us derive the more familiar Levi-Civita connection in the context of gravity.

\section{Palatini Formalism}
Palatini formalism allows cosmological Einstein gravity to be reformulated as a Yang--Mills--like\footnote {Although the MacDowell-Mansouri gravity employs Yang-Mills--like fields, the action is not an ordinary Yang-Mills one.}
theory of the de Sitter group $SO(d,1)$\footnote{We use the de Sitter group rather than the Poincar\'{e} one to allow for a non-vanishing cosmological constant.}.    This observation was first made in a physics context by MacDowell and Mansouri~\cite{MacDowell:1977jt}
whose goal was to find a geometric construction of simple supergravity. These ideas date
back far further in the mathematical literature to the work of Cartan on what is now known as the Cartan connection~\cite{MR0046972}. 
The starting point is to write hyperbolic space ${\mathbb H}^d$ as a homogeneous space $SO(d,1)/SO(d)$,
which we take as a local model for Riemannian geometry
on a manifold $M$. Then we introduce an $SO(d,1)$ Lie algebra valued Yang--Mills connection 
\be
\nabla=d+\left(\begin{array}{cc}0&e_n\\[3mm]e^m& \omega^m{}_n\end{array}\right)\equiv d+A\, ,
\ee
 whose curvature
\be
F=\nabla^2=\left(\begin{array}{cc}0&de_n+\omega_n{}^r\wedge e_r\\[4mm]
de^m+\omega^m_r\wedge e^r & d\omega^m{}_n
+\omega^m{}_r\wedge \omega^r{}_n+\ e^m\wedge e_n
\end{array}\right)\, ,
\ee
measures the error made in modeling inhomogeneous space with a homogenous one. 
Identifying the one-forms $e^m$ as orthonormal frames
\be
e_\mu{}^m \eta_{mn}e_\nu{}^n=g_{\mu\nu}\, ,\label{vielbeins}
\ee 
and $\omega^m{}_n$ as the spin connection, we recognize the curvature as
\be
F=\left(\begin{array}{cc}0&T_n\\[2mm]T^m&R^m{}_n+e^m\wedge e_n\
\end{array}\right)\, ,
\ee
whose entries are built from the Riemann curvature $R^m{}_n$ and torsion $T^m$.
Now examine the Yang--Mills transformation of the vielbeine computed from
$\delta_{\rm YM} A= [\nabla,\Lambda]$.  Setting
\be
\Lambda = \begin{pmatrix} 0  & \xi _n \\  \xi ^m &  -\lambda^m{}_n   \end{pmatrix} \, ,
\ee
we find
\bea
\delta_{\rm YM}e_\mu{}^m&=& \partial_\mu \xi^m +\omega_\mu{}^m{}_n\xi^n +\lambda^m{}_ne_\mu{}^n\nn\\[4mm]
&=&\xi^\nu \partial_\nu e_\mu{}^m + \partial_\mu \xi^\nu e_\nu{}^m +(\lambda^m{}_n+\xi^\nu\omega_{\nu}{}^m{}_n)e_\mu{}^n
+2\xi^\nu T_{\mu\nu}{}^m\, .\qquad
\eea
In the second line we have rewritten the Yang--Mills transformation
as the sum of  general coordinate and local Lorentz transformations with field dependent
parameters $\xi^\nu=e^\nu{}_n \xi^n$ and $(\lambda^m{}_n+\xi^\nu\omega_{\mu}{}^m{}_n)$
plus an unwanted term proportional to the torsion
\be
T_{\mu\nu}{}^m=\partial_\mu e_\nu{}^m 
- \partial_\nu e_\mu{}^m + 2 \omega_{[\mu}{}^m{}_n e_{\nu]}{}^n\, .
\ee
Constraining the torsion to vanish, $T^m=0$, as in
Palatini formalism, the Yang-Mills transformations coincide with geometrical general coordinate and local Lorentz transformations.  Furthermore, we may solve for the spin connection in terms of
the vielbein by
\be
\omega_{\mu}{}^m{}_n(e)=-\lambda^m{}_{n\mu}+\lambda_{\mu}{}^m{}_n-
\lambda_{\mu n}{}^m
\, ,\qquad
\lambda_{\mu\nu}{}^m\equiv \partial_{[\mu}e_{\nu]}{}^m\, .\label{Palatini}
\ee
Since this expression is tensorial,
the dependent field $\omega(e)$ transforms correctly under general coordinate
transformations. Moreover, either by explicitly varying the vielbein, or by requiring
the torsion constraint, it is easy to compute the local Lorentz transformation
of the dependent spin connection. These happen to exactly equal their Yang--Mills
counterparts, and are given by
\be
\delta \omega_{\mu}{}^m{}_n(e)=\xi^\nu\partial_\nu \omega_\mu{}^m{}_n+\partial_\mu \xi^\nu \omega_\nu{}^m{}_n 
-\partial_\mu \lambda^m{}_n + \lambda^m{}_r \omega_\mu{}^r{}_n
+\lambda_n{}^r\omega_\mu{}^m{}_r\, ,
\ee
when the vielbein undergoes general coordinate and local Lorentz transformations
with parameters $\xi^\mu$ and $\lambda^m{}_n$. This is the sum of a general coordinate
and local Lorentz transformation, which should be contrasted with the Yang--Mills transformation
of an {\it independent} $\omega$ when no constraint is imposed (the latter being subject to 
additional --unwanted-- curvature dependent terms).
Also, the Yang--Mills curvature $R(M)^m{}_n=R^m{}_n+e^m\wedge e_n$ now takes values in the $so(d)$ subalgebra of $so(d,1)$.
To analyze what has happened here we recall
how the connection $\nabla$ acts on 
$so(d)$ vectors $v^n$ 
\be
\nabla v^m = d v^m + \omega(e)^m{}_n v^n = dx^\mu e_\nu{}^m
(\partial_\mu v^\nu + \Gamma_{\mu\rho}^\nu v^\rho)\, ,
\ee
where $v^\mu$ is a section of the tangent bundle $TM$ and the 
terms in brackets are its Levi--Civita connection.  
Hence the $SO(d,1)$ Yang--Mills connection is equivalent to the
Levi-Civita connection on the tangent bundle so long as it obeys the torsion
constraint $T^m=0$.
We will denote the ${\mathbb R}^n$ 
bundle whose sections are related to those of $TM$ by the vielbein $e_\mu{}^m$ as ${\cal E}^m$.
Hence
\be
\Gamma({\cal E}^m)\ni v^m = e_\mu{}^m v^\mu\, \Rightarrow v^\mu\in \Gamma(TM)\, .
\ee
In physics, this isomorphism between bundles ${\cal E}^m$ and $TM$ is often 
referred to as ``flat'' and ``curved'' indices, and we will use it often in what follows.
We also ignore any difficulties encountered in finding global orthonormal frames~$e^m$,
since it suffices for all our computations to work in a single coordinate patch.

\section{The Tractor Connection}
\label{Tractor_Connection}
The methods of the previous section can be applied to more general geometries by studying larger space-time Yang--Mills groups.
Our goal, in this section, is to derive the tractor connection by modeling a conformal manifold $(M,[g])$ as a homogenous space of $SO(d+1,1)/P$ where
$
P = {\mathbb R}^*\otimes(SO(d)\ltimes {\mathbb R}^d) \, ,
$
consisting of conformal boosts, dilations, and rotations. 
Therefore we take as Yang--Mills gauge group $SO(d+1,1)$ which is also the starting point for
conformal gravity.   Here we review the key ideas of conformal gravity  which will lead to a physical derivation
of the tractor connection.

In the mathematics literature these ideas date back to work by Thomas~\cite{Thomas},
later systemized into the modern day tractor calculus in~\cite{Gover:2001, Bailey:1994, Gover:2009vc, Gover:2002ay, Cap:2002aj}. In the physics literature, the first generalization beyond gravity and simple supergravity was to conformal gravity and supergravity~\cite{Kaku:1977pa}.
This was later developed into a general technology for gauging space-time algebras~\cite{Kaku:1977pa,Kaku:1977rk,Townsend:1979ki}.

Just as the Cartan connection was an $SO(d)$-connection so that the structure group
equaled the denominator of the local quotient model for the geometry, we now aim to construct a
$P$-connection which will be closely related
to the tractor connection.
But as a starting point we study the $SO(d+1,1)$ Lie algebra valued connection which reads 
\be
\nabla=d \ +\ \left(\begin{array}{ccc}-b&-e_n&0\\[1mm]
                                 f^m&\omega^m{}_n&e^m\\[1mm]
                                 0&-f_n&b 
               \end{array}\right)\equiv d+A\, .\label{Yang--Mills connection}
\ee
Let us denote the $so(d+1,1)$-valued curvature
\be
F=\nabla^2=\left(
\begin{array}{ccc}
-R(D)&-R(P)_n&0\\[1mm]
R(K)^m&R(M)^m{}_n&R(P)^m\\[1mm]
0&-R(K)_n&R(D)
\end{array}
\right)\, ,
\ee
where
\bea
R(P)^m \ &=& d e^m + \omega^m{}_n \wedge e^n - b \wedge e^m\, ,\nn\\[3mm]
R(D)\ \ &=&db - f^m\wedge e_m\, ,\nn\\[3mm]
R(M)^m{}_n&=&d \omega^m{}_n + \omega^m{}-r \wedge \omega^r{}_n
-f^m\wedge e_n + f_n\wedge e^m\, ,\nn\\[3mm]
R(K)^m\ &=& d f^m +\omega^m{}_n\wedge f^n-f^m\wedge b \, .
\eea
Again the gauge field $e^m$ is assumed to be invertible and identified with the vielbein
as in~\eqn{vielbeins}. Again, $\omega^m{}_n$ will correspond to the spin connection,
but now we must deal with the scale and boost potentials $b$ and $f^m$.
In the original conformal gravity approach, requiring invariance of a four dimensional
parity even action principle, quadratic in curvatures, under Yang--Mills transformation necessitated 
the torsion $T^m=R(P)^m$ to vanish. This eliminated $\omega^m{}_n$ as an independent
field since it then depended on both $e^m$ and $b$. Moreover, the conformal boost potential $f^m$ appeared
only algebraically in the action so could be integrated out. This left an invariant action that depended,
in principle, only on the vielbein and conformal boost potential. In fact, the conformal boost $b$
actually decoupled at this juncture leaving the conformally invariant Weyl-squared theory
depending only on the metric. The tractor connection exactly equals the
$SO(d+1,1)$ Yang--Mills connection with a dependent
spin connection, the boost potential determined by its algebraic equation of motion and
the scale potential set to zero. 

Since a conformal calculus should be independent of dynamical details such as the 
choice of an action principle, let us rederive these results following the spacetime
algebra gauging method. First we examine the Yang--Mills transformations of the vielbein
and scale potential
\bea
\delta_{\rm YM} e_\mu{}^m&=&
\delta^{\rm GCT}_{e^\nu{}_m\xi^m} e_\mu{}^m 
+ \delta^{\rm SCALE}_{\lambda-\xi^\nu b_\nu}e_\mu{}^m+  \ \delta^{\rm LLT}_{\lambda^m{}_n+\xi^\nu \omega_{\nu}{}^m{}_n}
e_\mu{}^m 
+\xi^\nu R_{\mu\nu}{}^m(P)\, ,\nn\\[3mm]
\delta_{\rm YM} b_\mu&=&\ 
\delta^{\rm GCT}_{e^\nu{}_m\xi^m} b_\mu \ +\  \ 
\delta^{\rm SCALE}_{\lambda-\xi^\nu b_\nu}b_\mu
\, +\ \ \delta^{\rm CBOOST}_{\zeta^m-\xi^\nu f_\nu{}^m}b_\mu
\ \ +\  \ \xi^\nu R_{\mu\nu}(D)\, .\nn\\[3mm]\label{Yang Mills for e and b}
\eea
These expression have been obtained by rewriting the Yang--Mills
transformations
\be
\delta A = [\nabla,\Lambda]\, ,\qquad
\Lambda=\left(\begin{array}{ccc}-\lambda&-\xi_n&0\\[2mm]
\zeta^m&-\lambda^m{}_n&\xi^m\\[2mm]0&-\zeta_n&\lambda\end{array}\right)\, ,
\ee 
in terms of general coordinate, local lorentz, scale and conformal boost transformations
with field dependent parameters. Moreover general coordinate and local Lorentz transformations
act on curved and flat indices, respectively, in the usual way
\bea
\delta^{\rm GCT} v^\mu &=& \xi^\nu \partial_\nu v^\rho -\partial_\nu \xi^\mu v^\nu\, ,\nn\\[2mm]
\delta^{\rm GCT} v_\mu &=& \xi^\nu \partial_\nu v_\mu + \partial_\mu \xi^\nu v_\nu\, ,\nn\\[2mm]
\delta^{\rm LLT} v^m &=& \lambda^m{}_n v^n\, .\label{llt gct}
\eea
Scale and conformal boost transformations are defined on the vielbein (and in turn metric)
and scale potential as
\bea
\delta^{\rm SCALE} e_\mu{}^m &=&\lambda e_\mu{}^m\, ,\nn\\[2mm]
\delta^{\rm SCALE} g_{\mu\nu}\ &=& \lambda^2 g_{\mu\nu}\, ,\nn\\[2mm]
\delta^{\rm SCALE} b_\mu \ \ &=& \partial_\mu \lambda\, ,\label{scale}\\[6mm]
\delta^{\rm CBOOST} e_\mu{}^m &=&\ 0\, ,\nn\\[2mm] 
\delta^{\rm CBOOST} g_{\mu\nu}\ &=&\ 0\, ,\nn\\[2mm]
\delta^{\rm CBOOST} b_\mu\ \  &=&\  \zeta_\mu\, .\label{conformal boost}
\eea
Inspecting the Yang--Mills transformations~\eqn{Yang Mills for e and b}, we see that we must 
impose curvature constraints
\be
R_{\mu\nu}{}^m(P)=0=R_{\mu\nu}(D)\, .
\ee
This implies that Yang--Mills transformations now equal their geometrical 
counterparts
\be
\delta_{\rm YM}=\delta_{\rm GEOMETRY}=\delta^{GCT}+\delta^{LLT}+\delta^{SCALE}
+\delta^{CBOOST}\, .
\ee
The first constraint generalizes the usual torsion one $T^m=0$,  and is solved for the spin connection
by setting
\be
\omega(e,b)_\mu{}^m{}_n=\omega(e)_\mu{}^m{}_n-b^m e_{\mu n} - b_n e_\mu{}^m\, ,
\ee
where $\omega(e)$ is the usual Palatini expression given in~\eqn{Palatini}.
The $R(D)=0$ curvature constraint implies that the antisymmetric part of the 
conformal boost potential is the curl of the scale potential
\be
f_{[\mu\nu]}=\partial_\mu b_\nu - \partial_\nu b_\mu\, .\label{curl}
\ee
It is easy to check that the algebra of general coordinate, local Lorentz, local scale and
conformal boost transformations in~\eqn{llt gct},~\eqn{scale} and~\eqn{conformal boost}
closes on the independent fields $e_\mu{}^m$ and 
$b_\mu$. If we had kept the spin connection independent, its Yang--Mills
transformations would differ from the correct ones by a term proportional to the curvature 
$R(M)^m{}_n$. However the dependent spin connection $\omega(b,e)$ transforms
as desired
\bea
\delta^{\rm GCT} \omega(e,b)_\mu{}^m{}_n\ \ &=&
\xi^\nu\partial_\nu\omega(e,b)_\mu{}^m{}_n+\partial_\mu\xi^\nu\omega(e,b)_\nu{}^m{}_n\, ,\nn\\[2mm]
\delta^{\rm LLT} \omega(e,b)_\mu{}^m{}_n\ \ &=&-\partial_\mu \lambda^m{}_n
+\lambda^m{}_r\omega(e,b)_\mu{}^r{}_n-\omega(e,b)_\mu{}^m{}_r\lambda^r{}_n\, ,\nn\\[2mm]
\delta^{\rm SCALE}\omega(e,b)_\mu{}^m{}_n\ &=&0\, ,\nn\\[2mm]
\delta^{\rm CBOOST} \omega(e,b)_\mu{}^m{}_n&=& \zeta^n e_{\mu n}-\zeta_m e_\mu{}^m\, .
\eea
where these formul\ae ~are obtained by varying the vielbein and scale potential
as in~\eqn{llt gct},~\eqn{scale} and~\eqn{conformal boost}
(the easiest way to actually make this computation is to require the torsion constraint
$R(P)^m=0$ is invariant under variations).

A similar result holds for the other, so far dependent, field $f_{[\mu\nu]}$.
However it still remains to fix the symmetric part of the conformal boost potential. It cannot be
an independent field since its Yang--Mills transformation when expressed in terms of 
general coordinate, local Lorentz, scale and conformal boosts contains additional
unwanted terms proportional to the curvature $R(K)^m$. (There is no way to solve
algebraically the equation $R(K)^m=0$.) Therefore we must replace
the independent field $f_{(\mu \nu)}$ by an appropriate combination $f(e,b)_{(\mu \nu)}$.
In conformal supergravity, the symmetric part of the conformal boost potential is
determined by constraining the Ricci part of the curvature $R(M)_{\mu\nu}{}^m{}_n$.
Moreover, this constraint is determined solely by closure of the superconformal algebra,
the main point being that the  anticommutator of supersymmetry transformations
produces a translation $\{Q,Q\}\sim P$ whose Yang--Mills transformation
rule must be modified to produce a general coordinate transformation. In the conformal case, without supersymmetry, only the $[D,P]$ commutator produces a translation which yields no new information.

Instead therefore we must appeal to geometry to provide an appropriate constraint:
Although we have jettisoned $P$-Yang--Mills transformations
in favor of coordinate transformations, under the remaining
local symmetries, we still require that $F=\nabla^2$
transforms as a Yang--Mills curvature. In particular, this means that the middle slot,
$R(M)^m{}_n$ is invariant under scale transformations -- to see this just write down
the $\lambda$ dependent terms of $[F,\Lambda]$. It is of course true that 
for {\it any} choice of gauge fields $(e^m, b, \omega^m{}_n,f^m)$ that
\be
[\nabla,F]=0\, ,\label{Bianchi}
\ee
namely the Yang--Mills Bianchi identity. When $R(P)^m=0=R(D)$ it follows
as a consequence of~\eqn{Bianchi}
that $R(M)^m{}_n$ has the symmetries of the Riemann tensor
\be
R(M)_{[\mu\nu\rho]n}=0=R(M)_{\mu\nu mn}-R(M)_{mn\mu\nu}\, .\label{Riemann symmetries}
\ee
However, the only conformally invariant
tensor, built from two derivatives of the metric with the symmetries of the
Riemann tensor is the Weyl tensor. 

Requiring that $R(M)^m{}_n$
be the Weyl tensor
\be
R(M)^m{}_n=W^m{}_n\, ,
\ee
implies that the conformal boost potential equals the $\Rho$-tensor
\be
f_{\mu\nu}=\Rho(e,b)_{\mu\nu}\, ,
\ee
given by the trace-modified Ricci
\be
\Rho(e,b)_{\mu\nu}=-\frac{1}{n-2}\Big(R(e,b)_{\mu\nu}-\frac{1}{2(n-1)}\, g_{\mu\nu} R(e,b)\Big)\, .
\ee
Here, the Ricci tensor and scalar curvature are formed by tracing the 
$b$-dependent Riemann tensor 
$R(e,b)^m{}_n=d\omega(e,b)^m{}_n+ \omega(e,b)^m{}_r \omega(e,b)^r{}_n$
\be
R(e,b)_{\nu\rho}=R(e,b)_{\mu\nu\rho}{}^\mu\, ,\qquad
R(e,b)=R(e,b)_\nu{}^\nu\, .
\ee
Moreover, we stress that in the presence of a non-vanishing scale potential~$b$, 
the spin connection $\omega(e,b)$ is torsion-full so the Riemann tensor $R(e,b)^m{}_n$
does not obey the pair interchange and Bianchi of the first kind symmetries of~\eqn{Riemann symmetries} even though $R(M)(e,b)^m{}_n$ does.
This means that $\Rho(e,b)_{\mu\nu}$ is not symmetric, however its antisymmetric part is
exactly the curl of $b_\mu$ in agreement with~\eqn{curl}.
We will often denote the one-form $\Rho^m\equiv e^{\mu m} \Rho_{\mu\nu} dx^\nu$ and its trace 
by $\Rho=\Rho^\mu_\mu$. This expression for $f(e,b)_{\mu\nu}$ follows equivalently upon imposing 
\be
e^{\mu n}R(M)_{\mu\nu mn}=0\, .
\ee
Moreover it assures the correct transformations for the dependent combination $f^m=\Rho^m$
\bea
\delta^{\rm GCT} f(e,b)_\mu{}^m\ \ &=&
\xi^\nu\partial_\nu f(e,b)_\mu{}^m+\partial_\mu\xi^\nu f(e,b)_\nu{}^m\, ,\nn\\[2mm]
\delta^{\rm LLT} f(e,b)_\mu{}^m{}_n\ \ &=&\lambda^m{}_n f(e,b)_\mu{}^n\, ,\nn\\[2mm]
\delta^{\rm SCALE}f(e,b)_\mu{}^m\ &=&-\lambda f(e,b)_\mu{}^m\, ,\nn\\[2mm]
\delta^{\rm CBOOST}f(e,b)_\mu{}^m{}&=&
\partial_\mu \zeta^m +\omega(e,b)_\mu{}^m{}_n \zeta^n + \zeta^m b_\mu\, .
\eea

Before finally connecting these conformal gravity results to conformal geometry,
let us collect together some important data. The constrained connection reads
\be
\nabla=d \ +\ \left(\begin{array}{ccc}-b&-e_n&0\\[1mm]
                                 \Rho(e,b)^m&\omega(e,b)^m{}_n&e^m\\[1mm]
                                 0&-\Rho(e,b)_n&b 
               \end{array}\right)\, 
\ee
with independent vielbein and scale potential and 
dependent spin connection and conformal boost potential. 
All gauge transformations follow from the Yang--Mills ones
of the independent fields, but $P$-gauge transformations
are equivalent to the sum of general coordinate transformations
and field dependent local Lorentz, scale and conformal boost
transformations.  Since changes of coordinates  act on the
base manifold alone, only Lorentz scale and conformal boost gauge 
symmetries act fiber-wise. These generate the maximal parabolic
subgroup $P={\mathbb R}^*\otimes(SO(d)\ltimes {\mathbb R}^n)\subset SO(d+1,1)$
whose elements can be parameterized as
\be
{\cal W} = \left(
\begin{array}{ccc}
\Omega&0&0\\[2mm]
\Lambda^m{}_r \zeta^r&\Lambda^m{}_n&0\\[2mm]
-\frac{\zeta^r \zeta_r}{2\Omega}&-\frac1\Omega \zeta_r \Lambda^r{}_n{} &\frac1\Omega
\end{array}
\right)\, ,
\ee
and under which the connection transforms as
\be
\nabla \rightarrow {\cal W} \nabla {\cal W}^{-1}\, .
\ee
Therefore, we have constructed a bundle with parabolic $P$-connection. The curvature
and its transformation are
$$
F=\left(
\begin{array}{ccc}
0&0&0\\[2mm]
C(e,b)^m &\
 W(e,b)^m{}_n&0\nn\\[2mm]
0&\!-C(e,b)_n&0
\end{array}
\right)\qquad\qquad\qquad\qquad\qquad\qquad\quad\qquad
$$
\be
\qquad
\longrightarrow
\left(\!\!\!
\begin{array}{ccc}
0&0&0\\[2mm]
\frac1\Omega \Lambda^m{}_r(C(e,b)^r-W(e,b)^r{}_s\zeta^s) & \!\!\!\!\Lambda^m{}_r W(e,b)^r{}_s
\Lambda^s{}_n&0\nn\\[2mm]
0&\!\!\!-\frac1\Omega \Lambda_{nr}(C(e,b)^r-W(e,b)^r{}_s\zeta^s) &0
\end{array}
\right)\, ,
\ee
where the Weyl tensor $W(e,b)^m{}_n$ is the trace-free part of the Riemann tensor
$R(e,b)^m{}_n$ while the Cotton tensor $C(e,b)^m$ is the covariant curl
of the $\Rho$-tensor
\be
C(e,b)^m=d \Rho(e,b)^m + \omega(e,b)^m{}_n\wedge \Rho(e,b)^n\, .
\ee

Finally,
to connect conformal geometry
where one studies only metrics and
scale transformations (and diffeomorphisms)
we must eliminate the conformal boost as an independent
gauge symmetry and throw out the scale potential $b$.
This is easily achieved by examining the conformal boost
gauge transformation of the scale potential
\be
\delta^{\rm CBOOST} b_\mu = \zeta_\mu\, .
\ee
This transformation is an algebraic St\"uckelberg shift symmetry.
Therefore we may fix  the conformal boost gauge
\be
b_\mu = 0\, .
\ee
If we wish to remain in this gauge, we may only make
transformations respecting this condition. The scale potential
is invariant under local Lorentz transformations $\delta^{\rm LLT}
b_\mu=0$ so these are unaffected. Moreover, under general
coordinate transformations 
\be
\Big[\delta^{\rm GCT} b_\mu\Big]_{b=0}=0\, ,
\ee
so general coordinate transformations also respect vanishing of the 
scale potential. Scale and conformal boost transformations
must be treated more carefully however because
\be
\Big(\delta^{\rm SCALE}+\delta^{\rm CBOOST}\Big) \, b_\mu=\partial_\mu \lambda + \zeta_\mu\, .
\ee
Hence, whenever we make a scale transformation with parameter $\lambda$,
we must also perform a compensating conformal boost with parameter
\be
\zeta_\mu=-\partial_\mu \lambda\equiv \Upsilon_\mu\, .
\ee
We will often denote the one-form $d\lambda = \Omega^{-1} d\Omega = \Upsilon$
where $\Omega=\exp\lambda$ is the parameter for a finite, rather than infinitesimal,
scale transformation.

The $SO(d+1,1)$ connection $\nabla$ subject to constraints $R(P)^m=0=R(D)=
e_n\wedge {}^* R(M)^m{}_n$ and the gauge choice $b=0$ is the {\it tractor connection}.
It is the central operator of this thesis. Explicitly, the tractor connection equals
\be
\nabla = d + \left(\begin{array}{ccc}0&-e_n&0\\[1mm]
                                 \Rho(e)^m&\omega^m{}_n(e)&e^m\\[1mm]
                                 0&-\Rho_n&0 
               \end{array}\right)\, .
               \label{tractor connection}
\ee  

It can also be derived using BRST techniques~\cite{Boulanger:2004eh}. The important feature of the tractor connection is its transformation under  conformal transformations
by a scale $\Omega=\exp\lambda$,
\be
e^m\rightarrow \widehat e^{\, m}=\Omega\,  e^m
\ \ \Longrightarrow \ \ g_{\mu\nu} \rightarrow \widehat g_{\mu\nu} = \Omega^2 g_{\mu\nu}\, .\label{e}
\ee
Defining  $SO(d+1,1)$ conformal boost and scale matrices
\be
U=\exp\left(
\begin{array}{ccc}
0&0&\ 0 \ \\[2mm]
\Upsilon^m&0&\ 0 \\[2mm]
0&-\Upsilon_n&\ 0
\end{array}
\right)
=
\left(
\begin{array}{ccc}
1&0&\ \ 0\\[2mm]
\Upsilon^m&1&\ \ 0 \\[2mm]
-\frac12\Upsilon^r\Upsilon_r&-\Upsilon_n&\ \ 1\ 
\end{array}
\right)\label{U}
\ee 
and
\be
V=\exp\left(
\begin{array}{ccc}
\lambda&0& 0  \\
0&0& 0 \\
0&0& -\lambda
\end{array}
\right)
\equiv
\left(
\begin{array}{ccc}
\Omega&0& 0\\
0&1& 0 \\
0&0&\Omega^{-1}
\end{array}
\right)\, ,
\ee 
the tractor connection transforms simply as
\be
\nabla\rightarrow
VU\mathcal{D} U^{-1}V^{-1}\, .
\ee
This is the combination of a scale  and a compensating
conformal boost Yang--Mills gauge transformation with field-dependent parameter
\be-\Upsilon^m\equiv e^{\mu m}\partial_\mu \Omega\, .\ee 

Again, let us pause to survey what has been constructed. We started with
an $so(d+1,1)$ Yang--Mills connection on an $\mathbb{R}^{n+1,1}$ vector bundle 
over~$M$ that, upon solving curvature constraints and gauging away the
scale potential acts on vectors $T^M$
in the $so(d+1,1)$ fundamental representation restricted to the subgroup $P$ by
\be
\mathcal{D}_\mu \left(\begin{array}{c}T^+ \\[1mm]T^m\\[1mm] T^-\end{array}\right)=
\left(
\begin{array}{c}
\partial_\mu T^+ - T_\mu\\[1mm]
\nabla_\mu T^m - \Rho_\mu^m T^+ + e_\mu{}^m T^-\\[1mm]
\partial_\mu T^- - \Rho_{\mu\nu} T^\nu
\end{array}
\right)\, .
\ee
Here $T^\mu$ is a section of the tangent bundle $TM$ while $T^m$ is again a section of~${\cal E}^m$
and $\nabla_\mu T^m$ is equivalent to the Levi-Civita connection as explained above. 

The covariant, weight $w$, tractor index $M$
plays an analogous {\it r\^ole} to the flat $SO(d)$ indices in the
Riemannian context. 
Tractor indices can be raised, lowered and contracted
with the tractor metric 
$\eta_{MN}$ as in~\eqn{eta}. Moreover, since
\be
\label{weq}
\mathcal{D}T^M\rightarrow
\Omega^\w\ {\cal U}^M{}_N ({\mathcal{D}}+w \Upsilon)
T^N \, ,
\ee
when the weight $w=0$
we can use the tractor connection to produce new covariants and in
turn conformal invariants. We remark that the {\it r\^ole} of the boost
potential, which we gauge away, was precisely to cancel  the extra term proportional to
$w$ in~\eqn{weq}.

By now, the origins of the tractor connection, the central pillar of tractor calculus,  should be clear.  Since we now have a good understanding of tractor calculus, let us utilize it to construct physical theories implementing our principle of unit invariance. 

    \chapter{Bosonic Theories}
\label{Bosonic}
In this chapter, we apply the tractor technology developed in the previous chapters to implement our basic physics principle by constructing bosonic theories with manifest unit invariance, and then study their consequences.  Since we achieve unit invariance by introducing the scale $\sigma$, this necessitates the appearance of the scale tractor $I$ in our equations of motion.  Then the length of the scale tractor yields novel gravitational or soft mass terms built from curvatures.  These masses are manifestly unit invariant and are constant for conformally Einstein backgrounds. 

Upon picking a preferred unit system by holding the scale, $\sigma$, constant, the manifest unit invariance is broken and, consequently, masses are related to Weyl weights, which dictate how physical quantities transform under Weyl transformations. Surprisingly, the reality of the weights in this ``mass-Weyl weight relationship'' naturally leads to the Brietenlohner--Freedman bounds.  Moreover, at a special value of the weight, the scale tractor completely decouples leaving  behind strictly Weyl invariant theories. 

Our methods are very general and work for arbitrary spins.  For higher spins, we need a gauge principle that can handle massive and massless  particles in a single framework.  Tractors provide exactly such a framework.  Spin one and two are explicitly worked out as examples of this setup, which is then generalized to higher spins.  We start with a scalar field theory and later develop spin 1, spin 2, and general spin theories. 

 \section{Scalar Theory}
 \label{Scalar}
As advertised,  let us use tractors to  describe a massive scalar field in a curved
background in a manifestly unit invariant way. 
 Let $\varphi$ be a
weight $w$ scalar field so that under Weyl transformations 
\be \varphi
\mapsto \Omega^w \varphi\, .  \ee 
Since our aim is to construct a unit invariant theory, we have to inevitably use the scale tractor($I$), but further tractor operators are also needed to build a scalar equation of motion. 
The most promising is  the  Thomas $D$-operator because it contains a Laplacian in its bottom slot.   From the $D$-operator, we construct the equation of motion 
 \be 
 I_M D^M \varphi = 0\, ,\label{sceom} \, 
\ee
which can also be motivated by ideas coming from conformal scattering theory~\cite{ Gover:2002ay, Gover:2007, Graham}. 
The equation of motion follows from a Hermitian\footnote{Its hermicity can be explicitly checked by using~\eqn{parts} and \eqn{Ids}.} and manifestly unit invariant tractor action built from tractors 
 \be
S[g_{\mu\nu},\sigma,\varphi]=\frac12 \int \sqrt{-g} \, \sigma^{1-d-2w}
\varphi I_M D^M \varphi =
S[\Omega^2g_{\mu\nu},\Omega\sigma,\Omega^w\varphi]\, . 
\label{Saction}
\ee
Writing out~\eqn{sceom} at an arbitrary scale, we get 
\be
(\sigma^{-2} g^{\mu\nu} \wt \nabla_{\mu}\wt \nabla_{\nu}  + m_{grav}^2) \, \varphi  = 0 \, .
\label{Weom}
\ee
The first term is the Weyl compensated free scalar field theory  whose unit invariance was shown in Section \ref{Compensated_Scalar};  the second term is manifestly unit invariant gravitational mass term  given by
\be
m_{grav}^2  = I^M\!I_M=-\frac{2\sigma^2}{d}\left[\Rho +\nabla^\mu b_\mu -\frac{d-2}{2}\ b^\mu b_\mu\right]\, .
\label{g_mass}
\ee
Importantly, unit invariance is achieved only upon transforming the metric
$g_{\mu\nu}$, the scale $\sigma$ and the physical field $\varphi$ 
simultaneously.  Although the theory is unit invariant, it is not strictly Weyl invariant because of its $\sigma$ dependence.

The gravitational mass terms is in fact the Weyl compensated gravity action encountered earlier in Section~\ref{gravity}, which sets the mass scale.  Even though the mass term is unit invariant, it may not be constant because of its curvature dependence, which is unacceptable.  However,
if we work on a conformally Einstein manifold where
\be
\mathcal{D}_{\mu} I^M =0 \, ,
\ee
$I^2$ and thus the gravitational mass are both constant.  Notice that the gravitational mass term was absent in the Weyl compensated scalar theory presented in Section~\ref{Compensated_Scalar}.
We reemphasize that Weyl compensated flat theories miss some locally unit invariant theories that can be encapsulated efficiently by the tractor approach.  The scalar field theory 
is the first example in support of our claim. 

We choose the scale such that it is constant for the background metric of interest.  It allows us to pick a particular metric from a class of conformally Einstein metrics.  In that case $I^M$ takes the simpler form~\eqn{I},
so it is easy to use the expressions~\eqn{D} and~\eqn{eta} to evaluate
 the equation of motion~\eqn{sceom}.  We obtain  \footnote{We also refer the reader to
  Appendix~\ref{Tractor_components} for a tractor/tensor component
  dictionary.  }
(cf.\ \cite{Gover:2007})
\be
-\sigma\Big(\Delta+\frac{2\Rho}{d} w(w+d-1)\Big)\varphi=0\,
.\label{scomp} 
\ee 
Upon picking a special scale or a preferred unit system, we have broken the manifest unit invariance.  The above expression is {\it not}
unit invariant even if the scale $\sigma$, the field $\varphi$ and the metric transform simultaneously.  This is like breaking the manifest
Lorentz symmetry of  Maxwell's equations by picking Coulomb's gauge.
 
Since we are working on a conformally Einstein manifold, $\Rho$ is constant. At constant scale and constant $\Rho$, the
equation of motion~\eqn{scomp} describes a massive scalar field propagating in a curved background
\be
(\Delta-m^2)\varphi=0\, .
\ee
This allows us to read off the mass from~\eqn{scomp} and write the mass-Weyl weight relation
\be
m^2 = \frac{2\Rho}{d}\Big[\Big(\frac{d-1}2\Big)^2-\Big(w+\frac{d-1}2\Big)^2\Big]\, .\label{massweight}
\ee
This calibrates dimensionful masses with the dimensionless weights, and the transformation properties of the physical quantities can be encoded in term of masses rather than weights. 

\begin{figure}[h]
\centering
\includegraphics[width=100mm, height=10mm]{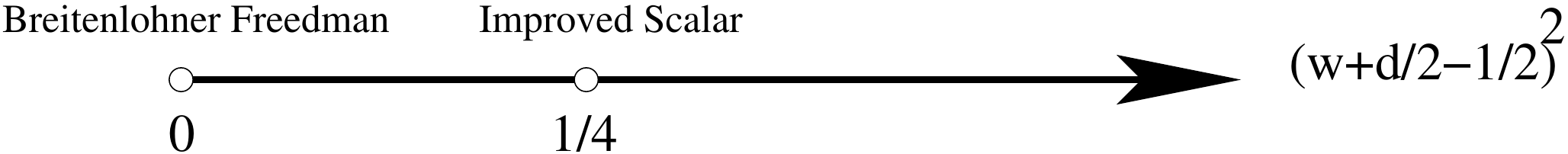}
\caption{ The Weyl weight $w$ can be reinterpreted as a scalar mass parameter. Generic values of $w$ (the thick line) give
massive theories, while $w=\frac12 -\frac d2$ and $w=1-\frac d2$ describe a scalar saturating the Breitenlohner--Freedman bound
(in Anti de Sitter space) and an improved scalar, respectively.\label{scalarmass}}
\end{figure}

In AdS, the mass-Weyl weight relationship produces  negative mass for some values of the Weyl weight, which seems completely disastrous. 
But, in fact, a slightly negative mass is allowed by Breitenlohner--Freedman bound~\cite{Breitenlohner:1982jf,Mezincescu:1984ev} for stable scalar
propagation in any space with constant curvature.  The Brietenlohner-Freedman bound is given by
\be 
m^2\geq \frac{\Rho}{2d}\,
(d-1)^2\, ,
\ee
which amounts to the reality of the tractor Weyl weights. 
The bound is saturated by setting the second term
in~\eqn{massweight} to zero, so that $w=\frac12 -\frac d2$. Since the Breitenlohner-Freedman bound
amounts to the reality of the weights, it is 
worth observing that, according to \eqn{massweight}, any real weight $w$ obeys this bound.
Figure~\ref{scalarmass} depicts the physical interpretation of the
various values of the Weyl weight~$w$.

As promised, we now analyze the scalar theory at special values of the Weyl weight to search for strictly Weyl invariant theory.
There is only one such value: $ w= 1- d/2$.
 At this weight, the scale tractor $I$ decouples and the
equation of motion~\eqn{sceom} can be rewritten as 
\be D^M \varphi =
0\, ,\label{spsc}
 \ee 
 because the top and middle slots of the Thomas
$D$-operator vanish as given in \eqn{SD}.  The same conclusion follows from a unit invariant
action principle given in \eqn {Saction} because the action
is independent of $\sigma$ when $w=1-\frac d2$.  Explicitly writing out ~\eqn{spsc}, 
it becomes 
\be 
\Big(\Delta-\frac R4\, \frac{d-2}{d-1}\, \Big)\varphi=0\, ,
\ee
which is the equation of motion for a conformally improved scalar field.  

Hence, in summary, the unit invariant equation of motion $I_M D^M\varphi =0$ describes a massive scalar field propagating in a curved background for generic $w$.  Upon picking a constant scale, the mass of the field is related to its Weyl weight, which in turn predicts the Breitenlohner-Freedman bound in AdS.  At the special value of $w=1-\frac d2$ the scale $\sigma$ decouples leaving a strictly Weyl invariant theory described by $D^M\varphi=0$.

\section{Vector Theory}
\label{spin1}
Conventionally, a vector field $V_\mu$ suffices to describe the dynamics of a spin-one particle.  However, a single vector field is not enough to write spin-one theory according to the
tractor philosophy espoused in Chapter~\ref{Conformal}.  Instead, we need to
arrange fields as $SO(d,2)$ multiplets transforming as tractors under
Weyl transformations. This necessitates the addition of auxiliary scalar
fields $(V^+, V^-)$ in order to build a tractor from $V_\mu$. So we start by 
introducing a weight $w$ tractor vector 
\be V^M=\begin{pmatrix}V^+\\ V^m
\\ V^-\end{pmatrix}\, , 
\ee 
and identify
 \be
V_\mu = e_\mu{}^m V_m\, . 
 \ee
The weight one vielbein transforms as $e_\mu{}^m\mapsto \Omega e_\mu{}^m$ and changes the weight of $V_\mu$ by one compared to $V^m$.  Under Weyl transformations, the components of the tractor-vector transform as  (see~\eqn{tractorvector}) 
\bea
V^+&\mapsto&\Omega^{w+1}V^+\, ,\nn\\[2mm] V_\mu\;
&\mapsto&\Omega^{w+1}(V_\mu+\Upsilon_\mu V^+)\, ,\nn\\[1mm]
V^-&\mapsto&\Omega^{w-1}(V^- -\Upsilon^\mu [V_\mu +\frac 12
  \Upsilon_\mu V^+])\, .\label{vecfo} 
\eea 
These formul\ae\ are
perhaps somewhat mysterious from a physical perspective. Firstly, what
do the extra fields $V^\pm$ mean? Secondly, how do we get rid of them?

The answer comes in the form of gauge invariance applicable to both massless and massive theories
that we soon describe in this section.  So, we pose 
\be
\delta V^M = D^M \xi\, ,\label{eee}
\ee
where the parameter $\xi$ is a weight $w+1$ scalar.  It is instructive to write these transformations out in components for independent fields
\bea \delta V^+ &=& (d+2w)(w+1)\, \xi\, ,\nn\\[2mm] \delta V_\mu \,
&=& (d+2w)\nabla_\mu \xi\, .\label{maxgauge} 
\eea 
Notice that we
recover the standard Maxwell type gauge transformation for the
vector~$V_\mu$, while the auxiliary $V^+$ is indeed a St\"uckelberg
field since its gauge transformation is a shift symmetry.  At $w = -1$, the auxiliary field, $V^+$ can be consistently 
set to zero allowing us to recover Maxwell equations from the Proca system. 

So much for $V^+$, what about $V^-$? Since the Thomas $D$-operator is null(\eqn{null}), the gauge transformations
obey a constraint
\be
D_M \delta V^M = 0 \, .
\ee
Therefore, we impose the same constraint  on our fields $V^M$
\be
D.V = 0 \,.
\label{DV}
\ee
This constraint is manifestly Weyl-covariant and can be algebraically solved for $V^-$:
\be
V^M = \begin{pmatrix}V^+\\[2mm]
V^m\\[2mm]
-\frac{1}{d+w-1} \left(\nabla.V -\frac 1{d+2w}\, \Big[\Delta -(d+w-1)\Rho\Big] V^+\right)
\end{pmatrix}\, ,
\ee
at least for $w\neq -\frac{d}{2}, 1-d$.  Therefore, the constraint eliminates $V^-$ as an independent field leaving
us with the necessary field content to describe both Proca and Maxwell systems. 

Next, we search for a tractor expression analogous to the Maxwell curvature but built out of the tractor-vector $V^M$. It must also be tractor gauge invariant.  We call it the ``tractor Maxwell curvature" and is given by
\be 
{\cal F}^{MN}=D^M V^N - D^N V^M\, .
\ee
Notice, it is {\it not} a curvature because $V^M$ is not a connection on the base manifold.   Its gauge invariance is manifest
because Thomas $D$-operators commute acting on scalars (for any
conformal class of metrics).   Evaluated at the constraint~\eqn{DV}, it is given by
\be
{\cal F}^{MN}=
\left(
\scalebox{.75}{\mbox{\begin{tabular}{ccc} $0$ & $(d+2w-2)(w+1) \wt V^n$&$-\frac{(d+2w-2)(w+1)}{d+w-1}\, \nabla.\wt V$ \\[3mm]
${\rm a/s}$& $(d+2w-2) F^{mn}$&
$\nabla_r F^{rm}
-(w\!+\!1)[2\Rho^m_r\wt V^r\! -\!\Rho \wt V^m \!+\!\frac 1{d+w-1} \nabla^m \nabla.\wt V]$
\\[3mm]
{ a/s}&{ a/s}&$0$\end{tabular}}}
\right)\, ,\label{FF}
\ee
Pay heed to the special arrangement of constituent expressions in this tractor connection; all possible spin one theories are incorporated  along with their constraints! More importantly, these expressions
transform into linear combinations of each other upon Weyl transformation.  

This setup should be compared with the familiar arrangement of electric and magnetic fields
in the standard Maxwell curvature.  Once we have the curvature tensor,  the equations of motion are given by
\be
\nabla_{\mu} F^{\mu\nu} = 0 \, \qquad \nabla_{[\rho} F_{\mu\nu]}= 0 \, .
\ee
We do something similar for our tractor setup; the only difference is that instead of contracting the Maxwell tractor curvature with a tractor covariant derivative, we contract it with a tractor vector.  Since we also want
tractor equations of motion to be manifestly unit invariant, the scale tractor is, once again, forced upon us by our symmetry principle.  We
propose the equation of motion to be
\be
 I_M {\cal F}^{MN}\equiv G^N
= 0.\label{maxeom} 
\ee
Upon expanding it, our equation of motion generalizes the scalar equation of motion ~\eqn{sceom}.  
\be
I\cdot D\,  V^N - I_M D^N V^M = 0\, .
\ee
The first term mimics the scalar equation of motion, while the second term has been forced by the gauge invariance.  
Now, we analyze the physics described by our tractor equations of motion. 

As before, we pick a preferred scale where $\sigma$ is constant and the background metric is 
Einstein. Next, we eliminate the higher derivatives that appear in the middle slot of  $G^M$ by differentiating and taking linear combinations
\be
G^m - \frac{1}{d+2w-2} \nabla^m G^+\equiv {\cal G}^m \,,\qquad G^+ = \sigma\frac{(d+2w-2)(w+1)}{d+w-1}\nabla \cdot \tilde{V} \,.
\ee
Having eliminated the higher derivates, miraculously, we find the Proca equation 
\be
{\cal G}^m=-\sigma \Big(\nabla_n F^{nm} +\frac{2\Rho}{d}(w+1)(d+w-2)\wt V^m\Big)\, . \label{Proca} \,
\ee
Here $ \wt V_\mu = V_\mu -\frac 1 {w+1} \nabla_\mu V^+$ is St\"uckelberg gauge invariant whose coefficient defines the mass of the system.
We define the mass term as the coefficient of the  $ \wt V_m$ producing a mass-Weyl weight relation 
\be
m^2=\frac{2\Rho}{d}\Big[\Big(\frac{d-3}{2}\Big)^2-\Big(w+\frac{d-1}{2}\Big)^2\Big]\label{maxmass} \, .
\ee
This mass-Weyl weight relation predicts a new Breitenlohner--Freedman bound
\be
m^2\geq\frac{2\Rho}{d}\, \Big(\frac{d-3}{2}\Big)^2\, ,\ee
for the Proca system. 

Now, we analyze the system at different values of the weight.  At generic $w$ we can gauge away the auxiliary St\"uckelberg field using~\eqn{maxgauge}, and the divergence of the equation of motion~\eqn{Proca} $\nabla_\mu {\cal G}^\mu=0$ implies $\nabla_\mu V^\mu=0$, so we obtain a wave equation for
a massive, divergence free vector field
\be
\left\{
\begin{array}{l}
\Big(\Delta +{\textstyle\frac{ {\ts 2}\Rho}{{\ts  d}}}\, [w(w+d-1) -1]\Big) V_\mu=0\, ,\\[5mm]
\quad \nabla^\mu V_\mu = 0\, .
\end{array}
\right.
\ee
 Examining the gauge transformations~\eqn{maxgauge} we see that $w=-\frac d2$ and $w=-1$
play special {\it r\^oles}. The case $w=-d/2$ is deceptive,
 since it appears to be a distinguished value.  In fact it actually amounts to
the massive Proca system\footnote{The technical details are as
  follows: Firstly, both $V^+$ and $V_\mu$ are inert under gauge
  transformations~\eqn{eee}.  Moreover $V^-$ decouples both from the
  field constraint and equation of motion $D\cdot V=0=G^M$; in fact if we assume invertibility of
  $(\Delta-\frac\sRho2 )$, then
 it
  can be gauged away using the gauge invariance $\delta V^-=D^-\xi =
  -(\Delta-\frac{\sRho}2)\xi$.  Then $\Rho V^+$ turns out to be proportional to
  $\nabla.V$ }
that
we found earlier for generic weights.

The case $w=-1$ is far more interesting. The auxiliary field $V^+$ is
inert under the transformations~\eqn{maxgauge}. Or in other words the
tractor quantity $X\cdot V$ is gauge invariant. Hence we may
consistently impose an additional constraint \be X_M V^M = V^+ = 0\, .  \ee
In that case, the equation of
motion~\eqn{Proca} simplifies to  \be G^n = -\sigma \nabla_m F^{mn}= 0\, , \ee
where $F_{\mu\nu}=2\nabla_{[\mu}V_{\nu]}$ is the usual Maxwell
curvature.  So $G^n=0$ is {\it precisely the system of Maxwell's
  equations in vacua!}  This is hardly surprising since at this value
of $w$, along with the additional $X\cdot V$ constraint, we have a
theory of a single vector $V_{\mu}$ along with its usual Maxwell gauge
invariance.  Notice, the tractor technology does not predict Weyl
(co/in)variance of Maxwell's equations in arbitrary dimensions since
the construction does involve choosing a scale. Rather the combination $-\sigma
\nabla_m F^{mn}$ belongs in the middle slot of a weight $-2$ tractor
vector.  Note also that the value $w=-1$ along with the constraints
$X\cdot V=D\cdot V=0$ implies the usual Weyl transformation rule for
the Maxwell potential~$V_\mu$, namely that $V_\mu$ is inert.

It turns out there is a further distinguished value of the weight
$w$. It is the canonical engineering dimension of a $d$-dimensional
field, namely~$w=1-\frac d2$ (just as for an improved scalar).
 In four
dimensions this value coincides with the $w=-1$ Maxwell one!  To see
that this value is special we evaluate~\eqn{FF} at this distinguished value and get
\be
{\cal F}^{MN}=
\left(
\scalebox{.75}{\mbox{\begin{tabular}{ccc} $0$ & $ 0 $ & $ 0 $ \\[3mm]
${\rm a/s}$& $ 0 $&
$\nabla_r F^{rm}
-(2\!-\! \frac{d}{2})[2\Rho^m_r\wt V^r\! -\!\Rho \wt V^m \!+\!\frac 2 d \nabla^m \nabla.\wt V]$
\\[3mm]
{ a/s}&{ a/s}&$0$\end{tabular}}}
\right)\, . \label{FF}
\ee
Notice that at $w=1-\frac d2$ every component of ${\cal F}^{MN}$ vanishes save for ${\cal F}^{m-}$.
Therefore there is no longer any need to introduce the scale $\sigma$ to obtain the field equations. Just as the scale tractor decoupled in~\eqn{spsc} for the improved
scalar field, it decouples again and we may simply replace~\eqn{maxeom} by vanishing of the tractor Maxwell curvature
\be
{\cal F}^{MN}=0\, .
\ee
This gives Weyl invariant equations of motion that depend only on the combination $\wt V_\mu$.
Without loss of generality, therefore, we can gauge away the St\"uckelberg field and are left with
a theory of a single vector $A_\mu\equiv \wt V_\mu\lvert_{V^+=0}$, with Weyl transformation law
\be
A_\mu\mapsto \Omega^{-\frac{d-4}2} A_\mu\, .
\ee
Writing out these apparently novel equations explicitly gives
\be
\Delta A_\mu - \frac 4 d \, \nabla^\nu \nabla_\mu A_\nu +\frac{d-4}{d}\Big(2\, \Rho_\mu^\nu A_\nu -\frac {d+2}2\,  \Rho A_\mu\Big)=0\, .
\ee
These equations have in fact been encountered before---they are
precisely the Weyl invariant, but non-gauge invariant vector theory of
Deser and Nepomechie~\cite{Deser:1983tm}. When $d=4$, they revert to
Maxwell's equations. Observe that the intersection of the conditions
$w=-1$ and $w=1-d/2$ is at $d=4$, 
precisely the value when the Maxwell
theory is Weyl invariant. It is rather pleasing that the simple
tractor equation~\eqn{maxeom} directly generates the curvature
couplings required for Weyl invariance.

\begin{figure}[h]
\begin{center}
\epsfig{file=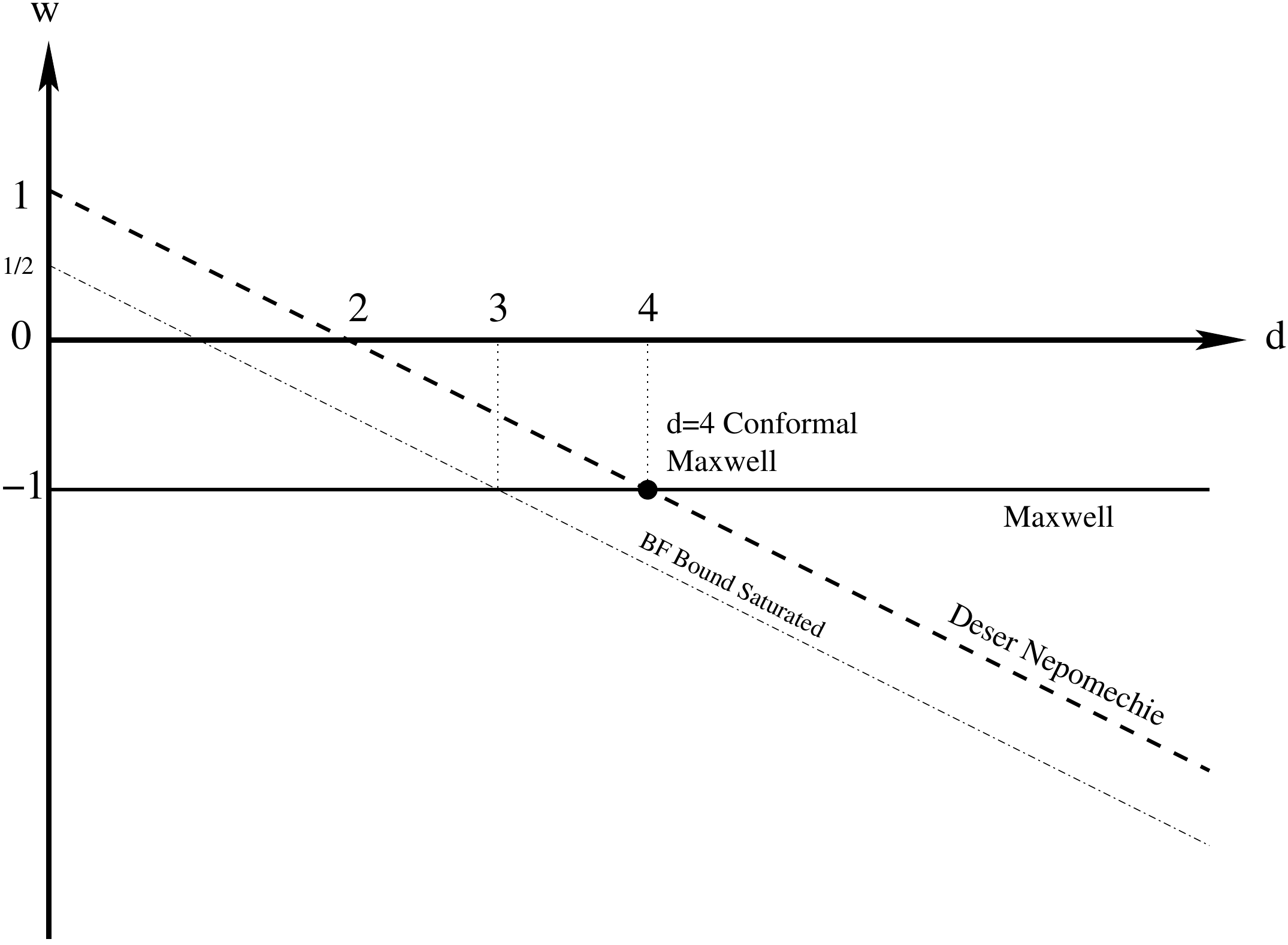,height=9cm}
\end{center}
\caption{A plot of the theories described by the tractor Maxwell system as a function of Weyl weight and dimension. Theories
saturating a vector Breitenlohner--Freedman bound appear at $w=\frac12-\frac d2$. \label{vectormass}}
\end{figure}

Recapitulating the situation, we started with a tractor expression for a manifestly unit invariant spin-one system with its gauge invariance and constraints. 
Upon picking a constant scale, we recovered the Proca system which at the special value of $w=-1$ produced Maxwell's equations.  At $w= 1-d/2$,
 the scale tractor decoupled completely, leaving behind the conformally invariant spin-one equations of Deser and Nepomechie. 
 The various theories we have described using the single
equation~\eqn{maxeom} are plotted in Figure~\ref{vectormass}.

\subsection{Spin One Tractor Action}
To complete the discussion of spin one systems, we construct a tractor action for the Proca system.  The tractor equations of motion $G^M$ do not arise as a consequence of varying an action. However,
the Proca equation~\eqn{Proca} follows from an action principle
$S[g_{\mu\nu},\sigma,V_\mu,V^+]$ which evaluated at a constant
curvature metric and constant scale reads 
\bea 
\label{Sform}
S&=&\int
\frac{\sqrt{-g}}{\sigma^{d+2w-2}}\Big\{ -\frac 1 4
F_{\mu\nu}F^{\mu\nu} +\frac{\sc P}{d}\, (w+1)(d+w-2)\ \wt V^{m}\, \wt V_{m}
\Big\}\nn\\[3mm] &=&\frac12 \int \frac{\sqrt{-g}}{\sigma^{d+2w-2}}
\Big\{V^+ {\cal G}^- + V^m {\cal G}_m \Big\} \, .\label{s0} \eea 
Here
${\cal G}^m$ is given in~\eqn{Proca}, and gives the Proca equation, while
\be {\cal G}^-=-\frac{2\Rho}{d}\frac{(d+w-2)(d+w-1)}{(w+1)(d+2w-2)}\,
G^+\, . \label{GG-} \ee 
The pair of equations of motion ${\cal
  G}^m=0={\cal G}^-$ are those that come from varying the action
\eqn{s0}.  Although the action is obviously gauge invariant with respect
to~\eqn{maxgauge}, Weyl invariance of this action is not manifest.
Therefore we construct an equivalent tractor action \be
S[g_{\mu\nu},\sigma,V^M] = \int\sqrt{-g} \, V_M {\cal H}^M\,
,\label{s2} \ee where the weight $-d-w$ tractor ${\cal H}^M$ is given
(at the choice of scale where $\sigma$ is constant) by 
\be {\cal H}^M
= \sigma^{1-d-2w}\begin{pmatrix}0\\[2mm] {\cal G}^m\\[2mm]{\cal
    G}^- \end{pmatrix}\, .\label{H} \ee 
Because
${\cal H}^M$ is a tractor vector of weight $-w-d$ the action enjoys the Weyl
invariance \be S[g_{\mu\nu},\sigma,V^M] =
S[\Omega^2g_{\mu\nu},\Omega\sigma,\Omega^wU^M{}_NV^N] \, .\label{s1}
\ee Moreover, the integration by parts formula~\eqn{parts}, along with
the invariance of \eqn{s0} (and hence \eqn{H}) with respect to the
gauge transformation \eqn{maxgauge}, implies the Bianchi identity \be
D_M {\cal H}^M=0\, \label{DH} .  \ee 
In summary, starting with weight $w-1$ field equations $G^M$
in~\eqn{maxeom}, which obey $I\cdot G=0=D\cdot G$, but do not arise
directly from varying an action, we have formed equations of motion
${\cal H}^M$ which do follow from an action and obey 
\be X\cdot {\cal
  H}=0=D\cdot {\cal H}\, \label{DHX}.  
\ee 
Now the displayed Weyl invariant equations determine $\cal H$ from its 
middle slot.  To write explicitly a tractor formula for ${\cal H}^{M}$, we note that
the tractor 
\be \stackrel{\circ}{\cal
  G}{\!}^M= \Big\{G^M-\frac 1{(d+2w-2)^2} D^M X\cdot
G\Big\}\, , \ee 
has middle slot equaling the Proca equation
$\stackrel{\circ}{\cal G}{\hspace{-1.2mm}}^m={\cal G}^m$.
Then  we can produce a tractor obeying~\eqn{DHX}
of any desired weight $k+w-1$ from the quantity $D_{N}\Big[\sigma^{k}X^{[N}  (\stackrel{\circ}{\cal
  G}{\!}^{M]}-Y^{M]} X\cdot \stackrel{\circ}{\cal
  G})\Big]$ where the scale dependent, weight~$-1$ tractor~$Y^{M}$ obeys~$X\cdot Y=1$ and is constructed explicitly in 
 Appendix~\ref{projectors} . In particular (excepting two exceptional weights) its middle 
slot is a non-zero multiple of a  power of $\sigma$ times ${\cal G}^{m}$.
  
  The projection methods outlined in Appendix~\ref{projectors} can be used to express the
integrand $\sqrt{-g} \, {\cal L}$ of the action~\eqn{s2} as a tractor scalar \bea
\sigma^{d +2w-2}{\cal L} \; =\;& -&\frac 1{4(d+2w-2)^2}\,
\widehat {\cal F}_{MN} \widehat {\cal F}^{MN} \nn\\[2mm]
&-&\frac1{2\sigma^2}\, (w+1)(d+w-2)\, (I\cdot I)\; \wt V^M \wt V_M\, .
\eea 
Here the hat on the tractor Maxwell curvature denotes a tractor
covariant (but scale dependent) projection onto its middle slot as
explained in Appendix~\ref{projectors}. The first term is reminiscent of the Maxwell
action, while the second reminds one of the Proca mass term.

\subsection{On-Shell Approach}

Although, we have already given a complete and unified description of the Maxwell and Proca systems using tractors, 
it is useful to have an on-shell approach thanks to its simplicity and easy applicability to higher spin systems.

As prelude, we review the on-shell approach to the Proca equation in components. 
As explained above, the Proca system is described by the pair of equations
\be
\left\{
\begin{array}{l}
\Big(\Delta +{\textstyle\frac{ {\ts 2}\Rho}{{\ts  d}}}\, [w(w+d-1) -1]\Big) V_\mu=0\, ,\\[5mm]
\quad \nabla^\mu V_\mu = 0\, .
\end{array}
\right.\label{system}
\ee
The latter, divergence constraint, implies that, in $d$-dimensions the system describes $d-1$ propagating modes
which obey the Klein--Gordon equation. However, this is only true at generic values of the mass and therefore at generic Weyl weights. 
 For special values of $w$ there may be ``residual'' gauge invariance implying a further 
reduction in degrees of freedom. Indeed, 
the (Maxwell) gauge transformation
\be
\delta V_\mu = \nabla_\mu \xi
\ee
leaves the divergence constraint invariant whenever the gauge parameter $\xi$ obeys
\be
\Delta \xi = 0.\label{5}
\ee
Then the identity
\be
[\Delta,\nabla_\mu]\xi = \frac{2\Rho}{d}\, (d-1) 
\nabla_\mu\xi
\ee
implies that the variation of the left hand side of the Klein--Gordon equation equals
$
-m^2\nabla_\mu \xi\, ,
$
which vanishes whenever the Maxwell mass (defined in~\eqn{maxmass}) does.
{\it I.e.}, when $m^2=0$, there is a residual gauge invariance which removes an
additional degree of freedom so that the system of equations~\eqn{system}
describes $d-2$ propagating photon modes. The divergence constraint is then
reinterpreted as the Lorentz choice of gauge.

The above discussion was completely standard, but let us see if it can
be reproduced using tractors: The Proca system is now described by a
weight~$w$ tractor subject to \be D\cdot V=0\, .\label{1} \ee We can
fix the St\"uckelberg gauge invariance~\eqn{maxgauge} by setting \be
X\cdot V=0\, .\label{2} \ee Then the pair of equations~\eqn{system}
correspond to tractor equations \bea I\cdot D\ V^M&=&0\,
,\label{3}\\[2mm] I\cdot V \;\;&=&0\, .\label{4} \eea Our task now is
to search for residual gauge symmetries of the system of
equations~(\ref{1}-\ref{4}).  Clearly we should study the
transformation \be \delta V^M = D^M \xi\, , \ee where $\xi$ has weight
$w+1$.  Equation~\eqn{1} is trivially invariant under this gauge
transformation, but~\eqn{4} requires
 \be 
 I\cdot D\ \xi = 0\, . 
\label{const}
 \ee
Notice that this is the tractor analog of~\eqn{5}. As a consequence,
equation~\eqn{3} is now invariant so the final condition is given by
varying~\eqn{2} which gives \be X\cdot D\ \xi = (d+2w)(w+1) \xi = 0\,
.  \ee 
Two values of $w$ make the above equation vanish: $ w= -\frac{d}{2},  -1$.
At $w=-d/2$, ~\eqn{const} reduces to $D^M \xi \equiv 0$ which is
uninteresting.  However, at $w=-1$ we find a genuine residual gauge
invariance. This value substituted in ~\eqn{maxmass} of course implies $m^2=0$ in agreement with the
above component computation.

\section{Spin Two}

The massive spin two system is more subtle than its spin one Proca
relative. Its massless case corresponds to (linearized) gravitons so
introducing a general background would lead us to the non-linear
Einstein theory of gravitation. Also, it is generally accepted that
massive spin two systems cannot be coupled to general curved
backgrounds consistently
(see~\cite{Aragone:1979bm,Kobayashi:1978mv,Kobayashi:1978xd,Buchbinder:1999ar,Buchbinder:2000fy,Deser:2001dt}). However,
these difficulties can be circumvented in cosmological backgrounds
modulo subtleties--special tunings of the mass parameters lead to the
partially massless spin two
theory~\cite{Deser:1983tm,Higuchi:1986py,Deser:2001pe}.  In the
following we specialize to constant curvature theories\footnote{A
  study of non-minimal gravitational couplings for massive spin two
  fields yields consistent propagation in Einstein
  backgrounds~\cite{Buchbinder:1999ar,Buchbinder:2000fy}. Therefore we
  strongly suspect that there exist non-minimal tractor couplings that
  extend the results of this Section to conformally Einstein
  metrics. We reserve this issue for future study.}. We first introduce the notion of partially masslessness and then construct various possible spin two theories---massive gravitons,
gravitons, partially massless spin two---using the simple on-shell
approach.

\subsection{Partial Masslessness}
\label{Mass}
The number of helicity states of a particle with spin $s$  in four dimensions is given by $2s+1$, which is the dimension of the representation of the rotation group $SO(3)$.  For instance, a spin one photon field, in four dimensions, should have three helicity states in its rest frame.  But we only observe two in nature since the third one is eliminated by the condition of gauge invariance.  These two helicity states ($\pm1$) form the representation of $SO(2)$, the little group of $SO(3)$.  So we can define the massless photon field as the one with two helicity states ($\pm 1$), thus, establishing a relationship between mass and helicity. 

\begin{figure}[h]
\centering
\includegraphics[width=150mm, height=40mm]{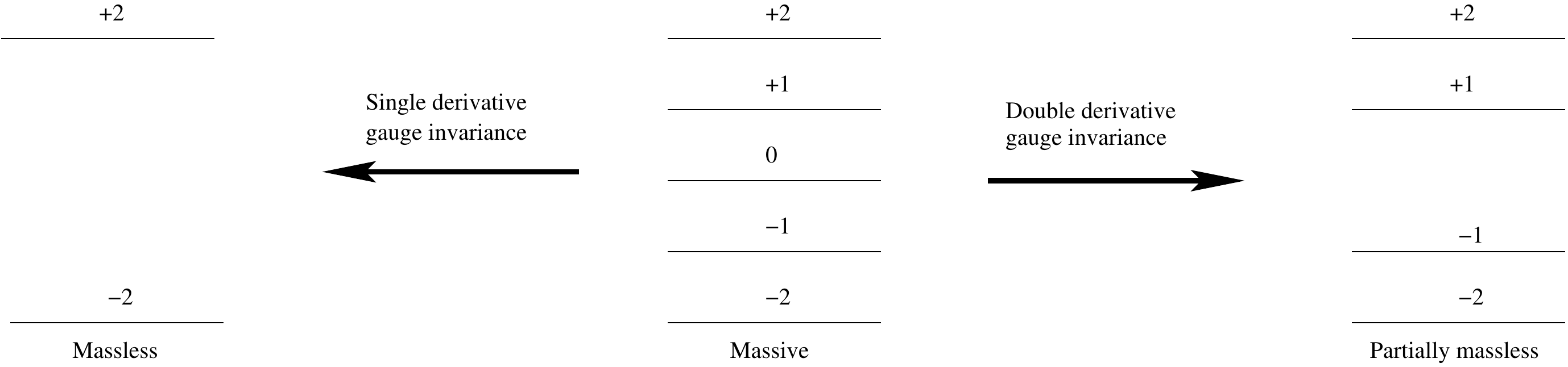}
\caption{Spin two helicities categorizing massive, massless, and partially massless spin two particles.  \label{Helicity}}
\end{figure}

Similarly, for a spin two particle in four dimension, we expect five different helicities or degrees of freedom.  In flat space, we only have a single derivative gauge invariance given by
\be
\delta V_{\mu\nu}= \nabla_{\mu} \xi_{\nu}
\ee
which removes three degrees of freedom leaving behind  two helicities ($\pm 2 $).  Hence, in flat space we either have five degrees of freedom or two degrees of freedom leading to either a massive or a massless spin two particle.  The curved case scenario is not so simple because an additional double derivative gauge invariance  
\be
\delta V_{\mu\nu} = \Big\{\nabla_{\mu}\nabla_\nu + \frac{2\Rho}{d}\ g_{\mu\nu}\Big\} \xi \, .
\ee
shows up leading to another possibility.  This gauge invariance removes one degree of freedom from the original five leaving behind the four degrees of freedom  ($\pm 2$, $\pm1$) which represents the partially massless particle.  In other words, we have three possibilities in curved space: massless, massive, and partially massless particle.  All possible spin two helicity states in curved spaces are given in Figure~\ref{Helicity}.


\subsection{On-Shell Approach}

The basic field is a weight $w$, symmetric rank two tractor $V^{MN}$.  A unified description of spin two is more complicated than spin one because we have
 a large number of unwanted fields, thus forcing us to introduce more
constraints.  As the spin two generalization of the spin one on-shell tractor equations~(\ref{1}-\ref{4}) we consider\footnote{The notation $D\cdot V^N$ is shorthand for $D_M
V^{MN}$.}
\bea
D\cdot V^N&=&0\, ,\label{DdotV}\\[1mm]
X \cdot V^N&=&0\, ,\label{XV} \\[1mm]
I\cdot D\ V^{MN}&=&0\, ,\label{IDV}\\[1mm]
I\cdot V^N &=& 0\, ,\label{IV}\\[1mm]
V_N^N&=&0\, .\label{trV}
\eea
The final trace relation is the only relation which is not a direct
analogue of a relation/equation in the spin one system of equations.
Written out in components these equations amount to the triplet of
equations. As usual special attention is required at certain
weights. We will suppress discussion of this as the analysis is a
straightforward generalisation of that above for lower spins.
\bea
&&\Big(\Delta +\frac{2\Rho}{d}[w(w+d-1)-2]\Big)\, V_{\mu\nu}=0\, ,\\[2mm]
&&\; \; \nabla.V_{\nu}=0\, ,\\[2mm]
&&\quad V_\nu^\nu=0\, ,
\eea
where the symmetric tensor $V_{\mu\nu}$ sits in the middle slot of $V^{MN}$.
These are precisely the on-shell equations of motion for a massive spin two graviton.
Using the usual Pauli--Fierz definition of the spin two mass for which
\be
\Big(\Delta -\frac{4\Rho}{d}-m^2\Big)V_{\mu\nu}=0\, ,\label{PF2}
\ee
we find the spin two mass-Weyl weight relation
\be
m^2=\frac{2\Rho}{d}\Big[\Big(\frac{d-1}{2}\Big)^2-\Big(w+\frac{d-1}{2}\Big)^2   \Big]\, .\label{spin2mass}
\ee
In agreement with the original Breitenlohner--Freedman bound~\cite{Breitenlohner:1982jf,Mezincescu:1984ev}, we get
\be
m^2\geq \frac{2\Rho}{d}\Big(\frac{d-1}{2}\Big)^2\, .
\ee
Note that the expression~\eqn{spin2mass} factors as $m^2=-\frac{2\sRho}{d}\, w(d+w-1)$. We will find that the value $w=0$ 
 corresponds to the gauge invariant, massless graviton limit where the Pauli--Fierz mass term is absent.

Let us now search for residual gauge invariances of the tractor equations of motion~(\ref{DdotV}-\ref{trV}).
First we consider transformations
\be
\delta V^{MN} = D^{(M}\xi^{N)}\, .\label{resid}
\ee 
where $\xi^N$ is a weight $w+1$ tractor vector. It is easy to see that equations~\eqn{DdotV},~\eqn{IDV},~\eqn{IV} and~\eqn{trV}
are invariant if
\be
I\cdot D \ \xi^N = 0 = I\cdot \xi = D\cdot \xi\, .\label{dots}
\ee
Invariance of~\eqn{XV}, however, requires
\be
(w+1)(d+2w) \xi^M + X^N D^M \xi_N=0\, .\label{theeq}
\ee 
Contracting this relation with $X_M$ yields
\be
2(w+1)(d+2w) X\cdot \xi=0\, .
\ee
It is not hard to verify that $D^{(M}\xi^{N)}\equiv 0$ at $w=-d/2$ so this value is uninteresting.
Hence either $w=-1$ or $X\cdot \xi=0$.  First we analyze the case $X\cdot \xi=0$. Using~\eqn{dots}
in conjunction with $X\cdot\xi=0$ we find
(using Appendix~\ref{Tractor_components}) that 
\be
X^N D^M \xi_N=-(d+2w)\xi^M\, .
\ee
Hence equation~\eqn{theeq} says
\be
w(d+2w) \xi^M=0\, .
\ee 
Thus we learn that~$\delta V^{MN}=D^{(M}\xi^{N)}$ is a residual gauge invariance\, 
of the onshell equations~(\ref{DdotV}-\ref{trV}) at weight $w=0$ with~$I\cdot D\, \xi^M=0=I\cdot\xi=D\cdot \xi=X\cdot\xi$. 
This weight corresponds to vanishing Pauli--Fierz mass and in components the residual gauge invariance reads
\be
\delta V_{\mu\nu}=\nabla_{(\mu}\xi_{\nu)}\, ,\quad \mbox{ where }\; \nabla.\xi=0=\Big(\Delta-\frac {2\Rho}d\ (d-1)\Big)\xi_\mu\, .
\ee
This is a residual, linearized diffeomorphism so, as promised,  the $w=0$ theory describes constant curvature gravitons. 

Now let us turn to the other case $w=-1$. There the equation~\eqn{theeq} becomes
\be
X^N D^M\xi_N=0\, .
\ee
A solution to  this equation is given by setting
\be
\xi^M=D^M\alpha
\ee
where $\alpha$ has weight one. Comparing with~\eqn{resid} we see that $\alpha=\frac 1d X\cdot \xi$
and have therefore found a new residual gauge invariance
\be
\delta V^{MN} = \frac 1d D^M D^N X\cdot \xi\, . 
\ee
In components this transformation reads
\bea
\delta V_{\mu\nu} = (d-2)\Big\{\nabla_{\mu}\nabla_\nu + \frac{2\Rho}{d}\ g_{\mu\nu}\Big\} \xi^+ \, ,\quad\mbox{ where } (\Delta+2\Rho)\xi^+=0\, .
\eea
This double derivative, scalar gauge invariance is one that has been seen before -- it is the on-shell residual gauge invariance
of a partially massive spin two field~\cite{Deser:1983tm,Deser:2001pe}. The weight $w=-1$ corresponds to a mass
\be
m^2=\frac{2\Rho}{d}\, (d-2)\, .
\ee
In four dimensions, this gives the well-known result $m^2=2\Lambda/3$. This value gives a positive squared mass
in de Sitter space where $\Rho$ and $\Lambda$ are positive, and obeys (as does any real weight $w$) the Anti de Sitter
Breitenlohner--Freedman bound~\eqn{BFs}.

\subsection{Off-Shell Approach} 

In Section \ref{spin1}, we provided a unified picture of the Proca and Maxwell equations in a single tractor equation
\be 
I_M {\cal F}^{MN} =0\, .
\ee
Similarly, massive gravitons, gravitons and partially massless spin two theories can also be unified in a single equation
\be
I_R \Gamma^{RMN}=0\, .
\ee
Here $\Gamma^{RMN}$ are quantities that will be termed tractor Christoffel 
symbols. The reason for this name will soon be clear, but the reader is warned that we are  not asserting that these are connection coefficients.
 Let us now explain this result in detail
by deriving it from first principles.

The model is described in terms of a weight $w$, rank two, symmetric tractor tensor.
The starting point, from which everything follows, is the gauge invariance
\be
\delta V^{MN}=D^{(M}\xi^{N)}\, .\label{2gauge}
\ee
There are various possibilities for the weight $w+1$ gauge parameter $\xi^M$. For example,
we could leave it unconstrained or ask it to satisfy relations built from $D^M$, $X^M$ and $I^M$.
The ``correct'' choice can be determined by comparison with the residual gauge symmetries discussed above and is\footnote{The model where
$D\cdot \xi=0$ is also interesting since it leads to the Weyl invariant spin two model for a trace-free symmetric rank two tensor
introduced by Deser and Nepomechie~\cite{Deser:1983tm}.}
\be
I\cdot \xi = 0\, .\label{paramc}
\ee
Hence in components
\be
\xi^M=\begin{pmatrix}\xi^+\\[1mm] \xi^m\\[1mm] \frac{\sf P}{d}\xi^+\end{pmatrix}
\ee
 and
\bea
\delta V^{++}&=&\;\;(d+2w)(w+1)\xi^+\, ,\nn\\[2mm]
\delta V^{m+}&=&\!\frac12 (d+2w) (w\xi^m+ \nabla^{m} \xi^+)\, ,\nn\\[1mm]
\delta V^{mn}&=&\;\; (d+2w)\Big(\nabla^{(m}\xi^{n)}+\frac{2\Rho}d \eta^{mn} \xi^+\Big)\, ,\label{2g}
\eea
while all other transformations are dependent on these ones. These are exactly the
gauge transformations of a St\"uckelberg approach to massive spin two excitations~\cite{Zinoviev:2001dt,Hallowell:2005np}.
Generically, they allow the auxiliary fields $V^{++}$ and $V^{m+}$ to be gauged away. 
At $w=0$, however, one obtains a massless graviton theory because the $\xi^+$ invariance
gauges away $V^{++}$, which in turn, makes $V^{m+}$ gauge inert.  Therefore,   $V^{m+}$ and can be gauge invariantly set to zero, 
leaving  linearized diffeomorphisms \be\delta V_{\mu\nu}=d\nabla_{(\mu}\xi_{\nu)}\, .\ee 
At $w=-1$, the auxiliary $V^{++}$ is inert and can be set to zero, while $V^{m+}$ is also inert
so long as $\xi^+$ gauge transformations are accompanied by the compensating transformation
$\xi^m=\nabla^m\xi^+$ which yields the partially massless gauge transformation
\be
\delta V_{\mu\nu}=(d+2w)\Big(\nabla_{\mu}\nabla_\nu \xi^{+}+\frac{2\Rho}d \, g_{\mu\nu} \xi^+\Big)\, .
\ee
These results are consistent with the ones found in the above on-shell approach.

Having established the correct gauge transformations, we can now develop dynamics for our theory.
Since we are working on a conformally flat background where~\eqn{D_commute} holds, the gauge transformations $\delta V^{MN}$ in~\eqn{2gauge} obey a constraint
\be
D_M\delta V^{MN} = \frac12 D^N \, \delta V^M_M\, .
\ee
Therefore we impose the same constraint on our fields $V^{MN}$
\be
D\cdot V^N -\frac 12 D^N V^M_M=0\, .\label{Vc}
\ee
This constraint implies that the independent field content of the model is
the physical spin two field $V_{\mu\nu}$ along with the auxiliaries $V^{m+}$ and $V^{++}$.

To build field equations we form the {\em tractor Christoffel symbols}
\be 2\Gamma^{RMN}= D^M V^{NR}+D^N V^{MR} - D^R V^{MN}\, ,\label{Gids}
\ee 
which, as mentioned above, are not claimed here to be related to
connection coefficients for any connection.
Unlike the tractor Maxwell curvature, these are not gauge invariant but transform as
\be
\delta \Gamma^{RMN} = \frac 12 D^M D^N \xi^R\, .\label{Gamgauge}
\ee
Using \eqn{D_commute} and \eqn{Vc}, it's easy to see that  they obey a trace and Thomas $D$ divergence identity
\be
\Gamma^{RM}{}_M=0=D_M \Gamma^{RMN}\, .\label{Cid}
\ee

The Christoffel gauge transformation~\eqn{Gamgauge} combined with the parameter
constraint~\eqn{paramc} imply that the tractor Christoffels contracted with $I^M$ are gauge invariant.
Hence we propose the gauge invariant equations of motion 
\be
G^{MN}=-2 I_R \Gamma^{RMN}=0\, .\label{2eom}
\ee
which are also unit invariant by construction. 
These equations  are, of course, not all independent. 
In fact we expect them to obey relations corresponding to the field constraint~\eqn{Vc} as well as 
a Bianchi identity coming from the gauge invariance~\eqn{2gauge}.
This is indeed the case; the Christoffel identities~\eqn{Cid} imply
\be
G^M_M=0=D_M G^{MN}\, .
\ee
These are alone not sufficiently many relations, since we would predict a pair of tractor vector relations.
However, a simple computation shows that $G^{MN}$ contracted with $I^M$ is parallel to the Thomas $D$ operator
\be
I_M G^{MN}=D^N\!{\cal X}\, ,
\ee
with ${\cal X}=I_M I_N V^{MN}$. 

It is easy to compare our proposed equations of motion with those found above via an on-shell approach.
Expanding out~\eqn{2eom} gives
\be
I\cdot D\ V^{MN} - 2 D^{(M} I\cdot V^{N)}=0\, .
\ee
It can be checked explicitly that choosing a gauge where $V^{++}=V^{m+}=0$ and using the field equations
implies that $X\cdot V^N=I\cdot V^N=D\cdot V^N=0$. In turn $I\cdot D\ V^{MN}=0$ 
which yields the on-shell equations~(\ref{DdotV}-\ref{trV}). 

Instead of choosing a gauge however, one can also compute by {\it tour de force} the component expressions for $G^{MN}$
and verify that they reproduce the known St\"uckelberg equations of motion for massive spin two in constant curvature backgrounds.
We devote the remainder of this Section to this calculation.

Our first step is to solve the field constraint~\eqn{Vc} for $(V^{--}, V^{m-},V^{+-})$ in terms of the independent
field components $(V^{mn}, V^{m+}, V^{++})$. The results are displayed in Table~\ref{sol_to_constraints}. We employ the symmetric
tensor algebra notation explained in Appendix~\ref{symmetric}. In this vein we have denoted
\be
V=V_{\mu\nu} dx^\mu dx^\nu\, \mbox{ and } V^\pm= V^\pm_\mu dx^\mu\, ,
\ee
and similarly for the components of $G^{MN}$.
Next we must compute various components of the tractor Christoffel symbols~\eqn{Gids}; the results are given in Table~\ref{tractor_christoffels}.

\newpage

\begin{landscape}
\begin{table}[h]
\begin{tabular}{|c|}
\hline \\[2mm]
$V^{+-} = \frac{1}{d(d+2w)}\left( [\Delta-(d+w)\Rho]V^{++}-(d+2w+2)\,{\bf div}\, V^+ +\frac{w^2+w+d+wd/2}{2}\, {\bf tr}\, V \right)$  \\[4mm] \hline 

\\[2mm]
$ V^{-}   =   \frac{1}{(d+w)(d+2w)} \Big ( [\Delta -\frac{\sRho}{d}(d(d+w)+2(w+1))] V^{+} -\frac{d+2w}{2}\, {\bf div}\, V+\frac{1}{d}\, {\bf grad}\, [\Delta -(d+w-2)\Rho]V^{++}$ \\ 
$-\frac{d+2w+2}{d}\, {\bf grad}\ {\bf div}\, V^+ +\frac{d(d+2w)+w(d+2w+2)}{4d}\, {\bf grad}\ {\bf tr}\,  V \Big )$ 
\\[4mm] \hline

\\[2mm]
$V^{--} = \frac{1} { d(d+w)(d+w-1)(d+2w)} \Big (  \frac{d(d+2w)}{2}\, {\bf div}^2 \,  V - \frac{1}{2}\{[(d+w)^2+w]\Delta + w(d+w)(d+w-1)\Rho\}\, {\bf tr}\,  V$   \\
$+ 2\{(w+1)\Delta + [(d+w)(d+w-1) +2(w+1)]\Rho\}\, {\bf div}\, V^+ -[\Delta^2+2\Rho\Delta - (d+w)(d+w-1)\Rho^2]V^{++} \Big ) $ \\[4mm] \hline

\end{tabular}
\caption{Solutions to the constraint $D.V^N= \frac{1}{2}D^NV$}
\label{sol_to_constraints}
\end{table}

\newpage

\begin{table}
\begin{tabular}{|c|}
\hline  \\[2mm]
$2\Gamma^{+++} = w(d+2w-2)V^{++}$    
\\[2mm] \hline

\\[2mm]
$2\Gamma^{++n} = (d+2w-2)(\nabla^nV^{++}-2V^{n+}) $
\\[4mm] \hline 

\\[2mm]
$2\Gamma^{r++}=-(d+2w-2)(\nabla^n V^{++}-2(w+1)V^{n+})$ 
\\[4mm] \hline

\\[2mm]
$2\Gamma^{+mn}=-(w+2)(d+2w-2)V^{mn}+2(d+2w-2)\nabla^{(m}V^{n)+}+\eta^{mn}2(d+2w-2)(\frac{\sRho}{d}V^{++}+V^{+-} )$
 \\[4mm] \hline

\\[2mm]
$2\Gamma^{r+n}= w(d+2w-2)V^{rn}+(d+2w-2)(\nabla^nV^{+r} - \nabla^rV^{+n})$ 
\\[4mm] \hline

\\[2mm]
$2\Gamma^{rmn}=(d+2w-2)(2\nabla^mV^{rn} - \nabla^rV^{mn} + 2\Rho^{mn}V^{+r} + 2 \eta^{mn}V^{-r})$
\\[4mm] \hline

\end{tabular}
\caption{Tractor Christoffels used to build equations of motion.}
\label{tractor_christoffels}
\end{table}

\end{landscape}

Just as for the tractor Maxwell system, some equations of motion involve higher derivatives that can be eliminated.  The field equation $G^{++}$ involves no higher derivatives while the higher derivatives in $G^{+}$ and $G$  can be eliminated by studying  linear combinations of field equations as well as their traces and divergences:
\bea
{\cal G}_o^{++} & =& G^{++}\, , \nn \\[2mm]
{\cal G}_o^{+}\   &=& G^{+} - \frac{1}{d+w}\, {\bf grad}\,  G^{++} \, ,\nn \\[2mm]
{\cal G}_o \;&=& G - \frac{2}{d+2w-2}\,{\bf grad}\, G^{+}\nn\\[3mm]
&+& \frac{2(d+2w-1)}{w(d+w-1)(d+2w-2)}\, {\bf g}\, \Big({\bf div}G^{+} - \frac{1}{d+2w}\Delta G^{++}\Big)\, .\nn\\
\eea
Even though all the higher derivatives have been eliminated in $({\cal G}_0,{\cal G}_0^+,{\cal G}_0^{++})$, these equivalent equations of motion do not directly follow from an action. Bianchi identities implied by the gauge invariances~\eqn{2g} will serve as the guiding principle in our pursuit of equations of motion $({\cal G},{\cal G}^+,{\cal G}^{++})$ that do come directly from an action principle \be S=\frac 12\int\sqrt{-g}\Big(V^{++}{\cal G}^{++}+V_m^+ {\cal G}^{m+}
+V_{mn}{\cal G}^{mn}\Big)\, .\ee  
The required Bianchi identities are
\bea
-{\bf div}\, {\cal G} + \frac{1}{2}w{\cal G}^+ & =& 0 \, ,\nn \\[2mm]
\frac{2\Rho}{d}\,{\bf tr}\,{\cal G} - \frac{1}{2}\,{\bf div}\,{\cal G}^+ + (w+1){\cal G}^{++} &=& 0\, .
\eea
These are solved by the following combinations of field equations
\bea
{\cal G}^{++} &=& \frac{4\Rho}{dw(d+2w-2)} \Big [ \frac{2(d+w-2)(d-1)\Rho}{(w+1)}{\cal G}_o^{++} - (d+w)\,{\bf div}\, {\cal G}_o^{+} \Big ]\, ,\nn \\ [4mm]
{\cal G}^+ \ &=&\frac{-8(d+w-1)(d+w)\Rho}{dw(d+2w-2)}\,  {\cal G}_o^+ \, ,\nn \\[2mm]
{\cal G} \;\;&=& {\cal G}_o - \frac{1}{4}\,{\bf g}\,{\bf tr} \,{\cal G}_o \nn\\[2mm]
&-&\frac{\Rho}{(d+2w-2)}\,  \Big [ 1- \frac2w\, \Big(1-\frac{w+1}{d+2w} -\,\frac{(d-1)^2}{d}\, \Big)\Big ]\, {\bf g}\, {\cal G}_o^{++} \, .
\nn \\[2mm]
\eea
Explicit expressions for ${\cal G}^{++}$, ${\cal G}^{+}$, and ${\cal G}$ are provided in the accompanying Table~\ref{companytable}. 
They are gauge invariant, derive from an action principle and obey the above identities. We finish this Section by showing they correctly
describe the massive spin two system in constant curvature along with its massless and partially massless limits.

Using the gauge invariance~\eqn{2g} the fields $V^{+}$ and $V^{++}$ can be gauged away (so long as $w\neq0,-1$) and the spin two equations of motion can be written in a simpler way in terms of the minimal covariant field content $V_{\mu\nu}$:
\be
{\cal G}_{\rm PF} \equiv  {\cal G}\lvert_{V^{+}=0, V^{++}=0} \ = G_{\rm Einstein} + G_{\rm mass}=0,\label{123}
\ee
where the linearized cosmological Einstein tensor is given by
\bea
G_{\rm Einstein} &=&\Big[ \Delta - \frac{4\Rho}{d}\, \Big]V - {\bf grad}\,{\bf div}V + \frac{1}{2}\, \Big[{\bf g}\,{\bf div}^2+{\bf grad}^2\,{\bf tr}\Big]V\nn\\[3mm]
 &-&\frac{1}{2}\ {\bf g}\,
\Big[\Delta+\frac{2\Rho}{d}(d-3)\Big]\,{\bf tr}\,  V \, .
\eea
and the Pauli--Fierz mass term~\cite{Pauli} is
\bea
G_{\rm mass} &=&  -m^2\Big[1- \frac{1}{2}\, {\bf g}\,{\bf tr}\, \Big]\,  V\, .
\eea
In these formul\ae\ the mass  $m^2$ is the same as given in~\eqn{spin2mass}.

Taking a  divergence of the cosmological Pauli--Fierz field equation~\eqn{123}, we learn the constraint:
\be
-m^2({\bf div}- {\bf grad}\,{\bf tr}) V = 0 \, . \\[2mm]
\ee
There is, a however, a further constraint obtained from the combination of a double divergence and trace of the field equation
\be
{\bf div}^2\, {\cal G}_{\rm PF} +\frac{m^2}{d-2}\,{\bf tr}\,  {\cal G}_{\rm PF} =\frac{d-1}{d-2}\, m^2\, [m^2-\frac{2\Rho}{d}(d-2)]\, {\bf tr}\, V=0
\ee
Hence when
\be
m^2\neq 0,\;\frac{2\Rho}{d}(d-2)\, ,
\ee
we immediately find
\be
{\bf tr}V = 0 = {\bf div}V\, .
\ee
These constraints imply that the field $V_{\mu\nu}$ describes the $(d+1)(d-2)/2$ degrees of freedom of a massive spin two excitation.
The same results can also be obtained from $ {\cal G}^{++} \lvert_{V^{+}=0,V^{++}=0}$ in conjunction with  ${\cal G}^{+} \lvert_{V^{+}=0,V^{++}=0}$.  
Plugging the constraints into~\eqn{123}, we recover~\eqn{PF2}--the massive on-shell spin two equation
\be
(\Delta - \frac{4\Rho}{d}-m^2)V = 0\, .
\ee
Finally, the special masses $m^2=0$ and $m^2=(2\Rho/d)(d-2)$ correspond to weights $w=0$ and $w=-1$, respectively.
In these cases, the above constraints become the respective Bianchi identities
\be
{\bf div} \, {\cal G}_{\rm PF}= 0\, ,\qquad \Big[{\bf div}^2+\frac{2\Rho}d \, {\rm tr}\Big] {\cal G}_{\rm PF}=0\, , 
\ee
corresponding to gauge invariances
\be
\delta V={\bf grad}\, \xi\, ,\qquad \delta V=\Big[{\bf grad}^2+\frac{2\Rho}{d}\Big]\, \xi^+\, .
\ee
These correspond to linearized diffeomorphisms and the partially massless gauge transformation of~\cite{Deser:1983tm}.
A detailed discussion of these theories is given in~\cite{Deser:2001pe}.   This concludes our demonstration that the simple tractor
equations describe the cosmological spin two system.

Having given an exhaustive discussion of cosmological spin two system, we analyze the tractor equations of motion \eqn{2eom} at a special weight  $w=1-d/2$.  At this weight, both of the field equations, $G^{++}$ and $G^{+m}$,  are zero and the scale completely decouples from $G^{MN}$.  This decoupling is the by now familiar phenomenon that we observed in scalar and spin one theories  leading to conformally invariant theories.  Similarly,  at the special weight, we find a {\it new} conformally invariant but gauge variant spin two theory with equation of motion given by
\be
\Gamma^{RMN}=0\, .
\ee
Hence, tractors are not only useful for unifying well known physical theories, but also for constructing new conformally invariant ones.
We now pass to analyzing systems with arbitrary higher spin values.

\newpage

\begin{landscape}

\begin{table}
\begin{tabular}{|l|}
\hline \\[2mm]
${\cal G}^{++}= \frac{8(d+2w-1)(d-1)\sRho^2}{d^2w(w+1)} \square V^{++} - \frac{16(d+w-2)(d-1)\sRho^3}{d^2(w+1)}V^{++}  
-\frac{16(d-1)(d+w-1)\sRho^2}{d^2w}{\bf div}V^+ - \frac{2\sRho}{d} \square {\bf tr} V +\frac{2\sRho}{d}{\bf div}^2 V + \frac{4(d-1)(d+w-2)\sRho^2}{d^2}{\bf tr} V$ \nn \\[4mm]  
 
\hline \\[2mm] 
${\cal G}^+=-\frac{8(d+w-1)\sRho}{dw} \Big ( -\frac{2(d-1)\sRho}{d} {\bf grad} V^{++} -\frac{w}{2}[{\bf div} - {\bf grad}\,{\bf tr}]V+[\square-{\bf grad}\,{\bf div}]V^+  
+ \frac{4\sRho(d-1)}{d}V^+ \Big )$ \nn \\[4mm] 

\hline \\[2mm]
${\cal G}= G_{linear Einstein} + G_{PF}-\frac{4(d+w-1)\sRho}{d}[{\bf grad}-{\bf g}{\bf div}]V^+ -\frac{2\sRho}{d}{\bf g}\square V^{++} + \frac{2\sRho}{d}{\bf grad}^2 V^{++}  
+\frac{4(d-1)(d+w-2)\sRho^2}{d^2}{\bf g}V^{++}$  \nn \\[4mm] \hline
\end{tabular}
\caption{Equations of motions for spin 2 system.}\label{companytable}
\end{table}

\end{landscape}

\section{Arbitrary Spins}

\label{arbitrary}

The methods explicated in detail for spins $s\leq2$ can be applied also to higher spin systems,
which can be massive, massless or partially massless with gauge invariances ranging from a single
derivative on a tensor parameter (depth one) to $s$ derivatives on a scalar parameter (depth~$s$)~\cite{Deser:2001pe}.
(A useful review on the extensive higher spin literature is~\cite{Bekaert:2005vh}.)
Again these models are all unified by a single tractor equation of motion. We begin 
with an on-shell approach.

\label{s>2}

\subsection{On-Shell Approach}

In components, the on-shell field equations for massive higher spin $s$ fields in constant curvature backgrounds\footnote{We ignore the possibility
of mixed symmetry higher spin fields, although these should be simple to handle using our approach.} are given by
\bea
&&\Big(\Delta +\frac{2\Rho}{d}[w(w+d-1)-s]\Big)V_{\mu_1\cdots\mu_s}=0\, ,\\[2mm]
&&\; \; \nabla.V_{\mu_2\ldots \mu_s}=0\, ,\\[2mm]
&&\quad V^\rho_{\rho\mu_3\ldots\mu_s}=0\, ,\label{osh}
\eea
It is not difficult to verify that these follow from the tractor equations of motion
\be
D\cdot V^{M_2\cdots M_s}=
X \cdot V^{M_2\cdots M_s}=
I\cdot D\ V^{M_1\cdots M_s}=
I\cdot V^{M_2\cdots M_s} =
V_R^{RM_3\cdots M_s}=0\, ,
\label{Gspin}
\ee
where $V^{M_1\cdots M_s}$ is a totally symmetric, rank~s, weight $w$ tractor tensor.
We also observe, that if instead of the standard definition of the mass, we define a parameter $\mu^2$
by the eigenvalue of the (Bochner) Laplacian so that
\be
\Delta V_{\mu_1...\mu_s} = \mu^2 V_{\mu_1...\mu_s}\, ,
\ee
then the mass-Weyl weight relation for an arbitrary spin $s$ is given by  
\be
\mu^2=\frac{2\Rho}d \Big[ \Big(\frac{d-1}{2}\Big)^2-\Big(w+\frac{d-1}{2}\Big)^2+s\Big]\, .
\label{MW}
\ee
This result predicts an arbitrary spin Breitenlohner--Freedman bound
\be
\mu^2\geq\frac{2\Rho}{d}\, \Big[\Big(\frac{d-1}{2}\Big)^2+s\Big]\, .\label{BFs}
\ee
Some of these bounds appear to be new ones despite the fact that unitarity bound may render them less interesting.  However, unitarity arguments sometimes render them less useful.  We also note that the three-dimensional topologically massive Maxwell system saturates this Breitenlohner--Freedman bound~\cite{Carlip:2008eq}.
As discussed in Section~\ref{Mass}, we can have more than one gauge invariance.  In fact, for an arbitrary spin systems in curved backgrounds, we have several gauge invariance which can be classified by the
number of derivatives acting on the gauge parameter.  We could have chosen the zero of the mass parameter to coincide with the appearance of a
single derivative gauge invariance which leads to
\be
m^2=\frac{2\Rho}d \Big[ \Big(\frac{d-5}{2}+s\Big)^2-\Big(w+\frac{d-1}{2}\Big)^2\Big]\, ,
\ee
implying the Breitenlohner-Freedman bound
\be
 m^2 \geq\frac{2\Rho}{d}\, \Big(\frac{d-5}{2}+s\Big)^2 \, .
\ee

Once again, equations \eqn{Gspin} enjoy residual gauge invariances, but now at weights $w=s-2,s-3,\ldots,0,-1$.
In tractors these read simply
\be
\delta V^{M_1\ldots M_s}= D^{(M_1}\cdots D^{M_t}\xi^{M_{t+1}\ldots M_s)}\, ,
\ee
where $ V^{M_1\ldots M_s}$ is of weight $w$ and $\xi^{M_1\ldots M_{s-t}}$ has weight $w+t$ and the parameter $t$ is called the depth
of a partially massless gauge transformation. Since these are on-shell residual transformations they
are also subject to
$$
X\cdot \xi^{M_1\ldots M_{s-t-1}}=
I\cdot \xi^{M_1\ldots M_{s-t-1}}=
D\cdot \xi^{M_1\ldots M_{s-t-1}}=0
$$
\be
I\cdot D\ \xi^{M_1\ldots M_{s-t}}=0=
\xi_R^{RM_1\ldots M_{s-t-2}}\, .
\ee
At depth $t$ the partially massless field $V^{M_1\ldots M_s}$ must have weight $w=s-t-1$ which 
in conjunction with~\eqn{MW} corresponds to masses
\be
\mu^2=-\frac{2\Rho}d \Big[ (s-t-1)(s-t-1+d)+t+1\Big]\, .
\ee 
These results reproduce those found earlier in~\cite{Deser:2001pe} by rather different methods.

\subsection{Off-Shell Approach}

The St\"uckelberg field content required to describe a massive spin~$s$ field in~$d$-dimensions
is equivalent to that of massless spin~$s$ field in~$d+1$ dimensions (see~\cite{Hallowell:2005np} for a detailed explanation\footnote{A related approach, in which AdS$_{d}$ higher spin fields are arranged in $O(d-1,2)$ multiplets with the aid of a compensating field can be found in~\cite{Vasiliev:2001wa}.}).
In~$d+1$ dimensions a massless spin~$s$ field is described by a totally symmetric rank~$s$ tensor
subject to the condition that its double trace vanishes. A counting of independent  field components therefore yields
\be
\begin{pmatrix}d+s\\ s\end{pmatrix}-\begin{pmatrix}d+s-4\\ s-4\end{pmatrix}\, .\label{numero}
\ee
The tractor description involves a weight $w$, totally symmetric, rank~$s$ tractor tensor $V^{M_1\ldots M_s}$ but again
field constraints are necessary.
Indeed, the same number of independent field components as in~\eqn{numero} solve the tractor field constraints\footnote{This follows from the binomial coefficient
identity
$$
\begin{pmatrix}d+s\\ s\end{pmatrix}-\begin{pmatrix}d+s-4\\ s-4\end{pmatrix}
=
\begin{pmatrix}d+s+1\\ s\end{pmatrix}-\begin{pmatrix}d+s\\ s-1\end{pmatrix}
+\begin{pmatrix}d+s-4\\ s-5\end{pmatrix}-\begin{pmatrix}d+s-3\\ s-4\end{pmatrix}
.
$$}
\be
D\cdot V^{M_2\ldots M_s}-\frac{s-1}{s}D^{(M_2} V_R^{M_3\ldots M_s)R}=0=V_{RS}^{RSM_5\ldots M_s}\, .\label{fc}
\ee
The first of these is consistent with our proposed gauge invariance
\be
\delta V^{M_1\ldots M_s}=D^{(M_1}\xi^{M_2\ldots M_s)}\, ,\label{sgauge}
\ee
where the parameter $\xi^{M_2\ldots M_s}$ is weight $w+1$ and obeys the parameter constraints
\be
I\cdot \xi^{M_1\ldots M_{s-2}} = 0 = \xi_R^{R M_3\ldots M_{s-1}}\, .
\ee
It is not difficult to write these out in components for the example of spin~3 (say) and check that they concur
with the general St\"uckelberg gauge transformations given in~\cite{Hallowell:2005np}. 

The field constraints~\eqn{fc} and gauge transformations~\eqn{sgauge} constitute the kinematics of our model. The dynamics
are determined by finding the gauge invariant higher spin generalization of the spin two equation of motion~\eqn{2eom}.
We conjecture this to be
\be
G^{M_1\ldots M_s}= I\cdot D\,  V^{M_1\ldots M_s} - s D^{(M_1}I\cdot V^{M_2\ldots M_s)}\, .
\ee
Its gauge invariance is trivially checked and it can also be rewritten in terms of higher spin tractor
Christoffel symbols\footnote{The generalized  Christoffel symbol approach to higher spins was pioneered in~\cite{deWit:1979pe, Damour:1987vm}.}. 
 Gauge invariance, the matching of counting of field components
and our explicit $s\leq2$ computations are already strong evidence in favor of this conjecture.

    \chapter{Fermionic Theories}
\label{Fermionic}
Following the spirit of the last chapter, in this chapter, we construct fermionic theories with manifest local unit invariance.  The double $D$-operator will be of central importance in this chapter  just as the Thomas $D$-operator played a prominent role in the last chapter.  The scale tractor will accompany all equations of motion.  The gravitational masses will be proportional to the square root of the length of the scale tractor:  m $\propto \sqrt{I.I}$. The Weyl weights, one again, will be related to masses producing a mass-Weyl weight relationship leading to the Breitenlohner-Freedman bounds analogous to the ones derived in the last chapter.  At special Weyl weights, the scale will again decouple, but the decoupling mechanism will be different from the one presented in the last chapter. 

Fermionic theories are subtle than their bosonic counterparts, and require some additional tractor technology.  To be more precise, we need spinor transformation rules under Weyl transformations, the action of the covariant derivative on spinors, and also tractor version of gamma matrices.  We start by developing the tractor technology needed to build fermionic theories and then use it to construct the Weyl invariant Dirac operator, the spin one-half theory followed by the spin three-half Rarita Schwinger theory.   

\section{Tractor Spinors}
\label{ss:Tractors}
The theory of spinors in conformal geometry is a well-developed subject to which tractor calculus can be applied~\cite{Helmut,Branson}.
A tractor spinor can be built from a pair of $d$-dimensional spinors. While the latter transform as $\frak{so}(d-1,1)$ representations, the tractor spinor
is a spinor representation of $\frak{so}(d,2)$; to avoid technical questions on the spinor type in $d$ and $d+2$ dimensions, we do not specify whether
the constituent $d$-dimensional spinors are Dirac, Weyl, or Majorana. From the  Dirac matrices
$\{\gamma^m,\gamma^n\}=2\eta^{mn}$ in $d$ dimensions,
we build $(d+2)$-dimensional Dirac matrices
\be\nn
\Gamma^+ = \begin{pmatrix}0 & 0\\ \sqrt{2} & 0 \end{pmatrix}, \qquad
\Gamma^- = \begin{pmatrix}0 & \sqrt{2}\\ 0 & 0 \end{pmatrix}, \qquad
\Gamma^m = \begin{pmatrix}\gamma^m & 0\\ 0 & -\gamma^m \end{pmatrix} \, ,
\ee 
subject to 
\be\nn
\{\Gamma^M,\Gamma^N\}=2\eta^{MN}\, .
\ee
From the gamma matrices, we can form the generators of rotations 
\be
M^{MN}= \frac{1}{2} \Gamma^{MN}, \qquad \Gamma^{MN} = \frac{1}{2}[\Gamma^M\Gamma^N - \Gamma^N\Gamma^M] \, ,
\ee
which will be useful for our later calculations. 

We have successfully derived a $d+2$-dimensional Clifford algebra from a $d$-dimensional one.  It is the first step towards building fermionic theories in tractor formalism.  But, to actually write these theories, we need a fermi field, so we define a weight~$w$ tractor spinor 
\be
\Psi = \begin{pmatrix} \psi \\ \chi \end{pmatrix},
\ee
built from a pair of $d$-dimensional spinors $\psi$ and $\chi$ transforming under $SO(d)$, while the spinor $\Psi$ itself transform as a $2d$-dimensional representation of $SO(d+1,1)$.
The top slot is of weight $w+1/2$ while the bottom slot has a weigh of $w-1/2$.  Infinitesimally, tractor gauge transformations are given by
\be
U^M{}_N = A^M{}_N + B^M{}_N\, ,
\ee
where
\be
A = \begin{pmatrix} \omega &0 & 0 \\ 0 & 0 & 0 \\ 0 & 0 & - \omega \end{pmatrix}, \qquad
B = \begin{pmatrix} 0 &0 & 0 \\ \gamma^m & 0 & 0 \\ 0 & - \gamma_n & 0 \end{pmatrix}, \qquad
\Upsilon_\mu = \partial_\mu \omega, \qquad \Omega = e^{\omega} \, .
\ee
For the spinor case, the infinitesimal transformation is given by 
\be
\delta \Psi = \frac{1}{4} U^{MN}\Gamma_{MN}\Psi
=  \begin{pmatrix}
 \frac{\omega}{2} & 0 \\[2mm] -\frac{\slashed {\Upsilon}}{\sqrt{2}} & \frac{-\omega}{2} 
 \end{pmatrix}   \begin{pmatrix}  \psi \\[4mm] \chi \end{pmatrix} \, , 
 \ee
which sends
\bea
\psi &\rightarrow& \psi + \frac{\omega}{2}\psi \, , \\[2mm]
\chi &\rightarrow& (1-\frac{\omega}{2}) \chi-\frac{\slashed {\Upsilon}}{\sqrt{2}} \psi  \, .
\eea
Exponentiating the infinitesimal transformations, we get the finite ones given by
\be
U = e^{\frac{1}{4}\Gamma_{MN}(A^{MN}+B^{MN})}
= \begin{pmatrix} \sqrt{\Omega} & 0 \\[2mm] -\frac{\slashed {\Upsilon}}{\sqrt{2\Omega}} & \frac{1}{\sqrt{\Omega} }    \end{pmatrix} \, .
\ee
This transformation matrix is similar to \eqn{Um} encountered earlier.  For a tractor spinor of an arbitrary weight w, 
\be
\Psi \mapsto \Omega^w U \Psi \, 
\ee
in response to the rescaling of the metric: $g \mapsto \Omega^2 g$.  Explicitly the transformations are  given by
\be
\begin{pmatrix}  \psi     \\  \chi \end{pmatrix} \rightarrow   \Omega^w  \begin{pmatrix}  \Omega^{1/2} \psi \\[2mm] \Omega^{-1/2} [\chi -\frac{\slashed {\Upsilon}}{\sqrt{2}} \psi ] \end{pmatrix} \, .
\label{spin_trans}
\ee

Finally, we need an expression for the tractor covariant derivative acting on a tractor spinor, which is defined by
\be\nn
\mathcal{D}_\mu \Psi=\begin{pmatrix}   \nabla_\mu \psi + \frac{1}{\sqrt{2}}\gamma_\mu \chi \\[2mm] \nabla_\mu \chi -\frac{1}{\sqrt{2}}\slashed{\Rho}_\mu \psi \end{pmatrix}\, ,
\ee
where $\nabla_\mu$ is the standard Levi-Civita connection acting on $d$-dimensional spinors.   For a detailed derivation of the covariant derivative on spinors, consult Appendix~\ref{Covariant_Spinor}.
We define another tractor operator built out of the canonical tractor $X^M$ and the gamma matrices
\be
\Gamma . X = \Gamma_M X^M = \begin{pmatrix} 0  & 0 \\ \sqrt{2} & 0 \end{pmatrix} \, .
\ee
As is apparent, it is the analog of the canonical tractor $X^M$ and will be used to project out the top slot of tractor spinors.  We have now assembled the ingredients required to compute the Thomas $D$-operator~\eqn{D} acting on spinors. (Details are again provided in Appendix ~\ref{Covariant_Spinor}. ) Of particular interest
is the ``Dirac--Thomas $D$-operator''
\bea\nn
\Gamma.D\,   \Psi &=& \begin{pmatrix}(d+2w-2) \slashed\nabla & \frac{1}{\sqrt{2}}(d+2w)(d+2w-2) \\[2mm] -\sqrt{2} \slashed \nabla^2 & -(d+2w)\slashed\nabla   \end{pmatrix}   \Psi \\[4mm]  &=& \begin{pmatrix}(d+2w-2) [ \slashed\nabla \psi + \frac{1}{\sqrt{2}}(d+2w) \chi ] \\[2mm] -(d+2w) \slashed\nabla \chi - \sqrt{2}[\Delta- \frac{\scriptsize \Rho}{2}(d-1)] \psi  \end{pmatrix} \, .\nn
\eea
Here we have denoted the contraction of tractor vector indices by a dot and it is worth bearing in mind that  the $d$-dimensional Weitzenbock
identity acting on spinors is  $\slashed \nabla^2 \psi = [\Delta- \frac{\footnotesize \Rho}{2}(d-1)]\psi$.

\subsection{The Dirac Operator}
It is well known that the massless Dirac equation is Weyl covariant in any dimension. We will show that this follows naturally from tractors.
Observe that we can use the canonical tractor to produce a weight $w+1$ tractor spinor from $\Psi$
\be\nn
\Gamma.X \, \Psi = \begin{pmatrix}0\\[2mm]\sqrt{2} \psi\end{pmatrix}\, ,
\ee
where $\psi$ has the transformation rule
\be
\psi\mapsto \Omega^{w+\frac12}\psi\, .\label{psitr}
\ee
Now acting with the Dirac--Thomas $D$-operator yields 
\be\nn
\Gamma.D \ \Gamma.X \ \Psi = 
(d+2w+2)\ \begin{pmatrix}
(d+2w)\psi\\[2mm]
-\sqrt{2}\slashed \nabla \psi
\end{pmatrix}\, .
\ee
Hence assigning $\Psi$ the weight $w=-\frac d2$ so that $\psi\mapsto \Omega^{\frac{1-d}{2}}\psi$, it follows from the tractor spinor gauge 
transformation rule~\eqn{spin_trans}, that
\be
\slashed \nabla \psi \mapsto \Omega^{-\frac{d+1}{2}} \slashed\nabla\psi\, ,
\ee
which proves the covariance of the Dirac operator. We are now suitably armed to construct fermionic tractor theories.

\section{Tractor Dirac Equation}

\label{Tractor Dirac  Equation}

 Fermionic theories pose some interesting puzzles for our tractor approach. Firstly, since the tractor approach is based on
arranging fields in $\frak{so}(d,2)$ multiplets, we might generically expect a doubling of degrees of freedom. This can be seen from
the previous Section where tractor spinors were constructed from pairs of space-time spinors. Secondly, the mass-Weyl weight
relationship~\eqn{massweight} relates the mass squared to the scalar curvature: $m^2 \propto \Rho$. However, massive spinor theories depend
linearly on the mass, and therefore the square root of the scalar curvature. It is not immediately obvious how this square root could arise.
We will solve both of these puzzles by employing several principles: to construct tractor-spinor and tractor-spinor-vector theories
\begin{enumerate}
\item We search for massive wave equations whose masses are related to Weyl weights by an analog of the scalar relationship~\eqn{massweight}.
\item We require that, in a canonical choice of scale, these theories match those found by the log-radial dimensional reduction of
$d+1$ dimensional massless Minkowski theories to $d$ dimensional constant curvature ones described in~\cite{Hallowell:2005np, Biswas:2002nk}
and Appendix~\ref{Doubled Reduction}.
\item We will impose as many constraints as consistent with the above requirements so as to find a ``minimal covariant field content''.
\end{enumerate}
These principles will become clearer through their applications, so let us provide the details.

In the previous chapter, we always started with a tractor with field content larger than needed.  Then, we imposed gauge invariances and constraints that were consistent
with the gauge conditions.  We also used the fact that (from an ambient viewpoint as described in~\cite{Gover:2002ay,Cap:2002aj} and further studied in~\cite{Gover:2009vc}),
the contraction of the scale tractor and Thomas $D$-operator $I.D$ generates bulk evolution, thus awarding $I . D$ a special significance in the construction of bosonic field theories.  We continue
the same story with spinors: we start with a weight~$w$ tractor spinor~$\Psi$ obeying

\bea\nn
I.D\ \Psi&=&0\, ,\nn\\[2mm]
\Gamma.D\ \Psi&=&0\, .\label{dirac1}
\eea
We can view the second equation as a scale covariant constraint eliminating the lower component of~$\Psi$.
Its solution is
\be\nn
\Psi=
\begin{pmatrix}
\psi\\[2mm]
-\frac{\sqrt{2}}{d+2w}\, \slashed{\nabla} \psi\
\end{pmatrix}\, .
\ee
In turn, the $I.D$ field equation, in the canonical scale $\sigma=$ constant, implies the massive wave equation\footnote{The value $w=-d/2$ is distinguished here, as in  fact is the value $w=-d/2+2$. In the first case we cannot solve the constraint in~\eqn{dirac1}. Also, in deriving~\eqn{wave}, we have dropped 
an overall factor $(d+2w-2)/(d+2w)$. However, below we give a second formulation of the system that still predicts~\eqn{wave} at $w=-d/2+2$.} 
\be\label{wave}
\Big[\Delta + \frac{2\Rho}{d}(w^2+wd+\frac d4)\Big]\,  \psi=0\, .
\ee
Defining the squared mass as the eigenvalue of $\Delta$ (note that  $\Rho$
is constant in an Einstein background) gives the spinorial mass-Weyl weight relationship
\be\label{massweyl}
m^2= -\frac{2\Rho}{d}\Big[\Big(w_\psi+\frac{d-1}2\Big)^2-\frac{d(d-1)}{4}\Big]\, ,
\ee
analogous to its bosonic counterpart~\eqn{massweight}. Here we have defined $w_\psi\equiv w+\frac12$ because
under Weyl transformations $\psi$ transforms according to~\eqn{psitr}.
Observe that reality of the weight $w$ for spaces with negative scalar curvature implies a Breitenlohner--Freedman type bound~\cite{Breitenlohner:1982jf, Mezincescu:1984ev} on the mass parameter  $m^2\geq\frac12\Rho(d-1)$. 
Before analyzing this system further, let us present an alternate formulation.

The Thomas $D$-operator is second order in its lowest slot. For Fermi systems, we would like to find a set of
first order field equations. To that end, we recall that the double $D$-operator \eqn{DD} is a first order operator.
In terms of double $D$-operator, we propose the Dirac-type equation. \be\label{dirac2}
I^M\Gamma^N D_{MN}\, \Psi=0\, .
\ee
This equation is similar in spirit to Dirac's proposal for writing four dimensional conformal wave equations by employing the six dimensional Lorentz generators~\cite{Dirac:1936}. Of course here, we also describe massive systems that are not invariant without coupling to scale.
In the canonical choice of scale \eqn{dirac2}  reads
\be\nn
-\sigma\begin{pmatrix}\slashed\nabla\psi +\frac{d+2w}{\sqrt{2}}\, \chi\\[2mm]-\slashed\nabla\chi+\frac{(d+2w)\!\!{\scalebox{.7}{ \Rho}}}{\sqrt{2}\, d}\, \psi\end{pmatrix}=0\, .
\ee

Firstly, when $d+2w\neq0,2$, it is easy to verify that these equations are equivalent to~\eqn{dirac1}. In general, they are more fundamental
because (at $d+2w\neq2$) the equation~\eqn{wave} follows as an integrability condition.
Moreover, even 
at $d+2w=2$ we can still define the double-$D$ operator by~\eqn{DD} and then have well-defined system (that implies the massive wave equation~\eqn{wave}).

The weight $d+2w=0$, has a special physical significance because at that weight we expect to find a scale invariant theory as seen in the previous chapter. Although it is not true that the scale decouples from the equation~\eqn{dirac2}, the modified equation
$$
\sigma^{-1}\Gamma.X I^M\Gamma^N D_{MN}\, \Psi=0\, ,
$$
is in fact independent of the scale at $w=-d/2$. It is then equivalent to the equation $\Gamma.X\  \Gamma^N D_{MN}\Psi=0$
(just as for scalars in Section~\ref{Scalar}). In components this amounts simply to the Dirac equation $\slashed\nabla\psi=0$.

In the above formulation, $w=-d/2$ is the only value at which multiplication by a factor $\Gamma.X$ yields a consistent system, at other values
the field~$\chi$ enters on the right hand side of the Dirac equation.
The presence of the second spinor~$\chi$ is undesirable, because it doubles the degrees of freedom of the $d$-dimensional theory.
We next explain how to obtain a tractor theory of a single $d$-dimensional spinor.   

Firstly observe that a massive Dirac equation is linear in the mass parameter, whereas according to~\eqn{massweyl} the constant scalar
curvature is proportional to the square of the mass. Therefore we need a tractor mechanism that somehow introduces the square root of
the scalar curvature while simultaneously relating the pair of spinors  $\chi$ and $\psi$. Examining our spinorial wave equations~\eqn{dirac2}
in their canonical component form, we see that a relationship $\chi=\alpha\psi$ means that this pair of equations are equivalent only when $\alpha^2=-\Rho/d$. This relationship
can be imposed tractorially using the projectors
\be\nn
\Pi_\pm \equiv \frac12\Big[ 1\pm \frac{\Gamma.I}{\sqrt{I.I}}\Big]\, .
\ee
(Recall that in a conformally Einstein background, $I^M$ is tractor parallel, so that $I.I$ is constant. Note that $I.I$ is positive
for negative scalar curvature.)
Hence we propose the tractor Dirac equations
\be\nn
I^M \Gamma^N D_{MN} \Psi = 0 = \Pi_+\Psi\, ,
\ee
We could equally well multiply the first of these equations by $\Gamma.X$ since its bottom slot is a consequence of the top one.
(The choice of $\Pi_+$ rather than $\Pi_-$ corresponds to the sign of the Dirac mass term.) In canonical components these
imply the massive curved space Dirac equation
\be\label{massdirac}
\Big[\slashed\nabla -\sqrt{\frac{-\Rho}{2d}}\ (d+2w) \Big]\psi=0\, .
\ee
Its  mass is again related to the weight of $\psi$ by~\eqn{massweyl}.

In summary, the irreducible tractor Dirac equation for a tractor spinor $\Psi$ (subject to the ``Weyl''-like condition\footnote{We cannot help but remark that this condition melds two of Weyl's seminal contributions to physics -- the Weyl spinor and Weyl symmetry.} $\Pi_+ \Psi=0$) is given by
$$
\Gamma.X I^M \Gamma^N D_{MN} \Psi=0\, .
$$ 
This equation of motion follows from a tractor action principle which we now describe.
To that end we need to 
introduce the tractor Dirac conjugate spinor, which is defined as
\be\nn
\overline\Psi \equiv \overline{\!\begin{pmatrix}\psi\\\chi\end{pmatrix}\!}=i \Psi^{\dagger} \Gamma^{\bar 0}=(\bar\chi \;\; \bar\psi)\, ,
\ee
where $\bar\psi$ and $\bar\chi$ are the standard $d$-dimensional Dirac conjugates of $\psi$ and $\chi$, and $ \Gamma^{\bar 0}$ obeys the following properties (because it derives from the product of the two timelike Dirac matrices of $\frak{so}(d,2)$):
\be \nn
(\Gamma^{\bar 0})^{2} = -1 \, , \qquad \Gamma^{\bar 0 \dagger} =- \Gamma^{\bar 0} \, , \qquad \Gamma^{M \dagger} = - \Gamma^{\bar 0} \Gamma^M \Gamma^{\bar 0} \, .
\ee
Then the required action principle is
\be
S=\frac 1{\sqrt{2}}\int \frac{\sqrt{-g}}{\sigma^{d+2w+1}}\overline\Psi \Gamma.X I^M \Gamma^N D_{MN} \Psi\, ,
\ee
where the tractor spinor $\Psi$ obeys $\Pi_+\Psi=0$. This action is hermitean. Since it is useful to possess the tractor machinery
required to vary actions of this type, let us prove this. Firstly, the double-$D$ operator $D_{MN}$ is Leibnizian. 
Moreover $\int \sqrt{-g} D_{MN} \Xi^{MN} = 0$ (up to surface terms) for any $\Xi^{MN}$ of weight zero. This allows us to integrate $D_{MN}$ 
by parts. Therefore,  to verify $S=S^\dagger$ we need to compute $D_{MN} \Big[\frac{1}{\sigma^{d+2w+1}}I^M \Gamma^N \Gamma.X \Psi\Big]$
which requires the following identities
\bea
D_{MN}\sigma\ \ \ &=&X_N I_M-X_M I_N\, ,\nn\\[2mm]
D_{MN}I^N\  &=&0\, ,\nn\\[2mm]
D_{MN} X^R \ &=&X_N \delta_M^R - X_M \delta_N^R\, ,\nn\\[2mm]
X^MD_{MN} &=&w X_N\, .
\eea
Orchestrating these, we find
\be\nn
S-S^\dagger=\frac{d+2w}{\sqrt{2}}\int \frac{\sqrt{-g}}{\sigma^{d+2w+1}}\ \overline\Psi (\Gamma.X \Gamma.I -\sigma)\Psi\, .
\ee

For generic weights $w$ this is non-vanishing, however using the condition $\Pi_+\Psi=0$ to conclude that $\Psi$ is in the image of $\Pi_-$
along with the facts that $\overline {\Pi_- \Psi}\equiv \overline\Psi \Pi_-$ and $\Pi_-(\Gamma.X \, \Gamma.I -\sigma)\Pi_-=0$, shows that $S=S^\dagger$. 
A similar  computation 
implies that the above action implies the field equations quoted.

Our final computation is to  write out the action principle in components. Rather than working at the canonical scale, lets us give the general result,
namely
\be
\nn
S=-\int\frac{\sqrt{-g}}{\sigma^{d+2w}}\ \bar\psi \left[\slashed\nabla - \frac12(d+2w)\, \Big(\slashed b +\sqrt{-\frac{2(\Rho+\nabla\cdot b -\frac{d-2}{2}\ b\cdot b)}{d}}\:  \Big)\right]\ \psi\, .
\ee
Each term has a simple interpretation.  The $\slashed b$ contribution covariantizes the leading Dirac operator with respect to scale transformations
so,   $\psi\mapsto \Omega^{w+\frac 12}\psi$ implies $[\slashed \nabla - \frac12 (d+2w) \slashed b] \psi \mapsto \Omega^{w+\frac12} [\slashed \nabla - \frac12 (d+2w) \slashed b]\psi$. These terms also follow from the standard Weyl compensator mechanism. The square root factor is the mass term which
equals (up to a factor $\sigma$) $\sqrt{I.I}$, and is therefore constant for conformally Einstein backgrounds.  The prefactor $(d+2w)$ calibrates the mass
to the square of the scale tractor and implies the mass-Weyl weight relationship~\eqn{massweyl}. When $w=-\frac d2$, the scale $\sigma$ decouples from
the action and we obtain the Weyl invariant curved space Dirac equation discussed in Section~\ref{ss:Tractors}.

\section{Tractor Rarita--Schwinger Equation}

\label{Tractor Rarita--Schwinger Equation}

In the spinor models we have encountered so far there have been choices for  mass terms: we could have used a ``gravitational mass term''~\eqn{g_mass} or a compensated mass term.  The gravitational mass term is proportional to $I.I$, while the compensated mass term is obtained by using the scale $\sigma$ to compensate a standard mass term (for example,
we could add a term $\frac12\int \frac{\sqrt{-g}}{\sigma^{d+2w}}\  \varphi^2$ to the scalar action principle). However, once we study models
with spins $s\geq1$, gauge invariances are necessary to ensure that only unitary degrees of freedom propagate. In Section \ref{arbitrary}, we showed
how higher spin gauge invariant tractor models described bosonic massless, partially massless~\cite{Deser:1983tm, Higuchi:1986py, Deser:2001pe, Deser:2001us, Deser:2001wx, Deser:2001xr, Deser:2003gw} and massive models in a single framework.
In particular, they implied  ``gravitational mass terms'' (rather than compensated ones) with masses dictated by Weyl weights. We now extend those 
results to the higher spin $s=3/2$ Rarita--Schwinger system. The following analysis closely mirrors the tractor Maxwell system studied in~\cite{Gover:2008sw, Gover:2009}
so we keep details to a minimum.

As field content, we take a weight $w$ tractor vector-spinor $\Psi^M$ subject to the gauge invariance
\be\nn
\delta \Psi^M = D^M \Xi\, ,
\ee
where $\Xi$ is a weight $w+1$ tractor spinor parameter. Since the Thomas $D$-operator is null, we may consistently
impose the field constraint 
\be\label{field}
D_M \Psi^M=0\, .
\ee
We assume that the background is conformally flat, so that Thomas $D$-operators commute\footnote{We leave an investigation
of whether non-minimal couplings could relax this restriction to future work. Any such study will be highly constrained by existing
results for gravitational spin~3/2 couplings, see~\cite{Deser:2001dt}.}. We now observe that the quantity
\be\nn
{\cal R}^{MNR}=3D^{[MN}\Psi^{R]}\, ,
\ee
is gauge invariant by virtue of the identity~\eqn{XDD} and use it to construct a set of tractor Rarita--Schwinger equations
coupled to the scale tractor
\be\label{rarita}
{\cal R}_M\equiv\Gamma_{MNR} I_S {\cal R}^{SNR}=0\, .
\ee
The final requirement we impose is the projective one found for spinors
\be\label{project}
\Pi_+\Psi^M=0\, , \qquad  \Pi _{-} \Xi = 0.
\ee
To verify that the set of equations~(\ref{field},\ref{rarita},\ref{project})
are the desired ones, we write them out explicitly in canonical components. This computation is lengthy but straightforward.
The field constraint~\eqn{field} and projective condition~\eqn{project} eliminate most of the field content leaving only
the top spinorial components~$\psi^+$, and middle vector slots~$\psi^m$, independent. Since the system will describe both massive and massless excitations, the spinor $\psi^+$ plays the {\it r\^ole} of a St\"uckelberg field.
The Rarita--Schwinger type equation ${\cal R}^M=0$ in~\eqn{rarita} then yields the independent field equations
\be\nn
\gamma^{\mu\nu\rho}\wt \nabla_\nu \psi_\rho +\sqrt{-\frac{2\Rho}{d}}\ \gamma^{\mu\nu} \Big([w+1]\psi_\nu-  \wt \nabla_\nu \psi^+\Big)=0\, .
\ee
Here the operator 
\be\nn
\wt \nabla_\mu = \nabla_\mu - \sqrt{\frac{-\Rho}{2d}}\gamma_\mu
\ee
is the modification of the covariant derivative acting on spinors found quite some time ago in a cosmological supergravity
context~\cite{Townsend:1977qa}. Its distinguishing property is that $[\wt\nabla_\mu,\wt \nabla_\nu]$ vanishes on spinors (but not vector-spinors).
The above equation of motion enjoys the gauge invariance
\bea\nn
\delta\psi_\mu &=& (d+2w) \wt \nabla_\mu \varepsilon\, ,\nn\\[2mm]
\delta\psi^+ &=& (d+2w) (w+1) \varepsilon\, .
\eea
We include the factor $(d+2w)$ to synchronize the component transformations  with the tractor ones $\delta \Psi^M=D^M\Xi$.
Notice  they imply that $\psi^+$ is an auxiliary St\"uckelberg field at generic $w\neq 1$, which can be gauged away 
leaving a massive Rarita--Schwinger field $\psi_\mu$. When $w$ does equal~$-1$, the field $\psi^+$ is gauge inert
and we may impose the additional constraint $\psi^+=0$ (in fact, a careful analysis shows that this field decouples completely
at $w=-1$). That leaves the massless Rarita--Schwinger equation  in AdS with standard gauge invariance
\be\nn
\gamma^{\mu\nu\rho}\wt \nabla_\nu \psi_\rho=0\, ,\qquad \delta \psi_\mu = \wt \nabla_\mu \varepsilon\, .
\ee
Returning to generic $w$, we may rewrite the above equation in the standard massive form
\be\nn
\gamma^{\mu\nu\rho}\nabla_\nu\psi_\rho + m\gamma^{\mu\nu}\psi_\nu = 0\, .
\ee
The integrability conditions for this system imply the usual constraints $\nabla^\mu\psi_\mu$ $=$ $0$ $=$ $\gamma^\mu\psi_\mu$,
and in turn $(\slashed \nabla - m)\psi_\mu=0$. The mass $m$ is here given in terms of weights by the mass-Weyl weight
relationship
\be\nn
m=\sqrt{-\frac{\Rho}{2d}}\ (d+2w)\, .
\ee
Via the spin~3/2 Weitzenbock identity, this implies a wave equation $(\Delta-\mu^2)\psi_\mu=0$
where $\mu^2$ obeys a Weyl weight relationship highly reminiscent of the spin~0 and~1/2 ones above
\be \label{m32}
\mu^2 = -\frac{2\Rho}{d}\Big[\Big(w_{\psi_m}+\frac{d-1}{2}\Big)^2-\frac{d(d-1)}{4}-1\Big]\, .
\ee 	
Here $w_{\psi_m}=w+\frac12$ because, in the St\"uckelberg gauge $X.\Psi=0$, we have
the Weyl transformation rule $\psi_\mu\mapsto \Omega^{w+3/2}\psi_\mu$. We end by observing,
that this result implies Breitenlohner--Freedman type bound for massive gravitini $\mu^2\geq \frac{\!\scalebox{.7}{\Rho}}{2d}\ [d(d-1)+4]$.
Following the procedure outlined at the end of Section~\ref{Tractor Dirac  Equation}, we write~\eqn{rarita} at arbitrary scale
\be
\nn
\gamma^{\mu\nu\rho}\nabla_\nu\psi_\rho + \frac{1}{2}(d+2w) \gamma^{\mu\nu}\Big(\slashed b + \frac{\sqrt{I^2}}{\sigma}\Big)\psi_\nu - (w+1)\gamma^{\mu} b \cdot \psi= 0\,\, .
\ee
To understand the above expression, let us define the Weyl-covariantized Rarita-Schwinger operator
\be
R^{\mu} \equiv  \gamma^{\mu\nu\rho}\nabla_\nu\psi_\rho + \frac{1}{2}(d+2w) \gamma^{\mu\nu}\slashed b \psi_\nu - (w+1)\gamma^{\mu} b \cdot \psi  \,  \, , \qquad \gamma \cdot \psi =0 \, .
\ee
The $b$ contribution covariantizes the Rarita-Schwinger operator with respect to scale transformations
such that $\psi_\mu \mapsto \Omega^{w+\frac32} \psi_\mu$ implies  $R^\mu\mapsto \Omega^{w-\frac32} R^{\mu}$, 
modulo the condition $\gamma\cdot\psi=0$. This operator also follows from the standard Weyl compensator mechanism. As before, the square root factor is the mass term,
and the  prefactor $(d+2w)$ calibrates the mass to the square of the scale tractor and implies the mass-Weyl weight relationship~\eqn{m32}.  In $d=2$, when $w=-\frac d2=-1$, the scale $\sigma$ decouples from
the equation of motion and we obtain the Weyl invariant curved space Rarita-Schwinger equation.

In fact, in arbitrary dimensions $d$ it is possible to write down a Weyl invariant Rarita--Schwinger system~\cite{Deser:1983tm}.
We can obtain that theory from our tractor one as follows:
Consider  a new field equation $\tilde R^{\mu} = R^{\mu} - \frac{1}{d} \gamma^{\mu} (\gamma \cdot R)$ $=$ $0$, or explicitly
\be
\tilde R^{\mu} = \slashed \nabla \psi^{\mu} -\frac{2}{d} \gamma^{\mu} \nabla \cdot \psi + \frac{d+2w}{2}[ \gamma^{\mu\nu}\slashed b \psi_\nu - \frac{2(d-1)}{d} \gamma^{\mu} b \cdot \psi]\, .
\ee
When $w= -\frac{d}{2}$, the scale dependence through the composite gauge field $b$ decouple completely, and we are left with the Weyl invariant Rarita-Schwinger system of~\cite{Deser:1983tm} generalized to arbitrary dimensions
\be \label{weyl32}
\slashed \nabla \psi^{\mu} -\frac{2}{d} \gamma^{\mu} \nabla \cdot \psi = 0 = \gamma\cdot \psi\, .
\ee
We can derive the same results efficiently using  tractors.  This requires  imposing two additional constraints
\be
X . \Psi = 0 \, ,\qquad \Gamma . X\  \Gamma . \Psi = 0 \,,
\ee
which in components read 
\be  
\psi^+ = \chi^+=\gamma \cdot \psi = 0.
\ee
As argued before, at $ w= \frac{-d}{2}$ the compensator field $\sigma $ can be safely eliminated without compromising the Weyl invariance.  At this special value of the weight, the tractorial expression describing Weyl invariant Rarita-Schwinger equation is\footnote{Note that there actually no pole in this expression in six dimensions as evidenced by the component expression~\eqn{weyl32}} 
\be
\tilde R^M = \Gamma . X [R^M - \frac{d-2}{d(d-6)}\Gamma^M (\Gamma . R)]=0 \, ,
\ee 
which in components exactly matches~\eqn{weyl32}.

    \chapter{Supersymmetry and Interactions}
\label{SUSY}
Given a tractor description of spinors and scalars, it is natural to search for 
a supersymmetric combination of the two. Here we study global supersymmetry.
In a curved background, globally supersymmetric theories require a generalization
of the constant spinors employed as parameters of supersymmetry transformations
in flat space. A possible requirement is to search for covariantly constant spinors,
although most backgrounds do not admit such special objects. Focusing on 
conformally flat backgrounds, a more natural condition is to require that the background
possess a Killing spinor $\varepsilon$ defined by
\be\nn
\nabla_\mu\varepsilon = -\sqrt{\frac{-\Rho}{2d}}\ \gamma_\mu\varepsilon\, .
\ee
As a consequence it follows that $\bar\varepsilon\varepsilon$ is constant.
This condition can be neatly expressed in tractors in terms of what we shall call a ``scale spinor''
\be\nn
\Xi=\begin{pmatrix}\varepsilon\\[1mm]\eta\end{pmatrix}\, ,\qquad \Pi_-\Xi=0\, .
\ee
Here $\eta$ is determined by the projective condition. The Killing spinor condition for $\varepsilon$ is
now imposed by requiring the weight $w=0$ tractor spinor $\Xi$ to be tractor parallel
\be\nn
{\cal D}_\mu\Xi=0\, .
\ee
From the scale spinor, we can form the scale tractor as
\be\nn
I^M=\frac{\sigma\, \overline\Xi\Gamma^M\Xi}{\, \overline\Xi \Gamma.X \Xi}\, ,
\ee
which justifies its name.

Having settled upon the global supersymmetry parameters, we specify the field content as a weight $w+1$ scalar~$\varphi$
and a weight $w$ tractor spinor~$\Psi$  subject to
\be\nn
\Pi_+\Psi=0\, .
\ee
We have chosen the tuning between weights of fermionic and bosonic fields in order to preserve supersymmetry.
The supersymmetry transformations are given by\footnote{There is no pole in the fermionic variation at $w=-d/2$; this can be checked explicitly from a component computation.}
\bea
\delta\varphi &=& \Re \Big(  \overline\Xi \Gamma.X \Psi\Big)\, ,\label{susyb} \\ 
\delta \Psi &=& \frac{1}{d+2w} \Big[(\Gamma.D-\frac1\sigma\,  \Gamma.X I.D)\varphi\Big]\, \Xi\, .\label{susyf}
\eea
Here we take $\varphi$ to be real, but make no assumption for reality conditions for the spinors. If the underlying $d$-dimensional
spinors are Majorana, there is no need to take the real part in the supersymmetry transformations of the bosons. For the independent
bosonic and fermionic field components, these transformations amount to
\be\nn
\delta\varphi=\Re(\sqrt2\bar\varepsilon\psi)\, ,\qquad\delta\psi=\Big[\Big(\slashed\nabla+\sqrt{\frac{-2\Rho}{d}}\ (w+1)\Big)\varphi\Big]\varepsilon\, .
\ee
The invariant tractor action for this system is the sum of the Bose and Fermi actions discussed in previous Sections
\be\nn
S=\int \frac{\sqrt{-g}}{\sigma^{d+2w+1}}\, \Big\{\overline\Psi \Gamma.X \, \Gamma^M I^N D_{MN}\Psi + \varphi I.D \varphi\Big\}\, .
\ee
To verify the invariance of this action one first uses the identity
\be\nn
\Gamma.X\,\Gamma^MI^N D_{MN}=\frac{\sigma}{d+2w-2} \Gamma.X\,  \Gamma.D\, ,
\ee
so that
\be\nn
\Gamma.X\  \Gamma^MI^N D_{MN}\delta\Psi=-\Big(\frac\sigma{(d+2w-2)(d+2w)}\Gamma.X \ \Gamma.D \ \Gamma.X \ \frac1\sigma\  I.D\varphi\Big)\Xi\, .
\ee
Then the identity
\be\nn
\Gamma.X\ \Gamma.D=-\frac{d+2w-2}{d+2w+2} \Gamma.D\ \Gamma.X +(d+2w)(d+2w-2)\, ,
\ee
yields
\be\nn
\overline\Psi \Gamma.X\  \Gamma^MI^N D_{MN}\delta\Psi=-(\overline\Psi \Gamma.X\Xi) \ I.D\varphi.
\ee
Comparing the last expression with the bosonic variation in~\eqn{susyb} completes our invariance proof.

\section{Interacting Wess-Zumino Model}

\label{Interactions}

To add interactions, we begin by closing the supersymmetry algebra off-shell with the aid of an auxiliary field.
In curved backgrounds, the square of a supersymmetry transformation yields an isometry as the generalization of
translations in flat space. Therefore we also need to explain how to handle isometries with tractors. On the bosonic field $\varphi$,
the supersymmetry algebra closes without any auxiliary field and the algebra of two supersymmetry transformations is given by
\be\nn
[\delta_1,\delta_2]\varphi = \Re(\overline\Xi_1\Gamma^{MN}\Xi_2) \, D_{MN} \varphi\, .
\ee
The adjoint tractor $\Re(\overline\Xi_1\Gamma^{MN}\Xi_2)$ is an example of what we shall call a ``Killing tractor''~\cite{KT,Gover:2002ay}. Let us make a brief aside to 
describe these objects:
Suppose that $\xi^\mu$ is any vector field. Then we can form a weight $w=1$ tractor
\be\nn
V^M=\begin{pmatrix}0\\[1mm] \xi^m\\[2mm]-\frac1d \nabla_\mu \xi^\mu\end{pmatrix}
\ee
subject to $X.V=D.V=0$. In turn we may build an adjoint tractor
\be\nn
V^{MN}=\frac1{d} D^{[M} V^{N]} =
\begin{pmatrix}
0 & \xi^n & -\frac 1d \nabla_\mu\xi^\mu\\[1mm]
\mbox{a/s} & \nabla^{[m}\xi^{n]}& 
\frac{1}{2d}\Big([\Delta+\Rho]\xi^m \!-\!\frac{d+2}{d}[\nabla^m \nabla_\mu +d\Rho^m_\mu]\xi^m \Big)
\\[3mm]
\mbox{a/s}&\mbox{a/s}&0
\end{pmatrix} \, .
\ee 
The operator 
\be\nn
\frac12 V^{MN} D_{NM} = \xi^\mu{\cal D}_\mu -\frac wd (\nabla_\mu\xi^\mu)\, ,
\ee
may be viewed as a tractor analog of the vector field $\xi^\mu\partial_\mu$.
Notice that acting on weight $w$ scalars, it gives the correct transformation law
for a conformal isometry
\be\nn
\delta \varphi = (\xi^\mu \partial_\mu  -\frac wd[\nabla_\mu\xi^\mu])\varphi\, .
\ee
It is not difficult to verify that $\Re(\overline\Xi_1\Gamma^{MN}\Xi_2)$ corresponds to $V^{MN}$ with
$\xi^\mu$ given by the Killing vector $\sqrt{2}\Re(\bar\varepsilon_1\gamma^\mu\varepsilon_2)$.
This shows that acting on $\varphi$, the supersymmetry algebra closes onto  
isometries\footnote{It could be interesting and natural in our framework to study extensions where the supersymmetry algebra closes onto conformal isometries.}. 

To close the algebra on the fermions we need first to understand how (conformal) isometries act on (tractor) spinors.
In the work~\cite{Gover:2009vc}, the double $D$-operator was related to the generators of ambient Lorentz transformations. This suggests
that, acting on tractors of arbitrary tensor type, we should  introduce the operator
\be\nn
\pounds = \frac 12 V^{MN}\Big[D_{NM}+ {\cal S}_{MN}\Big]\, ,
\ee
where ${\cal S}_{MN}$ are the ambient intrinsic spin generators. On spinors we have
\be\nn
{\cal S}_{MN}=\frac12 \Gamma_{MN}\, .
\ee
Indeed, the transformation rule $\delta\Psi = \pounds \Psi$ for a weight~$w$ tractor spinor $\Psi$ with top slot $\psi$ implies
\be\nn
\delta\psi = (\pounds_\xi  - \frac{w_\psi}{d}[\nabla_\mu\xi^\mu])\psi\, ,
\ee
where the Lie derivative on spinors is $\pounds_\xi\psi = (\xi^\mu\nabla_\mu + \frac14 \gamma^{\mu\nu}[\nabla_\mu\xi_\nu])\psi$.

To obtain a closed, offshell supersymmetry algebra for the fermions we need to introduce auxiliary fields. Since the off-shell bosonic and fermionic
field contents must balance, the details will depend on the dimensionality. Therefore, for simplicity, we now restrict ourselves to a four dimensional 
chiral multiplet with $(z,\psi,F)$ where $z$ and $F$ are complex scalars and $\psi$ is a Majorana spinor. We represent the scalars $z$ and $F$ by  
weight $w+1$ and $w$ tractor scalars with the same names while $\psi$ is the top slot of a weight $w$ tractor spinor $\Psi$  subject to $\Pi_+\Psi=0$.
Notice, this implies that independent spinor field content is characterized by 
\be\nn
\Gamma.X \Psi = \sqrt{2}\, \begin{pmatrix}0\\ \psi\end{pmatrix}\, .
\ee
It is therefore sufficient (and simplifying) to specify the transformation rules of $\Gamma.X\Psi$ in what follows
(note also that the operators $\Gamma.X$ and $\pounds$ commute). 
The supersymnmetry transformations of our tractor chiral multiplet then read
\bea
\delta z\quad\  &=& \overline\Xi \Gamma.X {\cal L} \Psi\, ,\nn\\
\delta (\Gamma.X {\cal L}\Psi) &=&\Gamma.X {\cal L} \Big( F + \frac{1}{d+2w}\Gamma.D z\Big)\Xi\, ,\nn\\
\delta F\quad &=& \frac{-1}{d+2w+2}
\, \overline\Xi \Gamma.D \Gamma.X {\cal L}\Psi\, . 
\eea
The rules for the complex conjugates are given by replacing ${\cal L}\mapsto{\cal R}$ where ${\cal L,R}=\frac12 (1\mp \Gamma^7)$, or explicitly
\be\nn
{\cal L,R}=\begin{pmatrix}L,R & \\ & R,L\end{pmatrix}\, ,\qquad L,R = \frac12 (1\mp \gamma^5)\, .
\ee
In components, these transformation rules agree with the usual ones for a massive Wess--Zumino model in an AdS background~\cite{Ivanov:1979ft,Dusedau:1985uf, Burges:1985qq}.
It is important to note however, since the tractor system treats the massive and massless systems on the same footing, the auxiliary field used here
differs from the standard one by terms linear in the complex scalar $z$.

Closure of the supersymmetry algebra on the scalars can be verified as described above. For the fermions, a tractor Fierz identity is required
\be\nn
{\cal R}\, \Xi_2\overline\Xi_1{\cal R}-(1\leftrightarrow2) = -\frac 18 (\overline\Xi_1\Gamma^{RS}\Xi_2) \ {\cal R}\, \Gamma_{RS}{\cal R} \, .
\ee
After some algebra it follows that
\be\nn
[\delta_1,\delta_2](\Gamma.X {\cal L} \Psi) = \pounds \, \Gamma.X {\cal  L} \Psi\, , \qquad [\delta_1,\delta_2] F = \pounds F\, ,
\ee
proving that the supersymmetry algebra closes. 

Armed with a closed supersymmetry algebra, an invariant action principle is easily obtained in  tractors:
\be
S = \int  \frac{\sqrt{-g}}{\sigma^{d+2w}}(\mathcal{L}_{\rm kin} + \mathcal{L}_{\rm int}) \, ,
\ee
where\\[2mm]
\bea
\mathcal{L}_{\rm kin} &=&  \frac{1}{2\sigma}\bar\Psi\Gamma. XI_M \Gamma_N D^{MN} \Psi - \frac{1}{\sigma}\bar z I . D z + | F|^2  \nn\\
&-& 2a(w+2) \big[(F - \bar F)(z- \bar z) +  a(2w+5)(z -\bar z)^2  \big]  \, , \\[2mm]
\mathcal{L}_{\rm int}  &=& \frac{1}{2}\bar\Psi \Gamma. X(\mathcal{L}W'' + \mathcal{R}\bar W'') \Psi\nn\\
&-&   FW' - \bar F \bar W' - 2a (w+1)(zW' + \bar z \bar W') -6a (W + \bar W)  \, .\nn\\
\eea
Notice the appearance of the weight $-1$ scalar $a$ in the weight  $2w$ lagrangian density.  At arbitrary scales $$ a =\frac{\sqrt{I.I}}{2\sigma}\, ,$$
while in a canonical choice of scale it is related to the four dimensional cosmological constant by $12a^2=-\Lambda$.

The action is split into a kinetic  and interacting pieces. The latter depends linearly on a weight $2w+1$ holomorphic potential~$W=W(z,\sigma,a)$. Since our tractor theories describe massive and massless excitations uniformly in terms of weights, the above splitting is not the canonical
one into a massless action plus potential terms, but rather uses the freedom of the function~$W$ to split the action into  free (generically massive) and interacting pieces. At $w=-2$. the Lagrangian density ${\cal L}_{\rm kin}$, expressed in components at the canonical choice of scale recovers the massless part of the supersymmetric AdS Wess--Zumino model quoted in~\cite{Dusedau:1985uf,Burges:1985qq}.


    \chapter{Conclusions and Future Outlook}
\label{Conclusions}
In this thesis, we treated  unit invariance, the freedom to locally choose a unit system, at par with diffeomorphism invariance and demanded that all physical theories respect this principle.  We showed that conformal geometry naturally harbored unit invariance in addition to diffeomorphism invariance and, therefore, is perfectly suited to describe physics in a unit invariant way.  In particular, we employed tractor calculus, the natural calculus on a conformal manifold, to construct physical theories with manifest unit invariance. A direct result of this approach is the unification of massive, massless, and even partially massless excitations in a single geometrical framework.  

Our construction clarifies the origins of masses in field theories, especially in curved spaces where it is necessary to survey all possible couplings to the background geometry and scale (rather than just the compensating mechanism alone) to obtain the theories we have described in the thesis.  Moreover, we learned that masses are related to Weyl weights producing an elegant mass-Weyl weight relationship.  Consequently, masses then measure the response of physical quantities under the changes of unit systems.  The reality of weights in this mass-Weyl weight relationship naturally lead to the Breitenlohner-Freedman bounds in spaces with constant curvature.

Our techniques would be helpful in the formulation of fundamental physics that remains to be understood.  For instance, tractors  can be employed to construct conformally invariant higher spin theories in curved backgrounds with conformal isometries.   Then relying on our ambient approach, conformal field theories  can  lead to a deeper understanding of  the AdS/CFT principle.  Tractors can also be used to naturally encapsulate the ideas arising in physics with two times.  Let us start by discussing the applications of tractor calculus to AdS/CFT correspondence. 

\section{AdS/CFT Correspondence}
In its simplest form, the AdS/CFT correspondence is a relation between a theory with gravity in d-dimensions  to a conformal field theory without gravity in $(d-1)$-dimensions.  Since the introduction of the AdS/CFT correspondence, conformal field theories have taken a central stage in the arena of physics. Hence, the construction of a conformally invariant field theory is of fundamental importance.  At a classical level, a clever mechanism already exists to construct such theories in constant curvature backgrounds.  First, the theory is coupled to a metric in a Weyl invariant way.  Having rendered the theory Weyl invariant, the metric is held constant while the fields are allowed to transform.  This clever maneuver turns the originally Weyl invariant theory in to a conformally invariant one.  However, there is an alternative, more efficient method based on tractor calculus.  The starting point is a tractor Weyl invariant theory via dilaton coupling.  Upon choosing a special value for the weight, the dilaton decouples leaving behind a conformally invariant theory.  Applying this method, we constructed several familiar conformally invariant theories in addition to a new conformally invariant spin two theory. 

Admittedly, our results are all classical, but in fact the greatest impact of our ideas may be to  quantization.
Classically, a scale invariant interacting theory can be constructed rather efficiently using our tractor techniques, 
but the scale symmetry does not survive the process of quantization and, therefore, is anomalous.  
We believe that the Weyl anomaly, although arising in the quantized theory, can be explained by classical geometry.  
Moreover, we surmise that anomaly is in part due to the Riemannian background where matter
dynamics take place.  If instead, we work with a conformal equivalence classes $[g_{\mu\nu},\sigma]$, the anomaly can be better understood.   In other words,  the anomaly has a classical analogue that can be described using our techniques with an extra scale field.  It's not so unusual for a quantum phenomenon to have a classical analogue.  An example is a classical equivalent of quantum tunneling process~\cite{Novello:1992tb}.

The ideas coming from the AdS/CFT correspondence~\cite{Maldacena:1997re,Witten:1998qj,Gubser:1998bc,Aharony:1999ti} where renormalization group flows can be formulated holographically~\cite{deBoer:1999xf} and scale or Weyl anomalies become geometric~\cite{Henningson:1998gx} further strengthen our belief that the anomaly can be resolved by our approach.  Indeed conformal geometry computations of Poincar\'e metrics~\cite{Fefferman} can be used to obtain physical information about these anomalies~\cite{deBoer:1999xf}. At the very least the tractor techniques provide a powerful machinery
for these types of computations, and optimally can provide deep insights into the AdS/CFT correspondence itself.

 \section{2T Physics}
The presence of conformal symmetry in physics, the mathematical structure of M theory \cite{Witten:1995ex}, and several other phenomenon hints at the existence of two time-like directions.  The ideas emerging from 2-T \cite{Bars:1998ph} physics suggests that there actually are (d+2)-dimensions of which we only see $d$-dimensional ``shadows."  With two times and an extra spatial dimension, we can arrange our $d$-dimensional theories as a manifestation of some theory in two higher dimensions.  Ideas of 2-T physics are in harmony to our  results where we found many $d$-dimensional excitations from a $d+2$-dimensional tractor framework.  For instance, we found that massless, massive, and even conformally invariant spin one theories sprang from the same source.  It is similar to cleverly arranging different two dimensional shadows on the wall, of a three dimensional object  that brings out the relationship among various shadows.

\begin{figure}[h]
\centering
\includegraphics[width=140mm, height=100mm]{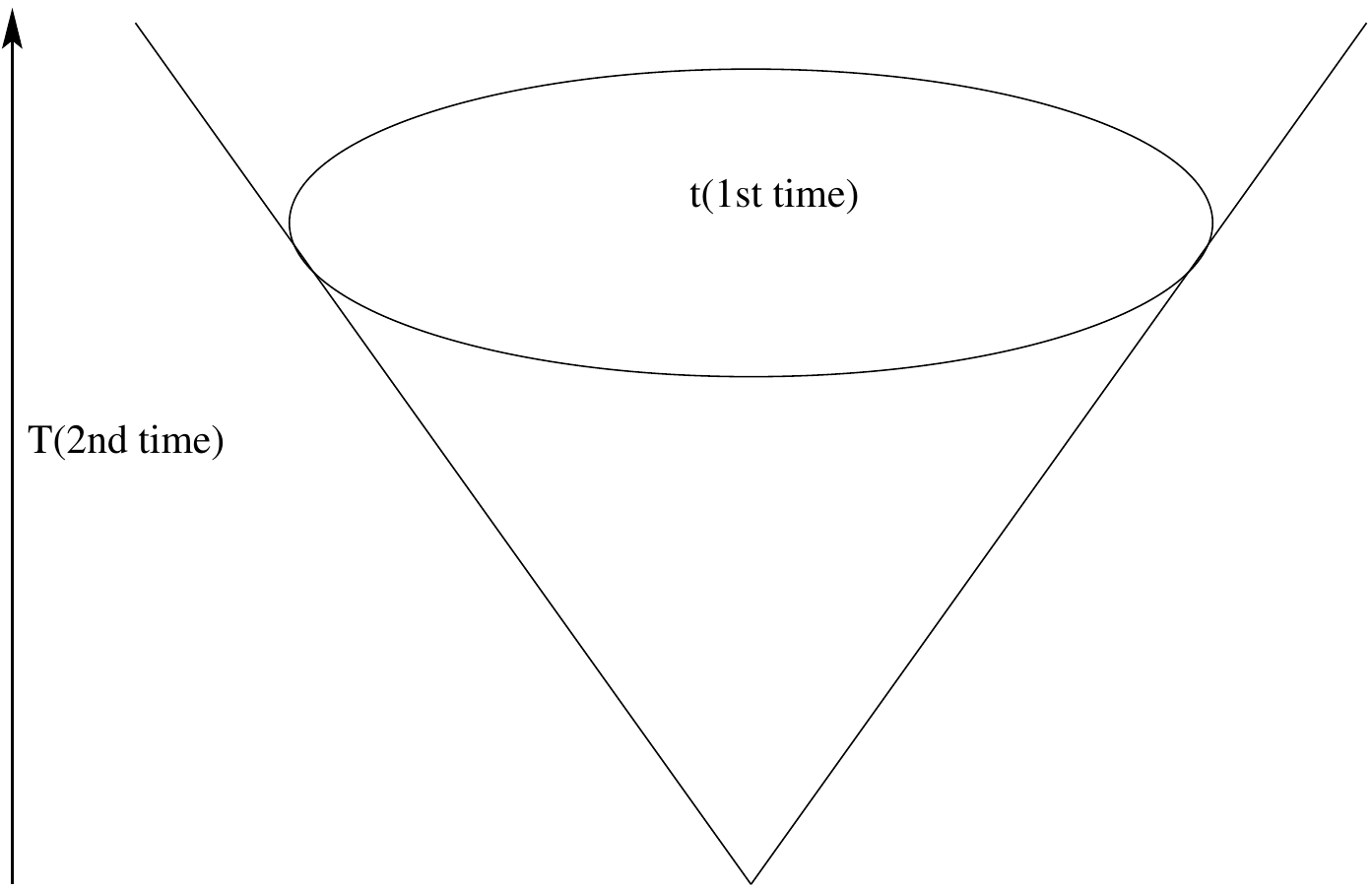}
\caption{The ambeint space with two times: one on the conformal manifold; the other in the ambient space. }
\label{2T}
\end{figure}

Physics with two times can be described by our tractor framework.  Ambiently, we naturally have two time dimensions in addition to an extra spatial dimension.  The first time is the conventional one, which lives on the $d$-dimensional manifold, while the other time lives in the ambient space as shown in Figure~\ref{2T}.  In fact, we speculate that the scale field, $\sigma$ is very much related to the second time.  
Physics with two times can be neatly described by a $d+2$-dimensional tractor vectors that naturally arranges all lower dimensional theories in a single geometrical framework. 

    %

    \appendix

    %
    %

\setcounter{equation}{0}
\renewcommand{\theequation}{\thechapter.\arabic{equation}}

\chapter{Compendium of Weyl Transformations}
\label{Compend}

In this Appendix, we list some elementary Weyl transformations
that follow from the metric transformation 
\be
g_{\mu\nu} \mapsto \Omega^{2} g_{\mu\nu}\, .\label{Weyl}
\ee
Throughout, we denote
\be
\Upsilon_\mu \equiv \Omega^{-1}\partial_\mu\Omega\, .
\ee
Firstly the volume form transforms as
\be
\sqrt{-g} \mapsto \Omega^d \sqrt{-g}\, .
\ee
Since the vielbein, Levi-Civita connection, and $rho$-tensor belong to the tractor connection, their transformations follow from (5) and (6). Explicitly,
 \bea
e_{\mu}{}^m \;&\mapsto& \Omega e_{\mu}{}^m , \nn\\[3mm]
\omega_\mu{}^m{}_n&\mapsto&\omega_\mu{}^m{}_n-\Upsilon^m e_{\mu n}+\Upsilon_n e_{\mu}{}^m\, ,\nn\\[3mm]
\Rho_{\mu n}\  &\mapsto&\Omega^{-1}\Big(\Rho_{\mu n}-\nabla_\mu \Upsilon_n +\Upsilon_n \Upsilon_\mu -\frac12 e_{\mu n} \Upsilon.\Upsilon\Big)\, .
\eea 
Similarly, the transformations of the Cotton--York and Weyl tensors follow from the transformation of the tractor curvature 
\be
{\cal F}_{\mu\nu}= [{\cal D}_{\mu}, {\cal D}_{\nu}]  \mapsto  U{\cal F}_{\mu\nu} U^{-1} \,\label{curvfo} \, ,
\ee
with
\be
{\cal F}_{\mu\nu}=
\begin{pmatrix}
0&0&0\\[2mm]
C_{\mu\nu}{}^m&W_{\mu\nu}{}^m{}_n&0\\[2mm]
0&-C_{\mu\nu n}&0
\end{pmatrix}\, .
\ee
Here the Cotton--York tensor equals
\be
C_{\mu\nu}{}^m=\nabla_\mu \Rho_\nu{}^m-\nabla_\nu \Rho_\mu{}^m\, ,
\ee
and explicitly equation~\eqn{curvfo} says 
\bea
W_{\mu\nu}{}^m{}_n &\mapsto& W_{\mu\nu}{}^m{}_n\, , \nn \\[2mm]
C_{\mu\nu}{}^m\  &\mapsto& \Omega^{-1}\Big(C_{\mu\nu}{}^m -W_{\mu\nu}{}^m{}_n \Upsilon^n \Big)\, .
\eea
{\it When} scalars, vectors, and one forms of weight~$w$ transform as 
\bea
f &\mapsto& \Omega^w f \, ,   \nn \\[2mm] 
v^{\mu} &\mapsto& \Omega^w v^{\mu}\, , \nn \\[2mm]
\omega_{\mu} &\mapsto& \Omega^w \omega_{\mu}  \, ,
\eea
their covariant derivatives transform as follows:
\bea
\nabla_\mu f\ &\mapsto& \Omega^w[(\nabla_\mu  + w \Upsilon_\mu) \,f ]\, ,\\[2mm]
\nabla_\mu v^\nu &\mapsto & \Omega^w[(\nabla_\mu  +w\Upsilon_\mu)\, v^\nu+ \Upsilon_\mu v^\nu - \Upsilon^\nu v_\mu +
\delta^\nu_\mu\ \Upsilon. v] \, ,\\[2mm]
\nabla_\mu \omega_\nu&\mapsto& \Omega^w[(\nabla_\mu  +w\Upsilon_\mu)\, \omega_\nu -\Upsilon_\mu \omega_\nu -\Upsilon_\nu \omega_\mu + g_{\mu\nu}\,  \Upsilon. \omega] \, .
\eea


\chapter{Tractor Component Expressions}

\label{Tractor_components}

In this Appendix we tabulate the component expressions for some of the more important tractor
quantities used in the text. Table~\ref{Tracod0} gives  the tractor covariant derivative acting on scalars, tractor vectors 
and rank two symmetric tractor tensors while Table~\ref{Tracod1} gives the tractor Laplacian on the same objects. Finally, Table~\ref{Tracod2} gives the Thomas
$D$-operator acting on scalars and tractor vectors.

\newpage
\begin{landscape}

\begin{table}[h]

\begin{tabular}{|c|}
\hline \\[3mm]
${\cal D} _{\mu}f = \nabla_{\mu} f$ 
 \\   [6mm] \hline

\\[3mm]
${\cal D}_\mu V^M=
\begin{pmatrix}
\nabla_\mu V^+ - V_\mu\\[2mm]
\nabla_\mu V^m +\Rho_\mu{}^m V^+ + e_\mu{}^m V^-\\[2mm]
\nabla_\mu V^- - \Rho_\mu{}^n V_n 
\end{pmatrix}$  
\\ [13mm] \hline

\\[3mm]
${\cal D}_\mu V^{(MN)}  =
\left(
\begin{array}{ccc}

\nabla_\mu V^{++}-2V_\mu{}^+ &
\nabla_\mu V^{+n}-V_\mu{}^n+ \Rho _\mu{}^nV^{++}+e_\mu{}^nV^{+-}  &
\nabla_\mu V^{+-}  -V_\mu{}^- - \Rho _{\mu r}V^{r+} \\[2mm]

{\rm Symm} &
\nabla_\mu V^{mn}+ 2 \Rho _\mu{}^{(m}V^{n)+}+2e_\mu{}^{(m}V^{n)-} &
\nabla_\mu V^{m-}+  \Rho _\mu{}^mV^{+-}+e_\mu{}^mV^{--}- \Rho _{\mu r}V^{rm} \\[2mm]
 
{\rm Symm} & 
{\rm Symm} &
\nabla_\mu V^{--}-2 \Rho _{\mu r}V^{r-}
\end{array}
\right) $
\\[15mm]  \hline

\end{tabular}
\caption{Tractor covariant derivative acting on scalars, tractor vectors, and rank two symmetric tractor tensors.}
\label{Tracod0}
\end{table}
\newpage

\begin{table}[h]
\begin{tabular}{|c|}
\hline \\[3mm]
${\cal D} ^2f = \Delta f $ 
\\ [6mm] \hline

\\[3mm]
${\cal D}^2V^M=
\left(
\begin{array}{c}
(\Delta- \Rho )V^+ -2\nabla^nV_n -dV^-\\[1mm]
\Delta V^m + 2 \Rho ^{mn}(\nabla_nV^+ - V_n)+\nabla^{m} \Rho V^+ +2 \nabla^mV^- \\[1mm]
(\Delta -  \Rho )V^- -2 \Rho ^{mn}\nabla_mV_n- \Rho ^{mn} \Rho _{mn}V^+ - V^n\nabla_n  \Rho 
\end{array}
\right) $
\\ [13mm]  

\hline \\[3mm]
${\cal D}^2 V^{(MN)} =$ 

$\left (  
{\tiny
\begin{tabular}{ c | c | c }

\multirow{3}{*} {$ \Delta V^{++} - 2(2\nabla. V^{+} - V_r{}^r + \Rho V^{++}+dV^{+-}) $}& 
$ \Delta V^{+n}-2\nabla.V^{n}- \Rho V^{+n}-4 \Rho_r^n V^{r+}$  &
$\Delta V^{+-}-2\nabla.V^{-}-2 \Rho V^{+-}-dV^{--} $\\

& $-(d+2)V^{n-}+ \Rho ^{sn} \nabla_sV^{++}+2\nabla^nV^{+-} $ & 
$+2 \Rho _{sr}V^{sr}- \Rho _r^s\nabla_sV^{+r}- \Rho _r^s \Rho _s^rV^{++}$ \\  

&$+\nabla_s( \Rho ^{sn}V^{++}) $  &
$-\nabla_s( \Rho _r{}^sV^{r+})$ \\ [3mm] \hline

\multirow{3}{*}{Symm}& 
$\Delta V^{mn}+2 \Rho ^{s(m}\nabla_sV^{+n)}+2 \Rho ^{sm} \Rho _s^nV^{++} $&
$ \Delta V^{m-}+ \Rho ^{sm}(\nabla_sV^{+-}-4V_s{}^--2 \Rho _{sr}V^{r+})$\\ 

 &+$4\nabla^{(m}V^{n)-}+4 \Rho ^{mn}V^{+-}+2\eta^{mn}V^{--} $   &
 $+2\nabla^mV^{--}- \Rho _r{}^s(\nabla_sV^{rm}+ \Rho _s^rV^{m+})- \Rho V^{m-}$\\
 
&$-4 \Rho _r{}^{(m}V^{|r|n)}+2\nabla_s( \Rho^{s(m}V^{+n)}) $& 
$+\nabla_s( \Rho ^{sm}V^{+-})-\nabla^s( \Rho _{sr}V^{rm})$ \\[3mm] \hline

 \multirow{2}{*}{Symm} &  &$\Delta V^{--}-2 \Rho _{rs}(\nabla^sV^{r-}+ \Rho ^{rs}V^{+-}- \Rho _t^sV^{tr})$  \\
 & Symm &$-2 \Rho V^{--}-2V^{r-}\nabla^s \Rho _{sr}$ \\
\end{tabular}}
\right)$ 
\\[20mm] \hline
\end{tabular} 

\caption{Tractor Laplacian acting on scalars, tractor vectors, and rank two index symmetric tractor tensors.}
\label{Tracod1}
\end{table}

\begin{table}[h]

\begin{tabular}{|c|}

\hline \\[3mm]
$
D^M f=
\begin{pmatrix}
(d+2w-2) w f\\[2mm]
(d+2w-2)\nabla^m f\\[2mm]
-(\Delta+w\Rho)f
\end{pmatrix} $ \\[15mm]

\hline \\[5mm]

$
D^MV^N \!=\!

 \left(\!\!
\scalebox{.95}{\mbox{$
 {\tiny
 \begin{array} {ccc}  
w(d+2w-2) V^+ &  w(d+2w-2)V^n  & w(d+2w-2)V^-   \\[3mm]
 (d+2w-2)(\nabla^mV^+ -V^m) &  (d+2w-2)(\nabla^mV^n+ \Rho ^{mn}V^++\delta^{mn}V^-) & (d+2w-2)(\nabla^mV^- - \Rho _{n}^m V^n)\\[3mm]
 -(\Delta+(w-1) \Rho )V^+ + 2\nabla.V + dV^- &
-(\Delta+w \Rho ) V^n - 2 \Rho ^n_m(\nabla^mV^+ - V^m)-V^+\nabla^n \Rho  -2 \nabla^nV^- & 
-(\Delta+(w-1)  \Rho )V^- +2 \Rho ^{mn}\nabla_mV_n+ \Rho ^{mn} \Rho _{mn} V^++ V^n\nabla_n  \Rho 
 \end{array}} $}}
\!\! \right) $ \\[15mm]
 
 \hline

\end{tabular}
\caption{Thomas D-operator acting on functions and tractor vectors.}\label{Tracod2}
\end{table}

\end{landscape}

\chapter{Projectors}

\label{projectors}

Equipped with a choice of scale $\sigma$, it is possible to convert any component expression 
into a tractor one using a projector technique. The expressions obtained this way are often
unwieldy, so that it is better to work directly in tractors from first principles. Nonetheless, we sketch here
a few details of the construction.

Given the a scale $\sigma$, we build the weight zero scale tractor 
\be
I^M=\frac1d\, D^M\sigma\, ,
\ee
from which we construct a null vector
\be
Y^M=\frac{1}{X\cdot I}\Big(I^M-\frac{I\cdot I}{2\, X\cdot I}\, X^M\Big)\, ,\qquad Y\cdot Y=0\, .
\ee
obeying
\be
Y\cdot X=1\, .
\ee

Armed with the null vectors $X^M$ and $Y^M$
we can now define the top, middle and bottom slots
of a tractor vector $V^M$ by 
\be V^+\equiv X\cdot V\ ,\qquad V^m\equiv
(V^M-Y^M X\cdot V-X^M Y\cdot V)\, , \qquad V^-\equiv Y\cdot V\, .
\ee 
Tautologically then,
\be
Y^M=\begin{pmatrix}1\\ \ 0\ \\ 0\end{pmatrix}\, ,
\ee
and the tractor metric decomposes as a sum of projectors
\be
\eta^{MN}=X^M Y^N + \Pi^{MN} + Y^M X^N\, .
\ee
For an arbitrary tractor vector $V^M$ we denote
\be
\widehat V^M \equiv \Pi^M_N V^N = \begin{pmatrix}0\\ V^m\\ 0\end{pmatrix}\, ,
\ee
and similarly for any tractor tensor. This method allows us to extract the components 
of any tractor.

As a simple example, we can relate the usual Maxwell curvature to the tractor one ${\cal F}^{MN}=D^M V^N-D^N V^M$
(see also~\eqn{FF})
using projectors
\be
\widehat{\cal F}^{MN}=\Pi^M_R\Pi^N_S {\cal F}^{RS} = 
\begin{pmatrix}0&0&0\\[1mm]
0&(d+2w-2)F^{mn}&0\\[1mm] 0&0&0\end{pmatrix}\, .
\ee

\chapter{Symmetric Tensor Algebra}

\label{symmetric}

Computations involving symmetric tensors with high or many  different ranks are greatly facilitated
using the algebra of gradient, divergence, metric, trace and modified wave operators first introduced by Lichnerowicz~\cite{Lichnerowicz:1964zz} and
systemized in~\cite{Damour:1987vm,Hallowell:2005np,Hallowell:2007zb} (see~\cite{Labastida:1987kw,Vasiliev:1988xc,Duval,Duval1} for other studies). For completeness, we review the key formul\ae\ here.
The key idea is to write symmetric tensors in an index-free notation
using commuting coordinate differentials, so that a symmetric rank $s$ tensor $\varphi_{(\mu_1\ldots \mu_s)}$ becomes
\be
\Phi = \varphi_{\mu_1\ldots \mu_s} dx^{\mu_1}\cdots dx^{\mu_s}\, .
\ee 
In this algebra,  it is no longer forbidden to add tensors of different ranks.
Then there are seven distinguished operators mapping symmetric tensors to symmetric tensors:
\begin{enumerate}
\item[${\bf N}$\;] --Counts  the number of indices \be {\bf N}\ \Phi = s \Phi \, .\ee
\item[${\bf tr}$\ ] --Traces over  a pair of indices \be {\bf tr}\  \Phi = s(s-1)\varphi^\rho{}_{\rho\mu_3\ldots \mu_s} dx^{\mu_3}\cdots dx^{\mu_s}\, .\ee
\item[${\bf g}$\;] --Adds a pair of indices using the metric
\be {\bf g}\  \Phi = g_{\mu_1\mu_2}\varphi_{\mu_3\ldots \mu_{s+2}} dx^{\mu_1}\cdots dx^{\mu_{s+2}}\, .\ee
\item[${\bf c}$\;] --The Casimir of the $sl(2)$ Lie algebra obeyed by the triplet $({\bf g,N}+\frac d2,{\bf tr})$ 
\be
{\bf c}={\bf g}\ {\bf tr} -{\bf N}({\bf N}+d-2)\, .
\ee
\item[${\bf div}$\ ] --The symmetrized divergence 
\be
{\bf div} \ \Phi = s \nabla^\rho \varphi_{\rho\mu_2\ldots \mu_s} dx^{\mu_2}\cdots dx^{\mu_s}\, .
\ee
\item[${\bf grad}$] --The symmetrized gradient
\be {\bf grad} \ \Phi = \nabla_{\mu_1}\varphi_{\mu_2\ldots \mu_{s+1}} dx^{\mu_1}\cdots dx^{\mu_{s+1}}\, .\ee
\item[$\square$\;] --The constant curvature Lichnerowicz wave operator
\be
\square=\Delta+\frac{2\Rho}{d}\, {\bf c}\, .
\ee
\end{enumerate}
The calculational advantage of these operators is the algebra they obey
\bea
[{\bf N},{\bf tr}]=-2{\bf tr}\, ,\quad
[{\bf N},{\bf div}]=-{\bf div}\, ,\quad
[{\bf N},{\bf grad}]={\bf grad}\, ,\quad
[{\bf N},{\bf g}]=2{\bf g}\, ,\nn
\eea
\bea
[{\bf tr},{\bf grad}]=2{\bf div}\, ,\quad
[{\bf tr},{\bf g}]=4{\bf N}+2d\, ,\quad [{\bf div,g}]=2{\bf grad}\, ,\nn\eea
\bea
[{\bf div,grad}]=\square-\frac{4\Rho}{d}\, {\bf c}\, .
 \eea
All other commutators vanish. In particular the Lichnerowicz wave operator is central!


\chapter{Doubled Reduction}

\label{Doubled Reduction}

There is a rather explicit relationship between the tractor theories we write down
(when the background is conformally flat) and  log radial reductions from massless theories in $(d + 1)$-dimensional flat Lorentzian spaces to massive ones in $d$-dimensional (anti) de Sitter spaces~\cite{ Biswas:2002nk, Hallowell:2005np}. The mathematical underpinning of this relationship is the connection between conformal and projective structures~\cite{Bailey:1994}. 
In fact, the independent field content of our tractor models, (which typically inhabit the top and middle slots of tractor fields) 
corresponds precisely to that of these log radial reductions. Therefore, for completeness, in this Appendix we present the
log radial reduction for spinor theories (see~\cite{Hallowell:2005np} for the bosonic case).

For spinor theories there are two possible reduction schemes  depending on how Dirac matrices in adjacent dimensions are handled.
One approach is to write the $(d+1)$-dimensional Dirac matrices as $\Gamma^M = (-i\Gamma^{d+1}\gamma^m,\Gamma^{d+1})$ 
where $\gamma^m$ then obey a $d$-dimensional Clifford algebra. Alternatively, beginning with the $d$-dimensional Dirac matrices
$\gamma^m$, the $(d+1)$-dimensional Dirac matrices are then  built by doubling, namely $\Gamma^M=(\sigma_z\otimes\gamma^m,\sigma_x\otimes \boldsymbol 1)$. Irreducibility of the spinor representations produced in these ways depends on the dimensionality~$d$ and metric signature, but
these details do not concern us here.  Either approach can be related to tractors, but the latter approach (which we adopt here) produces a doubled set of equations 
for which this relationship is simplest--we shall call it a ``doubled reduction''.

We start by writing the flat metric in log radial coordinates 
\bea
ds^{2}_{\rm flat} = dX^M G_{MN} dX^N = e^{2u}(du^2+ ds_{\rm dS} ) = E^A \eta_{AB} E^B \, . \label{logr}
\eea
Here, we consider the case where the underlying manifold is de Sitter for reasons of simplicity (the corresponding AdS computation  is not difficult either).
Notice that the indices $M, N,\ldots$  and $A, B,\ldots$  are not tractor indices but rather $(d + 1)$-dimensional curved and flat ones, respectively,  
while $\mu, \nu,\ldots$  and $m, n \ldots$ are $d$-dimensional. Note that in this background~$\Rho = \frac{d}{2}$. The $(d+1)$-dimensional vielbeine are 
\be
E^5 = e^udu \, , \qquad E^m = e^ue^m \, ,
\ee
where $e^m$ is the $d$-dimensional de Sitter vielbein.
The  de Sitter spin connection $\omega_{mn}$ obeys 
\be
de^m + \omega^{mn} \wedge e_n = 0 \, ,
\ee
in terms of which the $(d + 1)$-dimensional flat spin connection reads 
\be
\Omega^{mn} = \omega^{mn} \,, \qquad \Omega^{5n} = -e^n  \, .
\ee
We use the label $5$ to stand for the reduction direction so that 
$A = (m,5)$ and  $M = (\mu,u)$. The $(d + 1)$-dimensional Dirac matrices and covariant derivative acting on spinors are\bea
\Gamma^A &=&  \Big(  \begin{pmatrix} \gamma^m & 0 \\ 0 & - \gamma^m \end{pmatrix}\, , \,  \begin{pmatrix} 0 & \boldsymbol 1 \\ \boldsymbol 1 &0 \end{pmatrix} \Big) \, ,\nn\\[2mm]
\nabla_M &=& 
\Big( \begin{pmatrix}  \nabla_{\mu} & \frac{1}{2}\gamma_{\mu} \\ -\frac{1}{2}\gamma_{\mu} & \nabla_{\mu} \end{pmatrix} \, , \, \begin{pmatrix} \partial_u&0\\0&\partial_u \end{pmatrix} \Big)\, ,
\eea
where $\nabla_\mu$ on the right hand side of the second line is the de Sitter covariant derivative and $\gamma^m$ are the $d$-dimensional Dirac matrices. For future use we call
\be
\wt \nabla_\mu = \begin{pmatrix}  \nabla_{\mu} & \frac{1}{2}\gamma_{\mu} \\ -\frac{1}{2}\gamma_{\mu} & \nabla_{\mu} \end{pmatrix}\, .
\ee
Acting on spinors $[\wt \nabla_\mu,\wt \nabla_\nu]=0$.
The Dirac conjugate is defined as
$
\overline \Psi = i\Psi^\dagger \Gamma^0$.
We now have enough technology to start the doubled reduction.  As a warm up, we consider a  Dirac spinor
\be
\Psi = \begin{pmatrix}   \psi \\ \chi \end{pmatrix}\, ,
\ee
with action principle 
\be
S_{1/2}= - \int \sqrt{-G}\  d^{d+1}\!X\,  \overline \Psi \Gamma^M \nabla_M \Psi \, .
\ee
Upon making the field redefinition 
\be
\Psi =  e^{-\frac{ud}{2}}  \begin{pmatrix}   \psi \\ \chi \end{pmatrix} \, ,
\ee
all explicit $u$ dependence disappears and the action becomes 
\be
S_{1/2}= - \int du d^d x \sqrt{-g}\,  (\bar \psi, -\bar \chi) \begin{pmatrix} \slashed \nabla & \partial _u \\ \partial_u & - \slashed \nabla    \end{pmatrix}  \begin{pmatrix}   \psi \\ \chi \end{pmatrix} \, .
\ee
Varying this action, defining the weight by (and therefore the mass {\it a l\'a} Scherk and Schwarz~\cite{Scherk:1979zr}) $$\partial_u = w + \frac{d}{2}\, ,$$ 
and redefining $\chi$ by multiplying it with $\sqrt{2}$,  we recover the pair of equations of motion following from the tractor computation~\eqn{dirac2}
in a canonical choice of scale.
Having warmed up on spin~1/2, our next task is the  spin~3/2 doubled reduction.

\noindent
The $(d+1)$-dimensional  Rarita-Schwinger action is 
\be
S_{3/2}= - \int \sqrt{-G}\,  d^{d+1}\!X \, \overline \Psi_M \Gamma^{MNR} \nabla_N \Psi_R \, .
\ee
Using the  log radial coordinates~\eqn{logr}  and rescaling fields
\be
\Psi_M = e^{-\frac{u(d-2)}{2} } \begin{pmatrix}   \psi_M\\ \chi_M \end{pmatrix} \, ,
\ee
the action becomes devoid of explicit $u$ dependence and implies the doubled set of equations of motion
\be
\Big[
\begin{pmatrix} \gamma^{\mu\nu\rho}&0\\0&-\gamma^{\mu\nu\rho}\end{pmatrix}
\wt \nabla_{\nu} 
-  \begin{pmatrix}0& \gamma^{\mu\rho}\\\gamma^{\mu\rho}&0\end{pmatrix} (\partial_u-\frac{d}{2}+1)
 \Big]\begin{pmatrix} \psi_\rho\\ \chi_\rho\end{pmatrix} = 0\, .
\ee
(Note that the $\Psi_u$ variation simply implies the constraint that is a consequence of the above equation.)
Redefining the $\chi$ field as before, and equating $\partial_{\mu} =w + \frac{d}{2}$, the above equation agrees with the tractor Rarita-Schwinger one~\eqn{rarita} explicated in the canonical choice of scale.

\chapter{Homogeneous Spaces}
\label{Homogeneous}
In this Appendix, we explain the idea of a homogenous space and construct Euclidean space as a homogenous one.  
A homogeneous space X for a group G is a space on which G acts in a transitive way: $G \times X \mapsto X$.  The action of a group G is transitive on a set X if for any $x, y \in X$, there is some $g \in G$ such that $gx = y$.  Said plainly, the group action on an arbitrary point $x \in X$ must fill up X.  For instance, consider the group $\mathbb{R}^n$, which acts on itself by translations: for $v \in \mathbb{R}^n$, $T_v : \mathbb{R}^n \mapsto \mathbb{R}^n$ by $T_v(w) = v + w$. For convenience we pick $v = 0$ and can move to any point in $\mathbb{R}^n$  by a suitable translation filing up $\mathbb{R}^n$. Therefore, $\mathbb{R}^n$ acts transitively on itself.  The action of $O(2)$ on $\mathbb{R}^n$, on the other hand, is not transitive because we have multiple disjoint orbits:  for instance, we can not go from origin to a point  on a unit circle. However, if $O(2)$ acts on a particular circle, the action is transitive. 

It is also true that any transitive action of a group G on X is equivalent to the left action of G  on a coset space $G/H$, where H is called the stabilizer group. Therefore, $G/H $
is a homogenous space on which G acts transitively.  We omit the proof as it is not our main concern here.

As an example, we construct Euclidean space as a homogenous space. The transformations that keep the fundamental properties such as length and angles between vectors preserved, $ E^n \mapsto E^n$,  are rotations and translations.  These transformations send a point $x \in E^n$ to
\be
x \mapsto Ax + b \, ,
\ee
where $ A \in O(n)$ and $ b \in R^n$. Together, they generate the Euclidean group E(n)\footnote{It is the group $G$ for this particular geometry.}, which can be represented by $(n+1)\times (n+1)$ matrices of the form
\be
\begin{pmatrix} A & b \\ 0 & 1   \end{pmatrix} \,.
\ee
An arbitrary point $x$ in Euclidean space can then be represented as an $(n+1)$ component column vector:
\be
\begin{pmatrix} x \\1  \end{pmatrix} \,.
\ee
For convenience, let us pick a vector
\be
\vec{x} = \begin{pmatrix} 0 \\1  \end{pmatrix}\, ,
\ee
and study its transformations under the group action. The point transforms as 
\be
\vec{x} \mapsto \begin{pmatrix} A & b \\ 0 & 1  \end{pmatrix} \vec{x} =  \begin{pmatrix} b \\ 1  \end{pmatrix} \,.
\ee
If $b=0$, it is clear that $\vec{x}$ remains fixed with 
\be
H_{0}= \begin{pmatrix} A & 0 \\ 0 & 1   \end{pmatrix}\, ,\qquad A \in O(n) \, .
\ee

There is nothing special about the point we picked; we could have equivalently chosen any point in $E^n$ with the same result. It is rather easy to see why this holds.  We can bring an arbitrary point $x \in E^n$ to coincide with the origin by translating it by $-x$.  Explicitly, it is done by $t^{-1}_x$, where 
\be
t_x = \begin{pmatrix}  1 & x \\ 0 & 1 \end{pmatrix} \, .
\ee
The isotropy group of $x$ is then given by
\be
H_{x} = t_x H_{0} t^{-1}_x \,,
\ee
which is also isomorphic to $O(n)$.  Once we know the stabilizer group, H,  of the original group, E(n), we can redefine Euclidean space as a quotient of $ E(n) / O(n)$, which is the homogenous model for the Euclidean space.  

In summary, the idea of a homogenous space is to model a particular geometry. It is a group theoretic approach where a geometry is best understood by studying its symmetry group $G$ and a subgroup $H \subset G$, which captures the invariant features of the geometry.  Basically $H$ carries redundant information about the geometry.
Therefore,
\be
G/H =  \frac{\mbox{Rigid Motions}}{\mbox{Stabilizer of a point}}  \, ,
\ee
is a homogenous space on which G acts transitively.  We can also model inhomogeneous spaces by locally modeling them as homogenous ones.  We can use Euclidean space as a local model for Riemannian manifolds, a hyperbolic space for a Pseudo-Riemannian manifold, and a conformal sphere for conformal manifolds, the later being more important to us.  We will treat conformal manifolds as homogenous spaces which allows us to construct a tractor connection. 

\chapter{Tractor Identities}
\label{Identities}
In this Appendix, we provide some useful tractor identities that can  simplify tractor calculations.  Let us start by presenting two equivalent ways of defining the double-$D$ operator:
\be
\begin{array}{c}
D^{MN} = \frac{2}{d+2w-2}X^{[N}D^{M]}\, ,   \hspace{1in} [X^M, D^N]  = 2D^{MN} -(d+2w)h^{MN}  \, .
\end{array}
\ee
Although, we will primarily use the first definition, the second one is very useful for moving the canonical tractor past the Thomas $D$-operator.  The second definition when combined with the first one produces yet another way of expressing the double-$D$ operator
\be
D^{MN} = \frac{2}{d+2w+2}D^{[M}X^{N]}  \, .
\ee
Combining all three definitions of the double-$D$ operator yields a very powerful identity:
\be
D_MX_N = X_MD_N + (d+2w)( D_{MN} + \eta_{MN})  \, .
\ee

Next, we give identities involving the Thomas-$D$ and the double-$D$ operators acting on a weight one scalar $\sigma$, a weight one tractor vector $X$, and an arbitrary function of weight $w$:
\bea
D_M \, \sigma = d\, I_M \, ,\hspace{2in} D_MX_R = d\, \eta_{MR} \, ,\\[3mm]
X \cdot D = w(d+2w-2) \,, \hspace{1in}    D \cdot X = (d+w)(d+2w+2) \,,  \\[3mm]
D_{MN}\,\sigma  = 2 X_{[N} I_{M]} \, ,    \hspace{1.5in} D_{MN}X_R = 2 X_{[N}\eta_{M]R} \,, \\[3mm]
D_{MN}D^N = (w-1)D_M \, , \hspace{1.5in}  X^MD_{MN} = w X_N  \,.
 \eea
Some important cummutators involving the scale field $\sigma$ are
\bea
[D^m , \sigma] &=& (d+2w)I^M + 2I_ND^{NM} \,, \\[2mm]
[I.D,\sigma^k] &=& I^2\sigma^{k-1} k(d+2w+k-1) \,.
\label{Ids}
\eea

Next, we present identities most relevent in performing Fermi tractor calculations.  To that end, we recycle the Dirac slash notation and define it as the contraction of $\Gamma$ with any tractor operator T:
\be
\slashed T = \Gamma \cdot T \, .
\ee

Identities involving the commutator of tractor operators are provided below:
\bea
[\slashed D , \sigma] &=& 2\Gamma^MI^ND_{MN} +(d+2w) I \cdot \Gamma \,,  \\[3mm]
[\slashed X \,, \,I \cdot D] &=& 2 \Gamma^M I^ND_{MN} - (d+2w) I \cdot \Gamma \,,   \\[3mm]
[\slashed X\,, \,\slashed D] &=& 2\Gamma^M \Gamma^N D_{MN} -(d+2w)(d+2) \,, \\[3mm]
\{\slashed X\,, \,\slashed D\} &=& [\slashed X\,, \,\Gamma \cdot D]+2 D \cdot X  \,, \\[3mm] 
[\slashed X, D_M] &=& 2\Gamma^ND_{NM} - (d+2w)\Gamma_M  \,, \\[3mm]
[\slashed X , \Gamma_{RS}] &=& 4 X_{[R} \Gamma_{S]} \, .
\eea

Some other useful identities follow:
\bea
\Gamma^M \Gamma^N D_{MN}  &=&  2(w - \frac{1}{d+2w-2}  \slashed X \, \slashed D ) \, ,\\[3mm]
\Gamma^M I^N D_{MN}  &=& \frac{1}{d+2w-2}(\sigma \slashed D - \slashed X I \cdot D )\, ,   \\[3mm]
\Gamma^M I^N D_{MN}  \slashed D  &=& -\frac{1}{d+2w-4}\slashed X I \cdot D \slashed D \, ,  \\[3mm]
\Gamma^M I^N D_{MN}  \slashed X &=& \frac{1}{d}( \sigma(d+2) - \slashed X \slashed I ) \, , \\[3mm]
\Gamma^M I^N D_{MN} \Pi_{\pm} &=& -\Pi_{\mp} \Gamma^M I^N D_{MN} \, , \\[3mm]
\slashed X \Pi_{\pm} &=& 2\sigma - \Pi_{\mp}\slashed X \, , \\[3mm]
\slashed X \slashed D &=&(d+2w)(d+2w-2) - \frac{d+2w-2}{d+2w+2} \slashed D \slashed X \,.
\eea
Armed with the above tractor identities, many tractor calculations can be carried out rather easily. 


\chapter{Curvature Identities}
\label{curvature}
Covariant derivatives commute on scalars; such is not the case for spinors.  We present the commutator of covariant derivatives acting on spinor as well as a vector spinor.  
In addition, we also provide the Weitzenbock forumula for both spinors and vector spinors. 

The most useful identities are
\bea
[\nabla_\mu, \nabla_\nu] \psi &=& \frac{P}{d} \gamma_{\mu\nu} \psi \,, \\[2mm]
[\nabla_\mu, \nabla_\nu] \psi_{\rho} &=& \frac{P}{d} \gamma_{\mu\nu}\psi_{\rho} + \frac{4P}{d} g_{\rho [\mu}\psi_{\nu]} \,, \\[2mm]
\slashed \nabla^2 \psi  &=&[ \Delta- \frac{P}{2}(d-1)]\psi \,, \\[2mm]
\slashed \nabla^2 \psi_{\rho}  &=&[ \Delta- \frac{P}{2}(d-1) -\frac{2P}{d}]\psi_{\rho} = [ \Delta- \frac{P}{2d}(d^2-d+4)]\psi_{\rho} \, .
\eea
Since the $\Rho$ tensor is the trace adjustment of the Ricci tensor, we can rewrite Einstein's equations in terms of the $\Rho$ tensor, thus providing a relationship 
between the $\Rho$ tensor the cosmological constant .  The explicit formulas follow:
\bea
R_{\mu\nu} &=& (d-2)P_{\mu\nu} + P g_{\mu\nu} \,, \\[2mm]
R &= &2(d-1) P  \,, \\[2mm]
G_{\mu\nu} &=& (d-2)(P_{\mu\nu} - Pg_{\mu\nu}) \, .
\eea
When $G_{\mu \nu} = -\Lambda g_{\mu \nu} $, the cosmological constant 
\be
\Lambda = \frac{(2-d)R}{2d}\, =\,\frac{P(2-d)(d-1)}{d} \, .
\ee

\chapter{Tractor Spinors}
\label{Covariant_Spinor}
In this section we give explicit formulas for tractor covariant derivative acting on spinors.  Most of these formulas are given in~\cite{Branson}, but for the sake of completeness, we provide them below. 

\bea
\mathcal{D}_\mu \Psi &=& \partial_\mu \Psi + \frac{1}{4} \Gamma^{MN} \mathcal{A}_{\mu MN}\Psi \nn \\[2mm]
&=&\nabla_\mu \Psi + \begin{pmatrix} 0 & \frac{1}{\sqrt{2}}\gamma_\mu \\ -\frac{1}{\sqrt{2}}P_{\mu r}\gamma^r  & 0 \end{pmatrix}  \Psi \nn\,. \\[2mm]
\mathcal{D}_\mu \Psi&=&\begin{pmatrix}   \nabla_\mu \psi + \frac{1}{\sqrt{2}}\gamma_\mu \chi \\[2mm] \nabla_\mu \chi -\frac{1}{\sqrt{2}}\slashed P_\mu \psi \end{pmatrix}\, ,
\eea
where $ \nabla _\mu = \partial_\mu + \frac{1}{4}\omega_{\mu mn}\Gamma^{mn}$ is the standard covariant derivative acting on spinors. 
The tractor Laplacian acting on spinor yields 
\be
\mathcal{D}^2 \Psi = \begin{pmatrix} (\Delta - \frac{1}{2}P)\psi + \sqrt{2}  \slashed \nabla \chi   \\[2mm]
(\Delta - \frac{1}{2}P)\chi  - \sqrt{2}\slashed P_\mu \nabla^\mu   \psi   - \frac{1}{\sqrt{2}}   (\slashed \nabla P) \psi
\end{pmatrix} \, .
\ee
We can also calculate the Thomas-$D$ operator acting on spinors.  Slot by slot, the result is
\bea
D^+ \Psi &=& w(d+2w-2) \begin{pmatrix} \psi \\ \chi \end{pmatrix} \, ,\\[4mm]
D^m \Psi &=& (d+2w-2) \begin{pmatrix}   \nabla^m \psi + \frac{1}{\sqrt{2}}\gamma^m \chi \\[2mm] \nabla^m \chi -\frac{1}{\sqrt{2}}\slashed P^m \psi \end{pmatrix} \,,    \\[4mm]
D^- \Psi& =& \begin{pmatrix} -[\Delta +(w- \frac{1}{2})P] \psi - \sqrt{2}  \slashed \nabla \chi   \\[2mm]
-[\Delta +(w- \frac{1}{2})P]\chi  + \sqrt{2}\slashed P_\mu \nabla^\mu   \psi  + \frac{1}{\sqrt{2}}   (\slashed \nabla P) \psi
 \end{pmatrix} \, .
\eea

\section*{Tractor-Vector Spinor}
\label{Covariant_Tractor_Spinor}
The action of the tractor covariant derivative on a tractor spinor is a straightforward generalization of its action on spinors given in the previous Appendix.   
\bea
\mathcal{D}_\mu \Psi^R &=& \partial_\mu \Psi^R + \frac{1}{4} \Gamma^{MN} \mathcal{A}_{\mu MN} \Psi^R + \mathcal{A}_\mu{}^R{}_T \Psi^T \\[3mm]
\mathcal{D}^{n} \Psi^R &=& \begin{pmatrix}   \nabla^n \psi^R + \frac{1}{\sqrt{2}}\gamma^n \chi^R \\[2mm] \nabla^n \chi^R -\frac{1}{\sqrt{2}}\slashed P^n \psi^R \end{pmatrix} + \left(
\begin{array}{c}
- \Psi^n\\[1mm]
P^{nr} \Psi^+ + e^{nr}\Psi^-\\[1mm]
 - P^n{}_r \Psi^r
\end{array}
\right) \\[3mm]
&=&  \begin{pmatrix} \nabla^n \psi^+ + \frac{1}{\sqrt{2}}\gamma^n \chi^+ - \psi^n    \\[2mm]
\nabla^n \chi^+ - \frac{1}{\sqrt{2}}\slashed{P}^n \psi^+ - \chi^n  \\[2mm] \hline
\nabla^n \psi^r + \frac{1}{\sqrt{2}}\gamma^n \chi^r + P^{nr}\psi^+ + e^{nr}\psi^- \\[2mm]
\nabla^n \chi^r - \frac{1}{\sqrt{2}}\slashed P^n \psi^r + P^{nr}\chi^+ + e^{nr}\chi^- \\[2mm] \hline
\nabla^n \psi^- + \frac{1}{\sqrt{2}}\gamma^n \chi^- -P^n{}_r \psi^r \\[2mm]
\nabla^n \chi^- - \frac{1}{\sqrt{2}}\slashed{P}^n\psi^- -P^n{}_r \chi^r
\end{pmatrix} \, .
\eea
Once we know the action of tractor covariant derivative on a tractor spinor, it's an elementary exercise to computer the tractor laplacian. Explicitly, it is given by

\be
\mathcal{D}_n \mathcal{D}^n \Psi^R = \left(  \begin{smallmatrix} [\Delta - \frac{3}{2}P]\psi^+ + \sqrt{2}[\slashed \nabla \chi^+ - \gamma . \chi] -2\nabla . \psi - d\psi^- \\[2mm]
[\Delta - \frac{3}{2}P]\chi^+ -\sqrt{2}\slashed P^n[\nabla_n \psi^+ -\psi_n] -2\nabla . \chi - d\chi^- - \frac{1}{\sqrt{2}}(\slashed \nabla P)\psi^+ \\[2mm] \hline
[\Delta -\frac{3}{2}P]\psi^r + \sqrt{2}[\slashed \nabla \chi^r + \slashed P^r \chi^+ + \gamma^r \chi^-] + 2P^{nr}(\nabla_n\psi^+) +(\nabla^r P)\psi^++2\nabla^r\psi^- - P^{nr}\psi_n\\[2mm]
[\Delta -\frac{3}{2}P]\chi^r - \sqrt{2}\slashed P^n[\nabla_n \psi^r +  P_n{}^r \psi^+ + e_n{}^r \psi^-] + 2P^{nr}(\nabla_n\chi^+) +(\nabla^r P)\chi^++2\nabla^r\chi^-- P^{nr}\chi_n-\frac{1}{\sqrt{2}}(\slashed \nabla P)\psi^r \\[2mm]\hline
[\Delta -\frac{3}{2}P]\psi^- + \sqrt{2}[\slashed \nabla \chi^- - \slashed P_r \chi^r] - 2 P_{nr}\nabla^n \psi^r - (\nabla_r P)\psi^r - P_{nr}P^{nr}\psi^+ \\[2mm]
[\Delta -\frac{3}{2}P]\chi^- - \sqrt{2}\slashed P^n[\nabla_n \psi^- -  P_{nr} \psi^r] - 2 P_{nr}\nabla^n \chi^r - (\nabla_r P)\chi^r - P_{nr}P^{nr}\chi^+ -\frac{1}{ \sqrt{2}}(\slashed \nabla P) \psi^-
 \end{smallmatrix}  \right)
\ee

    %
    %


    %
    %


\addcontentsline{toc}{chapter}{References}
\bibliographystyle{utphys}
\begin{singlespacing}
  \bibliography{mybib}
\end{singlespacing}


\end{document}